\newcommand{\mat}[4]{\left(\begin{array}{cc}{#1}&{#2}\\{#3}&{#4}
\end{array}\right)}
\def\dmsol{\Delta m^2_{21}}
\def\dmatm{\Delta m^2_{31}}
\def\mlt{m^2_{\tilde{L}}}
\def\tr{\text{Tr}}

\def\BR{{\rm BR}}
\def\openone{\leavevmode\hbox{\small1\kern-3.3pt\normalsize1}}

\documentclass{JHEP3}

\usepackage{amssymb,amsmath,graphicx}
\usepackage{rotating,bm}
\usepackage{longtable}

\vspace{10cm}

\title{\vspace*{2cm} Running effects on neutrino parameters and $\bm{\ell_i
\rightarrow \ell_j \gamma}$ predictions in the triplet-extended
MSSM}

\author{F. R. Joaquim\thanks{On leave from the ``Centro de F\'{i}sica Te\'orica
de Part\'{i}culas (CFTP)'', Lisbon, Portugal. E-mail:
Filipe.Joaquim@cern.ch}
\\CERN, Theory Division\\ CH-1211 Geneva 23, Switzerland}

\preprint{CERN-PH-TH/2009-244}

\abstract{We investigate the renormalisation group effects induced
on neutrino mass and mixing parameters in a triplet-extended minimal
supersymmetric standard model where a vector-like pair of
hypercharge $\pm 1$ triplet superfields is added. We first rederive
the one-loop renormalisation group equation for the effective
neutrino mass operator and, for the case in which this operator
originates solely from the decoupling of the triplets, the
corresponding equations for neutrino masses, mixing parameters and
CP-violating phases. We compare our results with the ones obtained previously, and quantify
the importance of the RG induced
corrections to neutrino observables by means of numerical examples.
In the second part of the paper, we study the correlation of the
model's predictions for the lepton flavour violating processes $\ell_i
\rightarrow \ell_j \gamma$ with the measured neutrino mass squared
differences and mixing angles. We also emphasize the r\^ole played
by the unknown reactor neutrino mixing angle $\theta_{13}$ and the
Dirac CP-violating phase $\delta$. We point out that, if $\tan\beta$
is large, the results obtained in the commonly made approximations
may deviate significantly from the ones following from solving
numerically the relevant set of renormalisation group equations and
using the exact one-loop formulae for the decay rates.}


\begin{document}

\section{Introduction}
One of the most puzzling and longstanding problems in particle
physics concerns the explanation of the observed fermion mass and
mixing patterns. The interest around this subject has been renewed
with the confirmation that neutrinos are massive. This motivated
intense activity towards the search for an answer to the question on
how neutrinos acquire their tiny mass. From the theoretical point of
view, the idea that neutrino masses are suppressed by a large energy
scale has become the most popular one. This is
the basis of the well-known seesaw mechanism~\cite{seesawI} for
neutrino mass generation. The phenomenology of seesaw-inspired models
has been widely studied in the
literature~\cite{Numodels} with the goal of explaining the results
provided by neutrino oscillation
experiments~\cite{GonzalezGarcia:2007ib,Schwetz:2008er}.

If the mechanism generating neutrino masses operates at high-energy
scales, renormalisation group (RG) effects may induce important
corrections to the neutrino parameters. Several analyses devoted to
the study of the neutrino parameter running have supported this
expectation~\cite{Chankowski:2001mx}. The RG corrections to the
effective neutrino mass operator depend crucially on the properties
of the neutrino mass spectrum, on the absolute neutrino mass scale
and on the size of the $\tau$ Yukawa coupling. Therefore, the
running of the neutrino parameters is expected to be enhanced in
supersymmetric (SUSY) models with large $\tan\beta$. For instance,
neutrino mixing may be strongly augmented by the running from high
to low energies~\cite{Balaji:2000au} so that bimaximal neutrino mixing at
high scales can be made compatible with low-energy neutrino data by
including RG corrections~\cite{Antusch:2002hy}. Ultimately, with the
gradual increasing of neutrino data precision, even small RG effects
may turn out to be important for neutrino mass and mixing model
building.

Within seesaw-based scenarios, the RG flow above the decoupling
scale of the heavy {\em seesaw mediators} has to be properly
accounted for, including possible threshold effects due to the
presence of different mass scales. This has been done in
Refs.~\cite{Antusch:2005gp} in the framework of the so-called type I
seesaw mechanism~\cite{seesawI} where the heavy degrees of freedom
responsible for the suppression of neutrino masses are singlets
under the Standard Model (SM) gauge group. The results of such
studies are model dependent since the RG effects above and between
the thresholds depend on the Yukawa couplings of the heavy neutrino
singlets with the leptons and Higgs fields, which cannot be
reconstructed from low-energy data. Hence, the impact of the RG
corrections strongly depends on the structure of the unknown
fundamental couplings which is encoded in the effective neutrino
mass operator. The impossibility of reconstructing in a model-independent
way the high-scale neutrino sector parameters from low-energy measurements
of masses and mixing angles is the main problem of the type I seesaw
mechanism.

An alternative version of the seesaw mechanism (usually denoted as
type II or triplet seesaw) relies on the presence of heavy triplet
states~\cite{seesawII}. Its SM version requires a single scalar
triplet to generate mass for the three light neutrinos. In the
minimal supersymmetric standard model (MSSM) a vector-like pair of
hypercharge $\pm 1$ triplet superfields is demanded to ensure
anomaly cancelation and holomorphicity of the superpotential. The
analysis of the RG effects on neutrino parameters in the SM type II
seesaw has been presented in Ref.~\cite{Chao:2006ye}. A more
complete study covering both the SM and SUSY cases can be found in
Ref.~\cite{Schmidt:2007nq}. However, the RGEs for the
triplet-extended MSSM (TMSSM) derived in that paper differ from the
ones obtained in Refs.~\cite{Rossi:2002zb} and
\cite{Borzumati:2009hu}.

The couplings of the {\em seesaw mediators} to the SM lepton
doublets and/or Higgses may induce lepton flavour violation (LFV) in
the soft SUSY-breaking Lagrangian, even if the mechanism which
breaks SUSY is flavour blind~\cite{Borzumati:1986qx,Rossi:2002zb}.
If large enough, such LFV effects can drastically increase the rates
of LFV processes (like radiative charged-lepton decays $\ell_i
\rightarrow \ell_j \gamma$), which are otherwise suppressed to
levels well beyond the sensitivity of future experiments.
Complementarity studies between low-energy neutrino physics and LFV
decay searches have been carried out in the context of the SUSY type
I~\cite{Raidal:2008jk} and type
II~\cite{Rossi:2002zb,Joaquim:2006uz,Hirsch:2008gh} seesaw
mechanisms. The main difference between these two approaches is that
the triplet seesaw is much more predictive when it comes to
establishing a connection between low-energy neutrino physics and
LFV decay searches.

In this work we will investigate several aspects of the RG running
of neutrino parameters (masses, mixing angles and CP-violating
phases) in the TMSSM, where neutrino masses are suppressed via the
type II seesaw mechanism. The impact of the RG effects on
predictions for the branching ratios of the LFV decays $\ell_i
\rightarrow \ell_j \gamma$ will be also discussed and illustrated
with several examples. The layout of the paper is as follows: in
Section~\ref{sec2} we derive the RGE for the dimension-five
effective neutrino mass operator in the framework of an SU(5)
grand-unified model in which the heavy-triplet superfields are
naturally embedded. The general form of the RGEs for the neutrino
masses and mixing matrix is obtained in Section~\ref{sec3}.
Section~\ref{sec4} is devoted to the study of the neutrino parameter
running in the pure type II seesaw case. Namely, in
Section~\ref{sec41} we obtain the RGEs for the neutrino masses,
mixing angles and CP-violating phases (including approximate
analytical expressions), pointing out the discrepancies between our
results and those previously obtained in Ref.~\cite{Schmidt:2007nq}.
Some numerical examples are presented in Section~\ref{sec42}. In the
second part of the paper we discuss predictions of the considered
model for the LFV radiative decays $\ell_i \rightarrow \ell_j
\gamma$ taking into account the latest neutrino oscillation data. We
begin in Section~\ref{sec51} by presenting the rates of these decays
obtained with the help of the frequently made approximation which
neglects the RG running of the neutrino sector parameters, treats in
a simplified way the running of the slepton mass matrices and uses a
simplified formula for the $\ell_i \rightarrow \ell_j \gamma$ decay
rates. In Section~\ref{sec52} these results are compared with the
ones of an improved approximation in which the running of the
neutrino parameters is taken into account and with the results
obtained by solving numerically the full set of RGEs and by
computing the decay amplitudes using the exact one-loop formulae.
This allows us to quantify how much the approximate results of
Section~\ref{sec51} deviate from the more accurate approaches. In
particular, we show that in some cases the splitting of slepton
masses generated by the RG running can also be important.


\section{RGE for the effective neutrino mass operator}
\label{sec2}

Let us consider a supersymmetric N=1 Yang-Mills model with a
superpotential of the form:
\begin{equation}
\label{WYM}%
W=\frac{Y^{ijk}}{3!}{\Phi_i}{\Phi_j}{ \Phi_k}+ \frac{\mu^{ij}}{2}{
\Phi_i}{ \Phi_j}+\frac{\mathcal{O}^
{abcd}}{4!}{ \Phi_a}{ \Phi_b}{ \Phi_c}{\Phi_d}\,, %
\end{equation}
where the chiral superfields ${\Phi_i}$ contain a complex scalar
$\phi_i$ and a two-component fermion $\psi_i$, which transform as a
representation $R_i=R_i^1\otimes...\otimes R_i^n$ of the gauge group
$G=G_1\times ...\times G_n$. The first two terms in the above
superpotential are the ordinary Yukawa and mass terms and $\mathcal{O}$ is
a non-renormalisable operator suppressed by the inverse of a large
mass scale. Provided that higher-dimensional non-renormalisable
operators only appear in the superpotential, then the SUSY
non-renormalisation theorem still holds~\cite{Weinberg:1998uv}.
Consequently, the operator $\mathcal{O}$ can be renormalised taking
into account only wave-function renormalisation. Using the one-loop
anomalous-dimension matrices for the chiral
superfields~\cite{Barbieri:1982nz}
\begin{equation}
\label{gammas}%
\gamma_{i}^{(1)j}= \frac{1}{32\pi^2}\left[Y_{ipq}Y^{jpq}-4\delta_i^j
\sum_k g_k^2 C_k(i)\right]\;\;,\;\; Y_{ipq}\equiv (Y^{ipq})^\ast\,,
\end{equation}
and following for instance Refs.~\cite{Falck:1985aa}, one can write
the one-loop RGE for the operator $\mathcal{O}$ as\footnote{From now
on we will denote by $\dot{X}$ the derivative of the quantity $X$
with respect to $t=\ln(\Lambda/\Lambda_0)$, where $\Lambda$ is the
renormalisation scale and $\Lambda_0$ is a fixed but arbitrary
reference mass scale.}:
\begin{equation}
\label{rgeO}%
\dot{\mathcal{O}}^{abcd}=\mathcal{O}^{abcf}\gamma_{f}^{(1)d}+
(a\leftrightarrow d) +(b\leftrightarrow d)+(c\leftrightarrow d)\,.
\end{equation}
The quantities $g_k$ are the gauge coupling constants of the
sub-groups $G_k$ of $G$ and $C_k(i)$ denotes the corresponding quadratic
Casimir invariant of the irreducible representation of $\Phi_i$.

We now consider an extension of the MSSM where a vector-like pair of
triplet supermultiplets $T$ and $\bar{T}$ transforming under the
${\rm SU(3)}_c \times {\rm SU(2)} \times {\rm U(1)}_Y$ SM gauge
group as $T\sim(1,3,1)$ and $\bar{T}\sim(1,3,-1)$ is added. In a
grand-unified theory (GUT) these triplet states may
be part of the gauge group representation. For instance, in the
SU(5) GUT, $T$ and $\bar{T}$ are part of the $\bm{15}$ and
$\overline{\bm{15}}$ representations, respectively. In this case,
one has $\bm{15}=T\oplus S \oplus Z$ where $S\sim (6,1,-2/3)$ and
$Z\sim (3,2,1/6)$ under the SM gauge group.

\noindent Below the GUT scale, the superpotential reads:
\begin{eqnarray}
W& =& W_0 + W_T + W_{S,Z}\,,  \nonumber \\
W_0 & = & Y_e  e^c H_1  L
+Y_d d^c H_1  Q + Y_u  u^c Q  H_2  + \mu H_2 H_1\,, \nonumber \\
W_T&=& \frac{1}{\sqrt{2}}(Y_{T} L T L +\lambda_1 H_1 T H_1 +
\lambda_2 H_2 \bar{T} H_2)
+M_T {\rm Tr}[(i\sigma_2)T(i\sigma_2)\bar{T}] \label{WT}\,, \nonumber \\
W_{S,Z} & =& \frac{1}{\sqrt{2}}Y_S d^c S d^c + Y_Z  d^c  Z L + M_Z
Z\bar{Z}+M_S S\bar{S}\,, \label{su5b}
\end{eqnarray}
where $L$ ($Q$) is the lepton (quark) doublet supermultiplet and
$e^c$, $d^c$ and $u^c$ are the charged-lepton and quark singlet
supermultiplets. The MSSM superpotential is denoted by $W_0$ while
$W_T$ ($W_{S,Z}$) contains the couplings of $T,\bar{T}$ $(S,Z)$ with
the MSSM superfields, including the corresponding mass terms. We
adopt the SU(2) representation for the triplet
superfields\footnote{The representations of $S$ and $Z$ can be found
in Ref.~\cite{Rossi:2002zb}.}
\begin{equation}
T = (i \sigma_2) \frac{\sigma_i \,T_i}{\sqrt{2}}=
\mat{{T^0}}{ -\dfrac{T^{+}}{\sqrt{2}}}
{-\dfrac{{T^+}}{\sqrt{2}}}{-{T^{++}}} , ~~~~
\bar{T} =  (i \sigma_2) \frac{\sigma_i \,\bar{T}_i}{\sqrt{2}}=
\mat{{\bar{T}^{- -}}}{ -\dfrac{\bar{T}^{-}}{\sqrt{2}}}
{-\dfrac{\bar{T}^-}{\sqrt{2}}}{-{\bar{T}^{0}}}\,.
\end{equation}
To keep the discussion as general as possible, we will consider that
an effective neutrino mass operator of the
form~\cite{Weinberg:1979sa} :
\begin{equation}
\label{Wnu}%
W_\nu=\frac{1}{2}\mathcal{O}^\nu_{ij} \epsilon_{ab}\epsilon_{cd}
L_i^a H_2^b L_j^c H_2^d\,,
\end{equation}
is also present in the superpotential. In the above equation, $i,j$
and $a,b,c,d$ are family and SU(2) indices, respectively. Notice
that we neglect possible effective dimension-four K\"ahler operators
of the type $(L_i. H_2)(L_j. \bar{H}_1)/M^2$ or $(L_i.
\bar{H}_1)(L_j .\bar{H}_1)/M^2$ (where $M$ is some very large scale)
which could also give rise to an effective neutrino mass term after
electroweak symmetry breaking (EWSB). The phenomenology of
K\"ahler-generated neutrino masses (including the RG flow of
neutrino parameters) has been studied in Refs.~\cite{Casas:2002sn}.
Here we assume that these contributions are irrelevant because they
are suppressed by extra $1/M$ factors compared with the contributions
of the operators included in $W_\nu$.

The RGEs for all coupling and mass parameters in (\ref{su5b}) can be
found in Refs.~\cite{Rossi:2002zb} and \cite{Borzumati:2009hu}.
We have recomputed the full set
of RGEs and found complete agreement with the results obtained in the latter
reference\footnote{The RGEs obtained in Ref.~\cite{Borzumati:2009hu} show minor
differences with respect to those of \cite{Rossi:2002zb}. See \cite{Borzumati:2009hu}
for more details.}. Using
Eq.~(\ref{rgeO}) we obtain the one-loop RGE for
$\mathcal{O}^\nu$ in (\ref{Wnu}) as
\begin{equation}
\label{rgek}%
\dot{\mathcal{O}}^\nu_{ij}=2\,\gamma_{H_2}^{(1)H_2}\mathcal{O}^\nu_{ij}
+ \mathcal{O}^\nu_{ik} \gamma_{L_k}^{(1)L_j} + \gamma_{L_k}^{(1)L_i}
\mathcal{O}^\nu_{kj}\,.
\end{equation}
The one-loop anomalous dimensions for $L$ and $H_{1,2}$ can be computed
from Eqs.~(\ref{gammas}) and (\ref{su5b}), yielding the result:
\begin{eqnarray}
\label{gammaLiLj}%
16\pi^2\,\gamma_{L_i}^{(1)L_j}&=&\left[Y_e^\dag
Y_e^{}+\underline{3\,Y_T^\dag Y_T^{}}+\underline{3\,Y_Z^\dag
Y_Z^{}}\right] _{ij}-\left(\frac{3}{10}g_1^2 +
\frac{3}{2}g_2^2\right)\delta_i^j\,, \\
\label{gammaH1}%
16\pi^2\,\gamma_{H_1}^{(1)H_1}&=&\tr(Y_e^\dag Y_e^{}+3\,Y_d^\dag
Y_d^{})+\underline{3\,|\lambda_1|^2}-\frac{3}{10}g_1^2-
\frac{3}{2}g_2^2\,,\\
\label{gammaH2}%
16\pi^2\,\gamma_{H_2}^{(1)H_2}&=&3\,\tr(Y_u^\dag
Y_u^{})+\underline{3\,|\lambda_2|^2}-\frac{3}{10}g_1^2-
\frac{3}{2}g_2^2\,,
\end{eqnarray}
where the underlined terms are those absent from the MSSM. These new
contributions to the wave-function renormalisation of $L$ and
$H_{1,2}$ originate from the one-loop supergraphs shown in
Fig.~\ref{fig1}. It is straightforward to check
that setting $Y_{T,S,Z}=0$ and $\lambda_{1,2}=0$ in Eq.~(\ref{rgek}) one recovers the
MSSM RGE for $\mathcal{O}^\nu$ obtained in
Refs.~\cite{Chankowski:1993tx}.
\FIGURE[!ht]{ \label{fig1} \caption{One-loop (non MSSM) supergraphs
relevant for wave-function renormalisation of the lepton (left
diagram) and Higgs (right diagram) doublet superfields under the
superpotential (\ref{su5b}).}
\begin{tabular}{cc}
\includegraphics[width=5.8cm]{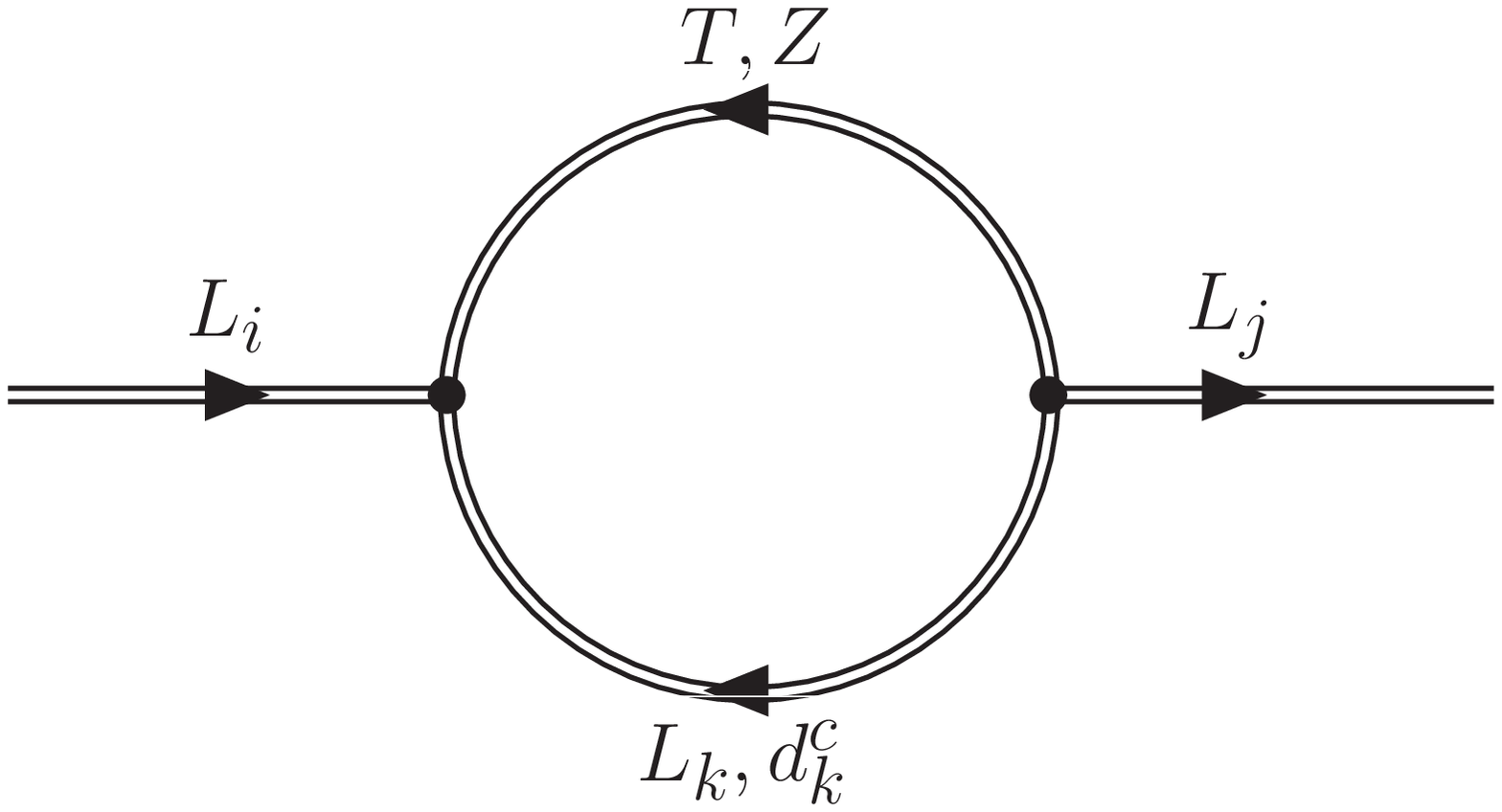}
&\hspace*{0.5cm} \includegraphics[width=5.8cm]{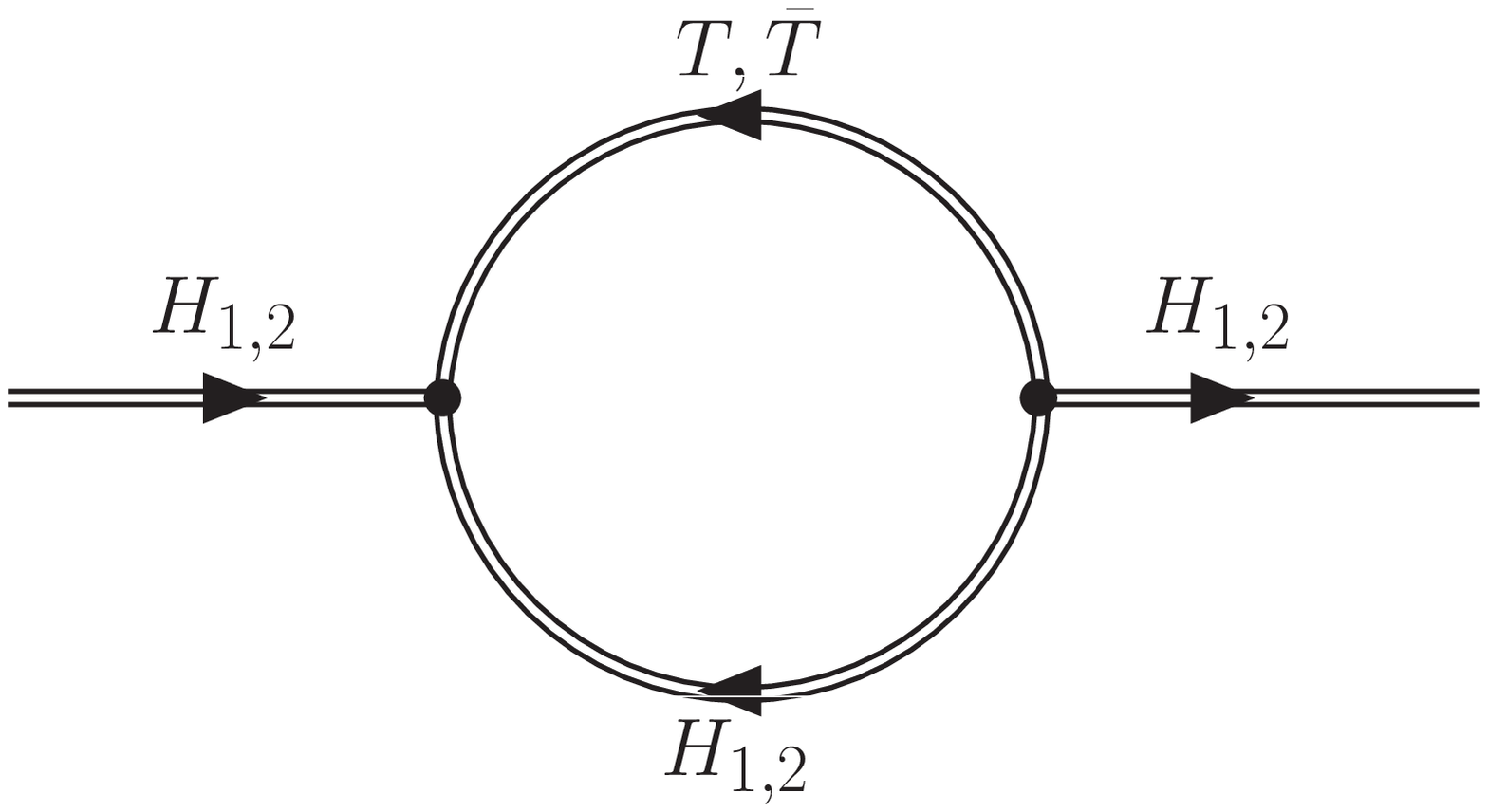}
\end{tabular}
}

After EWSB, the effective neutrino mass matrix
\begin{equation}
\label{meff1}%
m_\nu=\mathcal{O}^\nu v^2\sin^2\beta\,,
\end{equation}
is generated, where $v=174\,{\rm GeV}$ and $\tan\beta=\langle H_{2}^0
\rangle/\langle H_{1}^0 \rangle=v_2/v_1$. Its RGE can be derived
from Eqs.~(\ref{rgek})-(\ref{meff1}), leading to:
\begin{equation}
\label{rgemnu}%
16\pi^2 \dot{m}_\nu=\alpha_\nu m_\nu+ P_\nu^T m_\nu+m_\nu P_\nu\,,
\end{equation}
where
\begin{eqnarray}
\label{alphanu1}%
\alpha_\nu&=&6\,\tr(Y_u^\dag
Y_u^{})+\underline{6\,|\lambda_2|^2}-\frac{6}{5}g_1^2- 6 g_2^2\,,\\
\label{alphanu}%
P_\nu&=& Y_e^\dag Y_e^{}+\underline{3\,Y_T^\dag
Y_T^{}}+\underline{3\,Y_Z^\dag Y_Z^{}}\,.
\end{eqnarray}
When $T$, $S$ and $Z$ have different masses, the decoupling of these
states has to be performed. In practice, this corresponds to
switching off their interactions in the RGEs. At $\Lambda=M_T$ a new
contribution must be added to $\mathcal{O}^\nu$ due to the
decoupling of the triplet fields, in such a way that
\begin{equation}
\label{match} \mathcal{O}^\nu(M_T) \rightarrow \mathcal{O}^\nu(M_T)
+ \left.\frac{\lambda_2Y_T}{M_T}\right|_{\Lambda=M_T}\,,
\end{equation}
all quantities being taken at $\Lambda=M_T$. Below this scale, the
running of the effective neutrino mass matrix follows the RGE
(\ref{rgemnu}) with $Y_T=0$ and $\lambda_1=\lambda_2=0$. If $S$
and/or $Z$ is lighter than $T$, then it should be subsequently
decoupled without adding any contribution to $\mathcal{O}^\nu$ since
this operator arises solely from integrating out the SU(2)
triplets\footnote{In this work we neglect the one-loop threshold
corrections to the effective neutrino mass operator arising from the
decoupling of the heavy triplet states.}.
%

\section{Running of the neutrino masses and mixing matrix}
\label{sec3}

In this section we derive the RGEs for the neutrino masses, the
neutrino mixing matrix and the charged-lepton Yukawa couplings in
the framework of the model presented in the previous section. We
consider that at each scale $\Lambda$ the following relations hold
\begin{eqnarray}
U_\nu^T \,m_\nu\, U_\nu^{}={\rm diag}(m_1,m_2,m_3)\,&,&\; U_e^\dag
\,Y_e^\dag Y_e^{}\, U_e^{}={\rm diag}(y_e^2,y_\mu^2,y_\tau^2)\equiv
d_e^2\;,\nonumber\\
U_T^\dag \,Y_T^\dag Y_T^{}\, U_T^{}&=&{\rm
diag}(y_1^2,y_2^2,y_3^2)\equiv d_T^2\,,
\label{diags}%
\end{eqnarray}
where $U_e$ and $U_\nu$ are $3\times 3$ complex unitary matrices and
$m_i$ denotes the effective neutrino masses. The neutrino mixing
matrix $U$ at $\Lambda$ is then given by:
\begin{equation}
\label{Udef}%
U=U_e^\dag U_\nu^{}\,.
\end{equation}
To obtain the RGE for $U$, we adopt the procedure first introduced
in Ref.~\cite{Babu:1987im} for the renormalisation of the
Cabbibo-Kobayashi-Maskawa mixing matrix, and later used for the
neutrino case~\cite{Chankowski:1999xc,Casas:1999tg,Antusch:2003kp}.
We start by considering the ansatz
\begin{equation}
\label{UeUnudot}%
\dot{U}_e=U_e Q \;,\; \dot{U}_\nu=U_\nu R,
\end{equation}
where $Q$ and $R$ are anti-Hermitian matrices due to unitarity of $U_e$
and $U_\nu$. In order to
determine $Q$, one needs the RGE for the charged-lepton Yukawa
matrix $Y_e$:
\begin{equation}
\label{RGEYe}%
16\pi^2\dot{Y}_e=\alpha_eY_e +Y_e \,P_e\,.
\end{equation}
The expressions for $\alpha_e$ and $P_e$ are obtained by
considering the one-loop anomalous dimensions
$\gamma_{L_i}^{(1)L_j}$ and $\gamma_{H_1}^{(1)H_1}$ given in
Eqs.~(\ref{gammaLiLj}) and (\ref{gammaH1}), respectively, and the
same for the charged-lepton singlets $\gamma_{e_i^c}^{(1)e_j^c}=
2(Y_e^\ast Y_e^T)_{ij}-6\,g_1^2/5\,\delta_i^j$. This yields
\begin{equation}
\label{alphae}%
\alpha_e=\tr(Y_e^\dag Y_e^{}+3Y_d^\dag
Y_d^{})+\underline{3\,|\lambda_1|^2}-\frac{9}{5}g_1^2- 3
g_2^2\;\;,\;\; P_e= 3 Y_e^\dag Y_e^{}+\underline{3\,Y_T^\dag
Y_T^{}}+\underline{3\,Y_Z^\dag Y_Z^{}}\,,
\end{equation}
which, together with the diagonalisation of $Y_e^\dag Y_e^{}$ given
in (\ref{diags}), leads to the following result for $Q$
\begin{equation}
\label{Qij}%
16 \pi^2 Q_{ij}=(P_e^{\,\prime})_{ij}\,
\frac{y_{e_i}^2+y_{e_j}^2}{y_{e_j}^2-y_{e_i}^2}\;\;\;(i\neq j).
\end{equation}
In this equation, $P_e^\prime$ is given by
\begin{equation}
\label{Pepri}%
P_e^{\,\prime}\equiv U_e^\dag P_e U_e^{} = 3\, ( d_e^2+U_e^\dag
U_T^{}\, d_T^2\, U_T^\dag U_e^{} +U_e^\dag\,Y_Z^\dag
Y_Z^{}\,U_e^{})\,.
\end{equation}
Since $P_e^{\,\prime}$ is Hermitian and $\alpha_e$ is real, one has
$Q_{ii}=0$. Following the same procedure as for $Q$, but using now
Eqs.~(\ref{rgemnu})-(\ref{alphanu}) and (\ref{diags}), one can
derive the expression for the matrix $R$:
\begin{equation}
\label{Tij}%
16 \pi^2 R_{ij}=\frac{m_i^2+m_j^2}{m_j^2-m_i^2}\,(P_\nu^\prime)_{ij}
+ \frac{2\,m_i \,m_j}{m_j^2-m_i^2}\,(P_\nu^{\,\prime})_{ij}^\ast
\;\;\;(i\neq j)\,,
\end{equation}
where $P_\nu^\prime$ is now defined as:
\begin{equation}
\label{Pnupri}%
P_\nu^{\,\prime}\equiv U_\nu^\dag\, P_\nu^{}\, U_\nu^{} =
U^\dag\,d_e^2\,U^{}+3\,(U_\nu^\dag U_T^{}\,d_T^2\, U_\nu^{}
U_T^\dag+U_\nu^\dag\,Y_Z^\dag Y_Z^{}\,U_\nu^{})\,.
\end{equation}
Similarly as for $Q$, $R_{ii}=0$ since $P_\nu^\prime$ is
Hermitian and $\alpha_\nu$ is real. Finally, from Eqs.~(\ref{Udef})
and (\ref{UeUnudot}) (and taking into account that $Q$ is anti-Hermitian)
one obtains:
\begin{equation}
\label{Udot}%
\dot{U}=-QU+UR\,,
\end{equation}
with $Q$ and $R$ given as in Eqs.~(\ref{Qij}) and (\ref{Tij}). The first and second
terms on the right-hand side of the above equation
account for the contribution to the
running of $U$ coming from $\dot{U}_e$ and $\dot{U}_\nu$, respectively.

As for the neutrino masses $m_i$, their RGEs can be derived using
Eqs.~(\ref{rgemnu}) and (\ref{diags}), leading to:
\begin{equation}
\label{rgemi}%
16\pi^2
\dot{m}_i=\left[\alpha_\nu+2\,(P_\nu^\prime)_{ii}\right]m_i\,.
\end{equation}
For the charged-lepton Yukawa couplings $y_{e_i}$ $(e_i=e,\mu,\tau)$ the RGE reads
\begin{equation}
\label{rgeye}%
16\pi^2
\dot{y}_{e_i}=\left[\alpha_e+2\,(P_e^\prime)_{ii}\right]y_{e_i}\,.
\end{equation}
Notice that the presence of the new couplings $\lambda_1$ and
$Y_{T,Z}$ in $\alpha_e$ and $P_e^\prime$ may affect the running of
$y_{e,\mu,\tau}$.
%
%
\section{The pure type II seesaw case}
\label{sec4}
In general, the flavour structure and magnitude of the couplings
$Y_{S,Z}$ depend on the specific details of the SU(5) model
considered. Although not directly related with $m_\nu$ at
tree-level, these couplings affect the renormalisation of $Y_T$. It
is worth mentioning that, even if one imposes $Y_{S,Z}=Y_T$ at {\em
e.g.} $\Lambda=M_G$ (where from now on $M_G$ denotes the
grand-unification scale), the RG running will deviate $Y_{S,Z}$ from
the $Y_T$ trajectory. For simplicity, we will restrict ourselves to
the case in which the couplings $Y_{S,Z}$ are negligible when
compared with $Y_T$. This could for instance result from SU(5)
breaking effects, as discussed in Ref.~\cite{Rossi:2002zb}.
Therefore, from here onwards we set $Y_{S,Z}=0$. Moreover, in the
rest of this work we assume that the only contribution to the
effective neutrino mass operator $\mathcal{O}^\nu$ arises from the
decoupling of the heavy triplet states $T$ and $\bar{T}$. In other
words, we do not address alternative scenarios where extra
contributions to the effective neutrino operator are present.
Therefore, we have:
\begin{equation}
\label{matchII}
\mathcal{O}^\nu(M_T)=\left.\frac{\lambda_2Y_T}{M_T}\right|_
{\Lambda=M_T}\,,
\end{equation}
at $\Lambda=M_T$.

Although $\mathcal{O}^\nu=0$ for $\Lambda > M_T$, one can define a
{\em would-be} effective neutrino mass operator
$\mathcal{O^\nu}=\lambda _2 Y_T/M$ at any scale $\Lambda$. As a
consequence of the SUSY non-renormalisation theorem, the RGE of this
object coincides with the one given in (\ref{rgeO}). The running of
the effective neutrino mass and charged-lepton Yukawa matrices
follows the same RGEs as in (\ref{rgemnu}) and (\ref{RGEYe}),
respectively, with:
\begin{equation}
\label{PnuPeII}%
P_\nu= Y_e^\dag Y_e^{}+3\,Y_T^\dag Y_T^{}\;,\;P_e= 3 Y_e^\dag
Y_e^{}+3\,Y_T^\dag Y_T^{}\,.
\end{equation}
Since $m_\nu$ is now proportional to $Y_T$, the unitary
matrices $U_\nu$ and $U_T$ in Eqs.~(\ref{diags}) are identical and
\begin{equation}
\label{replac}%
Y_T^\dag Y_T^{}=\frac{ m_\nu^\dag m_\nu}{v_T^2}\;,\; y_i^2=
\frac{m_i^2}{v_T^2}\;,\;v_T=\frac{\lambda_2v_2^2}{M_T}\,.
\end{equation}
The quantity $v_T$ is the induced vacuum expectation value of the
neutral-scalar component of $T$: $\langle T^0 \rangle=v_T/\sqrt{2}$.
The RGE for the neutrino mixing matrix $U$ is shown in
Eq.~(\ref{Udot}) with $Q$ and $R$ defined as in Eqs.~(\ref{Qij}) and
(\ref{Tij}). Taking into account Eqs.~(\ref{diags}), together with the fact that
$U=U_e^\dag U_\nu$ and $U_\nu=U_T$, the matrices $P_e^\prime$ and $P_\nu^\prime$
are now given by
\begin{equation}
\label{PeprimeII}%
P_\nu^\prime=U^\dag d_e^2\,
U+3\,d_T^2\;\;,\;\;P_e^\prime=3\,d_e^2+3\,Ud_T^2\,U^\dag\,.
\end{equation}

In order to obtain the RGEs for the mixing angles and CP-violating
phases, we adopt the following parameterisation for $U$:
\begin{equation}
\label{Uparam}%
U=K_\varphi\, V \,K_\alpha\;,\; K_\varphi={\rm
diag}(e^{i\varphi_e},e^{i\varphi_\mu},e^{i\varphi_\tau})\;,\;K_\alpha={\rm
diag}(e^{-i\alpha_1/2},e^{-i\alpha_2/2},1)\,,
\end{equation}
where $\varphi_{e,\tau,\mu}$ are unphysical phases and
$\alpha_{1,2}$ are CP-violating Majorana phases. The unitary matrix $V$ is
parameterized in the standard way
\begin{equation}
\label{Udelta}%
V =\left(
\begin{array}{ccc}
c_{12} c_{13} & s_{12} c_{13} & s_{13} e^{-i \delta} \\
-s_{12} c_{23} - c_{12} s_{23} s_{13}  e^{i \delta} & \quad c_{12}
c_{23}  - s_{12} s_{23} s_{13} e^{i \delta} \quad
& s_{23} c_{13} \\
s_{12} s_{23} - c_{12} c_{23} s_{13} e^{i \delta} & -c_{12} s_{23} -
s_{12} c_{23} s_{13} e^{i \delta} & c_{23} c_{13}
\end{array}\right)\,,
\end{equation}
where $c_{ij}\equiv \cos\theta_{ij}$, $s_{ij}\equiv \sin\theta_{ij}$
and $\delta$ is the Dirac CP-violating phase. We identify the mixing
angles $\theta_{12}$ and $\theta_{23}$ as being the ones involved in
the solar and atmospheric neutrino oscillations, respectively, while
$\theta_{13}$ denotes the so-called CHOOZ or reactor neutrino mixing
angle.

Although physical observables do not depend on the phases
$\varphi_i$, these are crucial to obtain the RGEs for the physical
neutrino parameters (see discussion in Ref.~\cite{Antusch:2003kp}). From
(\ref{Udot}) and (\ref{Uparam}) we obtain
\begin{equation}
\label{eqUdot}%
i\,{\rm
diag}(\dot{\varphi}_e,\dot{\varphi}_\mu,\dot{\varphi}_\tau)\, V +
\dot{V} - \frac{i}{2}\,V\, {\rm
diag}(\dot{\alpha}_1,\dot{\alpha}_2,0)=-\widehat{Q}\,V +
V\,\widehat{R}\,,
\end{equation}
which can be used to extract the RGEs for the mixing angles and
phases. In the above equation, the matrices
$\widehat{Q}$ and $\widehat{R}$  are defined by
$\widehat{Q}=K_\varphi^\ast\,Q\,K_\varphi^{}$ and
$\widehat{R}=K_\alpha\,R\,K_\alpha^\ast$. Together with
Eqs.~(\ref{Qij}) and (\ref{Tij}), this leads to
\begin{eqnarray}
\label{Qhat}%
16 \pi^2 \widehat{Q}_{ij}&=&(\widehat{P}_e)_{ij}\,
\frac{y_{e_i}^2+y_{e_j}^2}{y_{e_j}^2-y_{e_i}^2}\;\;\;(i\neq j)\,,\\
\label{That}%
16 \pi^2 \widehat{R}_{ij}&=&\frac{m_i^2+m_j^2}{m_j^2-m_i^2}\,
(\widehat{P}_\nu)_{ij} + \frac{2\,m_i \,m_j}{m_j^2-m_i^2}\,
(\tilde{P}_\nu)_{ij} \;\;\;(i\neq j),
\end{eqnarray}
where
\begin{eqnarray}
\label{Pehat}%
\widehat{P}_e&=&K_\varphi^\ast\,P_e^\prime\,K_\varphi^{}= 3\,d_e^2+
3\,V d_T^2V^\dag\,,\\
\label{Pnuhat}%
\widehat{P}_\nu&=&K_\alpha\,P_\nu^\prime\,K_\alpha^\ast=V^\dag
d_e^2\, V+3\,d_T^2\,,\\
\label{Pnutilde}%
\tilde{P}_\nu &=&
K_\alpha\,(P_\nu^\prime)^\ast\,K_\alpha^\ast=K_\alpha^2\,V^T d_e^2\,
V^\ast K_\alpha^{\ast\,2}+3\,d_T^2\,.
\end{eqnarray}
The above results show that the RGE for the mixing matrix $U$ does
not depend on the unphysical phases $\varphi_{e,\mu,\tau}$, as
expected. Comparing the present model with the MSSM, it becomes
clear that the new contribution to the running of the neutrino
mixing matrix comes from the first term on the right-hand side of
Eq.~(\ref{eqUdot}), {\em i.e.} from the effect of the non-trivial
running of $U_e$ induced by the presence of the couplings $Y_T$.

At this point we would like to point out some discrepancies between
our results and those obtained in Ref.~\cite{Schmidt:2007nq}.
\\

\noindent 1) In Ref.~\cite{Schmidt:2007nq} it is claimed that the
RGEs for the mixing angles and CP-violating phases are independent of
the Majorana phases $\alpha_{1,2}$ at any scale $\Lambda > M_T$. This is
actually not the case, as can be seen from Eq.~(\ref{eqUdot}).
Although the first term on the right-hand side of this equation does
not depend on $\alpha_{1,2}$, the second term does (through the
contribution of the term proportional to $\tilde{P}$ in
$\widehat{R}$). In fact, the dependence of the RGE of $U$ on
$\alpha_{1,2}$ originates from the running of the effective neutrino
mass matrix, more specifically from the term proportional to
$Y_e^\dag Y_e$ in Eq.~(\ref{rgemnu}). On the other hand,
$\widehat{Q}$ is independent of the Majorana phases since it is
defined by $\widehat{P}_e$ (which does not depend on $\alpha_{1,2}$)
and the charged-lepton Yukawa couplings. Hence, $\dot{U}_e$ does not
show any direct dependence on $\alpha_{1,2}$.
\\

\noindent 2) Our results for the TMSSM RGEs agree with the ones of
Refs.~\cite{Rossi:2002zb} and \cite{Borzumati:2009hu}. However, we
find several discrepancies with those obtained in
Ref.~\cite{Schmidt:2007nq} where the conventions for the couplings
entering in $W_T$ are the same as the ones we are currently
adopting. We find that, in order to get complete agreement between
all results, the couplings $\lambda_{1,2}$ and $Y_T$ must be
replaced by $\sqrt{2}\lambda_{1,2}$ and $\sqrt{2}Y_T$ in all the
RGEs of Ref.~\cite{Schmidt:2007nq}, including the one for the
effective neutrino mass matrix. This affects the numerical
pre-factors in the second term of the right-hand side of
Eqs.~(\ref{Pehat})-(\ref{Pnutilde}), which in our case differ by a
factor of two from those of Ref.~\cite{Schmidt:2007nq}. The same
holds for all the terms proportional to $|\lambda_{1,2}|^2$
appearing in Eqs.~(\ref{alphanu1}) and (\ref{alphae}). We believe
that these discrepancies might be the result of an
inconsistent definition of the Feynman rules used in
Ref.~\cite{Schmidt:2007nq} for the vertices involving the triplet
states.
%
%
%

\subsection{RGEs for the neutrino masses, mixing angles and CP-violating phases}
\label{sec41}%

We can now use the master equation (\ref{eqUdot}) to obtain the RGEs
for the mixing angles $\theta_{ij}$ and CP-violating phases. For each
$\theta_{ij}$ we write the corresponding RGE in the form:
\begin{equation}
\label{RGEangles}%
\dot{\theta}_{ij}=\dot{\theta}_{ij}^\nu+\dot{\theta}_{ij}^e
\;,\;\dot{\theta}_{ij}^\nu=\sum_{b>a}{\rm Re}
\left[A_{ij}^{ab}\,\widehat{R}_{ab}\right] \;\;, \;\;
\dot{\theta}_{ij}^e =\sum_{b>a}{\rm Re}\left[B_{ij}^{ab}\,
\widehat{Q}_{ab}\right]\;\;,\;\;a,b=1,2,3\,,
\end{equation}
where $\dot{\theta}_{ij}^\nu$ and $\dot{\theta}_{ij}^e$ contain the
contributions coming from the running of $m_\nu$ and $Y_e$,
respectively. The coefficients $A_{ij}^{ab}$ and $B_{ij}^{ab}$
(shown in Table~\ref{Table1}) are determined by solving
Eq.~(\ref{eqUdot}). Similarly, for the full set of
phases
$\Phi=(\delta,\alpha_1,\alpha_2,\varphi_e,\varphi_\mu,\varphi_\tau)$,
\begin{equation}
\label{RGEphasegen}%
\dot{\Phi}=\dot{\Phi}^\nu+\dot{\Phi}^e\;\;,\;\;\dot{\Phi}^\nu=\sum_{b>a}{\rm
Im}\left[A_\Phi^{ab}\widehat{R}_{ab}\right]\;\;,\;\;\dot{\Phi}^e=\sum_{b>a}{\rm
Im}\left[B_\Phi^{ab} \widehat{Q}_{ab}\right]\;\;,\;\;a,b=1,2,3\,,
\end{equation}
where $\dot{\Phi}^\nu$ and $\dot{\Phi}^e$ include the terms stemming
from $\dot{m}_\nu$ and $\dot{Y}_e$. The coefficients $A_\Phi^{ab}$
and $B_\Phi^{ab}$ are given in Table~\ref{Table2}.
\TABLE[!ht]{ \label{Table1} \caption{Coefficients $A_{ij}^{ab}$
(first three rows) and $B_{ij}^{ab}$ (last three rows) which enter
the definition of the RGEs for the mixing angles $\theta_{ij}$,
given in Eq.~(\ref{RGEangles}).} \small{\begin{tabular}{lccc} \hline
\hline \noalign{\medskip} $ij$ &$A_{ij}^{12}$ &$A_{ij}^{13}$ &
$A_{ij}^{23}$
 \\
\noalign{\medskip} \hline \hline  \noalign{\medskip} %
$12$
&$1$
& $s_{12}\tan\theta_{13} e^{i\delta}$
& $-c_{12}\tan\theta_{13} e^{i\delta}$ \\ \noalign{\medskip} \hline
\noalign{\medskip}
$13$
&$0$
&$c_{12}e^{i\delta}$%
&$s_{12}e^{i\delta}$ \\
\noalign{\medskip} \hline \noalign{\medskip}
$23$  &$0$ &$-\dfrac{s_{12}}{c_{13}}$
&$\dfrac{c_{12}}{c_{13}}$ \\
\noalign{\medskip}\hline  \hline \noalign{\medskip} $ij$
&$B_{ij}^{12}$ &$B_{ij}^{13}$ & $B_{ij}^{23}$
 \\
\noalign{\medskip} \hline \hline  \noalign{\medskip} %
$12$
&$-\dfrac{c_{23}}{c_{13}}$
&$\dfrac{s_{23}}{c_{13}}$
& 0 \\
\noalign{\medskip} \hline  \noalign{\medskip}
$13$
&$-s_{23} e^{i\delta}$
&$-c_{23} e^{i\delta}$
&0 \\
\noalign{\medskip} \hline \noalign{\medskip}
$23$  &$c_{23}\tan\theta_{13} e^{i\delta}$
&$-s_{23}\tan\theta_{13} e^{i\delta}$
&$-1$ \\
\noalign{\medskip} \hline \hline \noalign{\medskip}
\end{tabular}
}}
\TABLE[!ht]{ \label{Table2} \caption{Coefficients $A_\Phi^{ab}$
(first six rows) and $B_\Phi^{ab}$ (last six rows) which enter the
definition of the RGEs for the CP-violating phases
$\Phi=(\delta,\alpha_1,\alpha_2,\varphi_e,\varphi_\mu,\varphi_\tau)$
given in Eq.~(\ref{RGEphasegen}).}
\renewcommand{\tabcolsep}{0.8pc}
\begin{tabular}{lccc} \hline  \hline \noalign{\medskip}
$\Phi$ &$A_\Phi^{12}$ &$A_\Phi^{13}$ & $A_\Phi^{23}$
 \\
\noalign{\medskip} \hline \hline  \noalign{\medskip} %
$\delta$
&$\dfrac{1}{c_{12}s_{12}}$
& $-\dfrac{s_{12}V_{22}}{c_{13}c_{12}s_{23}}- \dfrac{V_{22}^\ast
e^{i\delta}}{s_{13}c_{13}c_{23}}$
& $\dfrac{V_{21}^\ast e^{i\delta}}{s_{13}c_{13}c_{23}}-
\dfrac{c_{12}V_{21}}{c_{13}s_{12}s_{23}}$ \\ \noalign{\medskip}
\hline  \noalign{\medskip}
$\alpha_1$
&$2\cot\theta_{12}$
&$\dfrac{2V_{31}}{c_{13}c_{23}}+ \dfrac{2V_{21} }{c_{13}s_{23}}$
&$\dfrac{2V_{32}}{c_{13}c_{23}}-
\dfrac{2\,c_{12}V_{21} }{c_{13}s_{23}s_{12}}$ \\
\noalign{\medskip} \hline \noalign{\medskip}
$\alpha_2$  &$2\tan\theta_{12}$ &$\dfrac{2V_{31}}{c_{13}c_{23}}-
\dfrac{2\,s_{12}V_{22} }{c_{13}s_{23}c_{12}}$
&$\dfrac{2V_{32}}{c_{13}c_{23}}+
\dfrac{2V_{22} }{c_{13}s_{23}}$ \\
\noalign{\medskip} \hline \noalign{\medskip}
$\varphi_e$
&$\dfrac{1}{c_{12}s_{12}}$
&$\dfrac{V_{31}}{c_{13}c_{23}}- \dfrac{s_{12}V_{22}
}{c_{13}s_{23}c_{12}}$
&$\dfrac{V_{32}}{c_{13}c_{23}}- \dfrac{c_{12}V_{21}
}{c_{13}s_{23}s_{12}}$ \\
\noalign{\medskip} \hline \noalign{\medskip}
$\varphi_\mu$
&0
&$\dfrac{V_{22}}{c_{13}s_{23}}$
&$\dfrac{V_{21}}{c_{13}s_{23}}$ \\
\noalign{\medskip} \hline \noalign{\medskip}
$\varphi_\tau$
&0 &$\dfrac{V_{31}}{c_{13}c_{23}}$  &$\dfrac{V_{32}}{c_{13}c_{23}}$ \\
\noalign{\medskip}\hline  \hline \noalign{\medskip} $\Phi$
&$B_\Phi^{12}$ &$B_\Phi^{13}$ & $B_\Phi^{23}$
 \\
\noalign{\medskip} \hline \hline  \noalign{\medskip} %
$\delta$
&$\dfrac{c_{23}V_{32}}{c_{13}s_{12}s_{23}}- \dfrac{V_{32}^\ast
e^{i\delta}}{s_{13}c_{12}c_{13}}$
& $-\dfrac{s_{23}V_{21}}{c_{23}c_{12}c_{13}}- \dfrac{V_{21}^\ast
e^{i\delta}}{s_{13}c_{13}s_{12}}$
& $-\dfrac{1}{s_{23}c_{23}}$ \\
\noalign{\medskip} \hline  \noalign{\medskip}
$\alpha_1$
&$\dfrac{2c_{23}V_{32}}{s_{12}c_{13}s_{23}}$
&$\dfrac{2s_{23}V_{22}}{s_{12}c_{13}c_{23}}$
&$-\dfrac{2}{s_{23}c_{23}}$ \\
\noalign{\medskip} \hline \noalign{\medskip}
$\alpha_2$  &$\dfrac{2c_{23}V_{31}}{c_{12}c_{13}s_{23}}$
&$\dfrac{2s_{23}V_{21}}{c_{12}c_{13}c_{23}}$
&$-\dfrac{2}{s_{23}c_{23}}$ \\
\noalign{\medskip} \hline \noalign{\medskip}
$\varphi_e$
&$\dfrac{c_{23}V_{32}}{s_{12}s_{23}c_{13}}- \dfrac{V_{21}
}{c_{13}c_{12}}$
&$\dfrac{s_{23}V_{22}}{s_{12}c_{23}c_{13}}- \dfrac{V_{31}
}{c_{13}c_{12}}$
&$-\dfrac{1}{s_{23}c_{23}}$ \\
\noalign{\medskip} \hline \noalign{\medskip}
$\varphi_\mu$
&$-\dfrac{V_{13}^\ast}{c_{13}s_{23}}$
&0
&$-\cot\theta_{23}$ \\
\noalign{\medskip} \hline \noalign{\medskip}
$\varphi_\tau$
&0 %
&$-\dfrac{V_{13}^\ast}{c_{13}c_{23}}$  %
&$\tan\theta_{13}$ \\
\noalign{\medskip} \hline \hline \noalign{\medskip}
\end{tabular}
}
Regarding the neutrino masses $m_i$, the RGEs can be obtained
replacing $P^\prime_\nu$ by $\widehat{P}_\nu$ in Eq.~(\ref{rgemi}),
leading to:
\begin{equation}
\label{midot}%
16\pi^2 \dot{m}_i=\left[\alpha_\nu+2\,\left(V^\dag d_e^2\,
V\right)_{ii}+6\,y_i^2\right]m_i\,.
\end{equation}
As expected, $\dot{m}_i$ does not depend on the Majorana phases
$\alpha_{1,2}$. Moreover, the third term on the right-hand side of
the above equation is independent of the neutrino mixing angles
and CP-violating phases. In contrast, the second term (also present
in the MSSM case) does depend on those quantities.

The complete expressions for the RGEs of the neutrino mixing angles
and phases can be obtained by inserting the coefficients given in
Tables~\ref{Table1} and \ref{Table2} into Eqs.~(\ref{RGEangles}) and
(\ref{RGEphasegen}), respectively. In general, the final result is
too lengthy to be presented here. However,
following the procedure of Ref.~\cite{Antusch:2003kp}, we will
expand the RGEs to the leading order in the (small) mixing angle
$\theta_{13}$. Let us first concentrate on the neutrino contribution
to the RGEs of the mixing angles, {\em i.e.} the terms
$\dot{\theta}_{ij}^\nu$ in Eq.~(\ref{RGEangles}). These depend on
$V$, on the neutrino masses and on the charged-lepton Yukawa
couplings $y_{e_i}^2$. In view of the strong hierarchy $y_e \ll y_\mu
\ll y_\tau$, we will keep only the terms proportional to $y_\tau^2$.
In this limit, and in the zeroth order in $\theta_{13}$, we find
\begin{eqnarray}
\label{RGE13ap}%
\dot{\theta}_{13}^\nu &\simeq&
-\frac{y_\tau^2}{32\pi^2}\frac{m_3}{\dmatm}\sin(2\theta_{23})
\sin(2\theta_{12})\left[ -m_1\cos(\alpha_1-\delta)+\frac{m_2
\cos(\alpha_2-\delta)}{ (1-r)}+\frac{r\,m_3\cos\delta}{
(1-r)}\right]\,,\nonumber\\
\label{RGE12ap}%
\dot{\theta}_{12}^\nu &\simeq& -\frac{y_\tau^2}{32\pi^2}\frac{|m_1
e^{i\alpha_1}+m_2\, e^{i\alpha_2}|^2}{r\dmatm}
s_{23}^2\sin(2\theta_{12})\,,\nonumber\\
\label{RGE23ap}%
\dot{\theta}_{23}^\nu &\simeq& -\frac{y_\tau^2}{32\pi^2}
\left[c_{12}^2 \frac{|m_3+m_2\,
e^{i\alpha_2}|^2}{\dmatm\,(1-r)}+s_{12}^2\frac{|m_1+m_3\,
e^{i\alpha_1}|^2}{\dmatm}\right]\sin(2\theta_{23})\,,
\end{eqnarray}
where
\begin{equation}
\label{rdef}%
r=
\frac{\dmsol}{\dmatm}\;,\;\Delta m^2_{ij}\equiv m_j^2-m_i^2\,. 
\end{equation}
The above expressions make explicit the discrepancies between our results
and those of Ref.~\cite{Schmidt:2007nq}. From Eqs.~(\ref{RGE23ap})
it is clear that the running of the neutrino mixing angles {\em does
depend} on the Majorana phases through the renormalisation of the
effective neutrino mass matrix $m_\nu$. The above contributions to
the RGEs originate from the wave-function renormalisation of the
lepton doublets, namely from the term proportional to $Y_e^\dag
Y_e^{}$, present below and above the decoupling scale of the
triplets. Therefore, Eqs.~(\ref{RGE23ap}) are valid
both in the effective and in the full theory. Not surprisingly, the
results for $\dot{\theta}_{ij}^\nu$ agree with those obtained for
the MSSM in Ref.~\cite{Antusch:2003kp}. Hence, the expansions given
in Ref.~\cite{Schmidt:2007nq} are {\em not valid} above the mass
scale of the triplets since they do not account for the dependence
of the RG running on the Majorana phases, which may play a crucial r\^ole in the
running of the neutrino parameters.

The approximate expressions for the RGEs of the Majorana phases
$\alpha_{1,2}$ read
\begin{eqnarray}
\label{RGEalpha1ap}%
\!\!\!\!\!\!\!\!\!\!\dot{\alpha}_{1}^\nu &\simeq &
-\frac{y_\tau^2}{4\pi^2} \left\{ \frac{m_1 m_2
s_{23}^2c_{12}^2}{r\,\dmatm}\sin(\alpha_1-\alpha_2)+\frac{
m_3\cos(2\theta_{23})}{\dmatm}\left[
m_1s_{12}^2s_{\alpha_1}+\frac{m_2\,c_{12}^2s_{\alpha_2}}{(1-r)}
\right]\right\}\,,\\
\label{RGEalpha1ap}%
\!\!\!\!\!\!\!\!\!\!\dot{\alpha}_{2}^\nu &\simeq &
-\frac{y_\tau^2}{4\pi^2} \left\{ \frac{m_1 m_2
s_{23}^2s_{12}^2}{r\,\dmatm}\sin(\alpha_1-\alpha_2)+\frac{
m_3\cos(2\theta_{23})}{\dmatm}\left[
m_1s_{12}^2s_{\alpha_1}+\frac{m_2\,c_{12}^2
s_{\alpha_2}}{(1-r)}\right]\right\}\,,
\end{eqnarray}
in contrast with the result $\dot{\alpha}_{1,2}^\nu\simeq 0$
obtained in Ref.~\cite{Schmidt:2007nq} at zeroth order in
$\theta_{13}$. There, the lowest-order term in the expansion of
$\dot{\alpha}_{1,2}$ was found to be of first order in
$\theta_{13}$, giving rise to the conclusion that the RG effects on
$\alpha_{1,2}$ are small. The reason why the above terms were not
obtained in \cite{Schmidt:2007nq} has to do with the fact that they
vanish for $\alpha_{1,2}=0$, which is the only limit in which the
expressions given in that reference are valid.

The neutrino contribution to the RGE of the Dirac CP-violating phase
$\delta$ can be expressed in the form~\cite{Antusch:2003kp}
\begin{equation}
\label{RGEdeltap}%
\dot{\delta}^\nu=\frac{y_\tau^2}{32\pi^2}\frac{\dot{\delta}^
\nu_{(-1)}}{\theta_{13}}+
\frac{y_\tau^2}{8\pi^2}\dot{\delta}^\nu_{\scriptsize{(0)}}\,,
\end{equation}
where $\dot{\delta}^\nu_{\scriptsize{(-1)}}$ and
$\dot{\delta}^\nu_{\scriptsize{(0)}}$ are given by
\begin{eqnarray}
\label{ddotm1}%
\!\!\!\dot{\delta}^\nu_{\scriptsize{(-1)}} &=&
\frac{m_3\sin(2\theta_{12})\sin(2\theta_{23})}{\dmatm} \left[
m_1\sin(\alpha_1-\delta)+m_2\,\frac{\sin(\alpha_2-\delta)}{1-r}
+\frac{rm_3}{1-r}\sin\delta\right]\,,\\
\dot{\delta}^\nu_{\scriptsize{(0)}} &=&
\frac{m_1m_2\,s_{23}^2}{r\dmatm}\sin(\alpha_1-\alpha_2)+
\frac{m_3\,\cos(2\theta_{23})}{\dmatm} \left[
m_1s_{12}^2\sin{\alpha_1}+\frac{
m_2\,c_{12}^2\sin{\alpha_2}}{1-r}\right] +\nonumber \\
\label{ddot0}%
&+& \frac{m_3 c_{23}^2}{\dmatm} \left[
m_1c_{12}^2\sin(2\delta-\alpha_1)+
\frac{m_2}{1-r}s_{12}^2\sin(2\delta-\alpha_2)\right]\,.
\end{eqnarray}
Again, this differs from Ref.~\cite{Schmidt:2007nq} since
$\dot{\delta}^\nu_{\scriptsize{(0)}}$ does not vanish (even in the
limit of vanishing Majorana phases) and
$\dot{\delta}^\nu_{\scriptsize{(-1)}}$ does depend on
$\alpha_{1,2}$. In the limit $\alpha_{1,2}=0$, our result for
$\dot{\delta}^\nu_{\scriptsize{(-1)}}$ agrees with the one of
\cite{Schmidt:2007nq}.

Let us now consider the charged-lepton contribution to the RGEs of
the mixing angles and CP-violating phases denoted by
$\dot{\theta}_{ij}^e$ and $\dot{\Phi}_i^e$ in Eqs.~(\ref{RGEangles})
and (\ref{RGEphasegen}). At zeroth order in $\theta_{13}$ we obtain
\begin{eqnarray}
\label{dott12eap}%
\hspace*{-1.5cm}\dot{\theta}_{12}^e &\simeq&
\frac{3r\dmatm}{32\pi^2v_T^2} \frac{y_e^4- y_\mu^2 y_\tau^2+ y_e^2
(y_\mu^2-y_\tau^2)\cos(2\theta_{23})}
{(y_\mu^2-y_e^2)(y_\tau^2-y_e^2)}\sin(2\theta_{12})\simeq
-\frac{3r\dmatm}{32\pi^2v_T^2} \sin(2\theta_{12})\,,\,\,\,\,\,\,\,\,\\
\label{dott23eap}%
\dot{\theta}_{23}^e &\simeq&- \frac{3\dmatm}{32\pi^2v_T^2}
\frac{y_\mu^2+y_\tau^2}
{y_\tau^2-y_\mu^2}(1-rc_{12}^2)\sin(2\theta_{23})\simeq -
\frac{3\dmatm}{32\pi^2v_T^2}(1-rc_{12}^2)\sin(2\theta_{23})\,,\\
\label{dott13eap}%
\dot{\theta}_{13}^e &\simeq&- \frac{3r\dmatm}{32\pi^2v_T^2}
\frac{y_e^2(y_\tau^2-y_\mu^2)}
{(y_\mu^2-y_e^2)(y_\tau^2-y_e^2)}\sin(2\theta_{12})
\sin(2\theta_{23})\cos\delta\simeq 0\,,
\end{eqnarray}
where the final results correspond to the limit
$y_{e,\mu}\rightarrow 0$. For the Majorana phases we find
$\dot{\alpha}_{1,2}^e\simeq 0$, at zeroth order in $\theta_{13}$,
while for $\dot{\delta}^e$ we have
\begin{equation}
\label{dotdeltaeap}%
\dot{\delta}^e\simeq\frac{3r\dmatm}{32\pi^2 v_T^2}
\frac{y_e^2\,(y_\tau^2-y_\mu^2)} {(y_\mu^2- y_e^2) (y_\tau^2-y_e^2)}
\theta^{-1}_{13}\sin(2\theta_{12}) \sin(2\theta_{23})\sin\delta\,.
\end{equation}
Comparing our results for
the charged-lepton contribution to the running of the mixing angles
and CP-phases (in the $y_{e,\mu}\rightarrow 0$ limit) with those of
Ref.~\cite{Schmidt:2007nq}, we find a general agreement.
The only difference, which is the consequence of the
discrepancies in the RGE of $Y_e$ (see the paragraph preceding
Section~\ref{sec41}), is the overall factor
of 3 in Eqs.~(\ref{dott12eap})-(\ref{dotdeltaeap}) which
replaces the factor of 3/2 of \cite{Schmidt:2007nq}.

From Eqs.~(\ref{dott12eap})-(\ref{dott13eap}) one immediately
concludes that $\dot{\theta}_{12,13}^e$ are mainly controlled by the
factor $r\dmatm/v_T^2=y_2^2-y_1^2$, with $\dot{\theta}_{13}^e$
further suppressed by $y_e^2/y_\mu^2 \ll 1$. On the other hand,
$\dot{\theta}_{23}^e$ is essentially governed by
$\dmatm/v_T^2=y_3^2-y_1^2$. Therefore, since $r$ is small we expect
larger RG effects on $\theta_{23}$ than on the remaining mixing
angles. This is in contrast with what happens for the contributions
$\dot{\theta}_{ij}^\nu$ shown in (\ref{RGE13ap}), where the running
is typically enhanced for $\theta_{12}$ with respect to
$\theta_{13,23}$ due to an $1/r$ factor present in
$\dot{\theta}_{12}^\nu$. Since the
overall signs of $\dot{\theta}_{12,23}^\nu$ and
$\dot{\theta}_{12,23}^e$ are the same, both the
neutrino and charged-lepton contributions to $\dot{\theta}_{12,23}$
tend to affect the RG flow of these mixing angles in the same way.
As for the running of $\theta_{13}$, the main contribution to
$\dot{\theta}_{13}$ comes from $\dot{\theta}_{13}^\nu$ which, for a
given value of $m_i$, may be positive or negative depending on the
values of $\alpha_{1,2}$ and $\delta$.

The RGEs for the neutrino masses $m_i$ and the charged-lepton Yukawa
couplings $y_{e_i}$ can be obtained from Eqs.~(\ref{rgemi}) and
(\ref{rgeye}) with $\alpha_\nu$, $\alpha_e$ and $P_{\nu,e}^\prime$
given as in (\ref{alphanu}), (\ref{alphae}) and (\ref{PeprimeII}),
respectively. Here we focus on the running of the parameter $r$
defined in (\ref{rdef}). The value of $r$ is crucial in model
building since, although not affected by overall factors in the
effective neutrino mass matrix, it is sensitive to the flavour
structure of $m_\nu$. At low energies $|r(m_Z)|\simeq 0.03$ (see
Section~\ref{sec42}).

We consider two types of neutrino mass spectra,
\begin{eqnarray}
\label{NOIO}%
{\rm Normally-ordered\,(NO)}:\,m_1 < m_2 < m_3\,,\\
{\rm Inversely-ordered\,(IO)}:\,m_3 < m_1 < m_2\,,
\end{eqnarray}
in such a way that $r$ is positive (negative) for the NO (IO) case.
Using Eqs.~(\ref{alphanu}), (\ref{rgemi}) and (\ref{PeprimeII}), we
can write the RGE for $r$ as~\footnote{Notice that the following
equations are equivalent to each other. We choose to write them
differently for the NO and IO cases to better identify the terms
which are proportional to $m_{1,3}^2/\dmatm$.}:
\begin{eqnarray}
\label{rhodot}%
{\rm NO}:\,4\pi^2 \dot{r}&=&-r\Delta P^\prime _{32} +
\frac{m_1^2}{\dmatm}\left(\Delta P^\prime _{21}
-r \Delta P^\prime _{31}\right)\,,\\
{\rm IO}:\,4\pi^2 \dot{r}&=&\Delta P^\prime _{21}(r-1) +
\frac{m_3^2}{\dmatm}\left(\Delta P^\prime _{21} -r \Delta P^\prime
_{31}\right)\,,
\end{eqnarray}
with $\Delta P^\prime _{ij}\equiv(P_\nu^\prime)_{ii}-
(P_\nu^\prime)_{jj}$. To better distinguish the two main sources of
RG effects, we express $\dot{r}$ in the form:
\begin{equation}
\label{rdecomp}%
r=\dot{r}_e+\dot{r}_{\!_T}\,,
\end{equation}
where $\dot{r}_e$ and $\dot{r}_{\!_T}$ denote the terms depending on
the charged-lepton and $Y_T$ Yukawa couplings $y_{e_i}$ and $y_i$,
respectively. The contributions $\dot{r}_e$ originate from the first
term of $P_\nu^\prime$ (see Eq.~(\ref{PeprimeII})) and therefore
will depend on $y_{e,\mu,\tau}^2$ and on the neutrino mixing
parameters. At leading order in $\theta_{13}$ (and keeping only the
terms proportional to $y_\tau^2$), we obtain for both the NO and IO
neutrino mass spectra:
\begin{eqnarray}
\label{rhodotnuNO}%
{\rm NO}:\,\dot{r}_e&=&-\frac{y_\tau^2}{16\pi^2} \left\{
\frac{m_1^2}{\dmatm}
\left[r+3r\cos(2\theta_{23})-2(2-r)\cos(2\theta_{12})s_{23}^2\right]+
\right.\nonumber
\\
&+&\left.r\left[1+3\cos(2\theta_{23})-2\cos(2\theta_{12})s_{23}^2\right]
\right\}\,,\\
\label{rhodotnuIO}%
{\rm IO}:\,\dot{r}_e&=&-\frac{y_\tau^2}{16\pi^2} \left\{
\frac{m_3^2}{\dmatm}
\left[r+3r\cos(2\theta_{23})-2(2-r)\cos(2\theta_{12})s_{23}^2\right]+
\right.\nonumber
\\
&+&\left.4(1-r)s_{23}^2\cos(2\theta_{12})\right\}\,.
\end{eqnarray}
It is worth noticing that in the limit of quasi-degenerate neutrinos
($m_1 \simeq m_2 \simeq m_3 \gg \dmatm$) and/or large $\tan\beta$,
the RG effects on $r$ due to $\dot{r}_e$ are enhanced. For
hierarchical (HI) neutrino masses (NO with $m_1 \ll \dmsol$), we do
not expect major effects due to the suppression factor of $r$
present in the second term of (\ref{rhodotnuNO}). Yet, in the inverted-hierarchical (IH)
situation (IO with $m_3 \ll \dmsol$) there is an unsuppressed term
in Eq.~(\ref{rhodotnuIO}) which, depending on the value of
$\tan\beta$, may lead to a considerable running of $r$.

The remaining contribution to $\dot{r}$ (denoted by $\dot{r}_T$ in
Eq.~(\ref{rdecomp})) does not depend directly on the neutrino mixing
angles since it originates from the second term of $P_\nu^\prime$,
which is diagonal. From Eqs.~(\ref{rhodot}) we obtain
\begin{equation}
\label{rhodotT}%
\dot{r}_{\!_T}=-\frac{3\,(y_3^2-y_1^2)}{4\pi^2}\,r(1-r)=-\frac{3\,\dmatm}
{4\pi^2v_T^2}\,r(1-r)\,,
\end{equation}
for both the NO and IO cases. Contrarily to the results
obtained for $\dot{r}_e$, the above equation is exact in the sense
that it does not rely on any expansion nor on any special limit of
the couplings. An immediate conclusion that can be drawn from
(\ref{rhodotT}) is that, although $\dot{r}_{\!_T}$ is always
negative, $|r|$ decreases (increases) when going from low to high
energies, for a NO (IO) neutrino mass spectrum. Clearly, this is
only true in the limit of negligible $\dot{r}_e$.
%

\subsection{Numerical examples}
\label{sec42}%

\TABLE[!ht]{ \label{Table3} %
\caption{Best-fit values (with 1$\sigma$ errors) and 3$\sigma$
allowed intervals for the neutrino oscillation parameters from
global data including solar, atmospheric, reactor (KamLAND and
CHOOZ) and accelerator (K2K and MINOS)
experiments~\cite{Schwetz:2008er}.}
\renewcommand{\tabcolsep}{1.1pc}
\begin{tabular}{lcc} \hline  \hline \noalign{\medskip}
Parameter & Best-fit & $3\sigma$
\\
\noalign{\medskip} \hline \hline  \noalign{\medskip} %
$\Delta m^2_{21}\: [10^{-5}\,{\rm eV}^2]$
& $7.65^{+0.23}_{-0.20}$
& 7.05--8.34  \\ \noalign{\medskip} \hline
\noalign{\medskip}
$|\Delta m^2_{31}|\: [10^{-3}\,{\rm eV}^2]$
&$2.40^{+0.12}_{-0.11}$
& 2.07--2.75 \\
\noalign{\medskip} \hline \noalign{\medskip}
$\sin^2\theta_{12}$
& $0.304^{+0.022}_{-0.016}$
& 0.25--0.37\\
\noalign{\medskip} \hline \noalign{\medskip}
$\sin^2\theta_{23}$
& $0.50^{+0.07}_{-0.06}$
& 0.36--0.67\\
\noalign{\medskip} \hline \noalign{\medskip}
$\sin^2\theta_{13}$
& $0.01^{+0.016}_{-0.011}$
& $\leq 0.056$\\
\noalign{\medskip} \hline \noalign{\medskip}
$|r|$
& 0.032
&0.027--0.038\\
\noalign{\medskip} \hline \hline \noalign{\medskip}
\end{tabular}
}

At low energies, the neutrino mixing angles and mass squared
differences are constrained by solar, atmospheric, reactor (KamLAND
and CHOOZ) and accelerator (K2K and MINOS) neutrino oscillation experiments. The
results of a global analysis~\cite{Schwetz:2008er} of the data
provided by these experiments are summarised in Table~\ref{Table3}
where the best-fit values and $3\sigma$ intervals for $\theta_{ij}$,
$\dmsol$ and $|\dmatm|$ (as well as for $|r|$) are presented.

In the following, we will show some numerical examples with the aim of quantifying
the RG effects on the neutrino mass and mixing
parameters due to the presence of the heavy triplets $T$ and
$\bar{T}$. We adopt the bottom-up approach, {\em i.e.} we start at
$\Lambda=m_Z$ with the best-fit values of the low-energy neutrino
parameters and evolve the full set of the MSSM RGEs up to
$\Lambda=M_T$. At this scale, we extract $Y_T$ according to
Eq.~(\ref{matchII}) and run the TMSSM RGEs to the
GUT scale $M_G\simeq 2\times 10^{16}\, {\rm GeV}$. The RG effects
in the neutrino mixing matrix $U$ and in the neutrino masses $m_i$ are
governed by Eqs.~(\ref{Udot}) and (\ref{rgemi}), respectively.
We will only consider the cases of HI and IH neutrino mass
spectra, for which the contributions to the running coming from the
neutrino sector are, in general, suppressed with respect to what
happens in the quasi-degenerate limit. In addition, only the
results for $r$ and the mixing angle
$\theta_{23}$ will be shown since, as discussed in the previous section,
the RG effects induced by $Y_T$ are less important for $\theta_{12,13}$.

\FIGURE[!ht]{ \label{fig2} \caption{Left plot: Values of $r$ and
$|r|$ at $\Lambda=M_G$ for the HI (lower branch) and IH (upper
branch) cases as a function of the largest $y_i$ coupling ($y_3$ for
HI and $y_2$ IH). The blue-filled regions correspond to the
variation of $r$ and $|r|$ in the interval $M_T=10^9$ GeV
(black-dashed line) to $10^{14}$ GeV (black-solid line) and
$\tan\beta=5$. The blue-solid (dashed) line shows the result for
$\tan\beta=50$ and $M_T=10^9\, (10^{14})$ GeV. The horizontal pink
bar denotes the low-energy $3\sigma$ allowed range for $|r|$ as
given in Table~\ref{Table3}, while the best-fit value is indicated
by the horizontal dash-dotted line. Right plot: the same as in the
left plot but for the mixing angle $\theta_{23}$. For both plots we
used $\theta_{13}(m_Z)=0$ and the best-fit values for the remaining parameters
($\theta_{12}$, $\theta_{23}$, $\dmsol$ and $|\dmatm|$).
All CP-violating phases are set to zero.}
\begin{tabular}{cc}
\includegraphics[width=7.35cm]{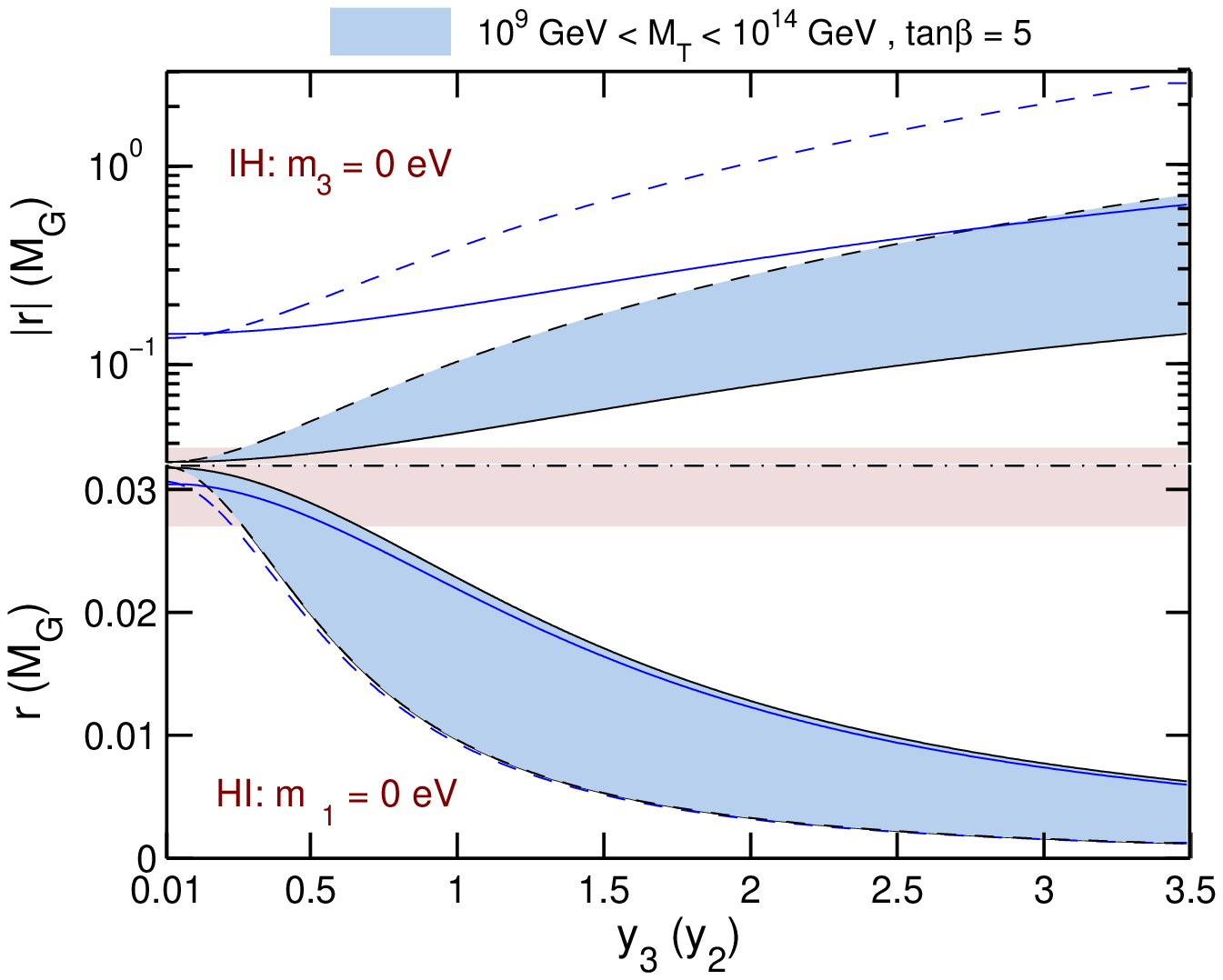} &
\includegraphics[width=7.1cm]{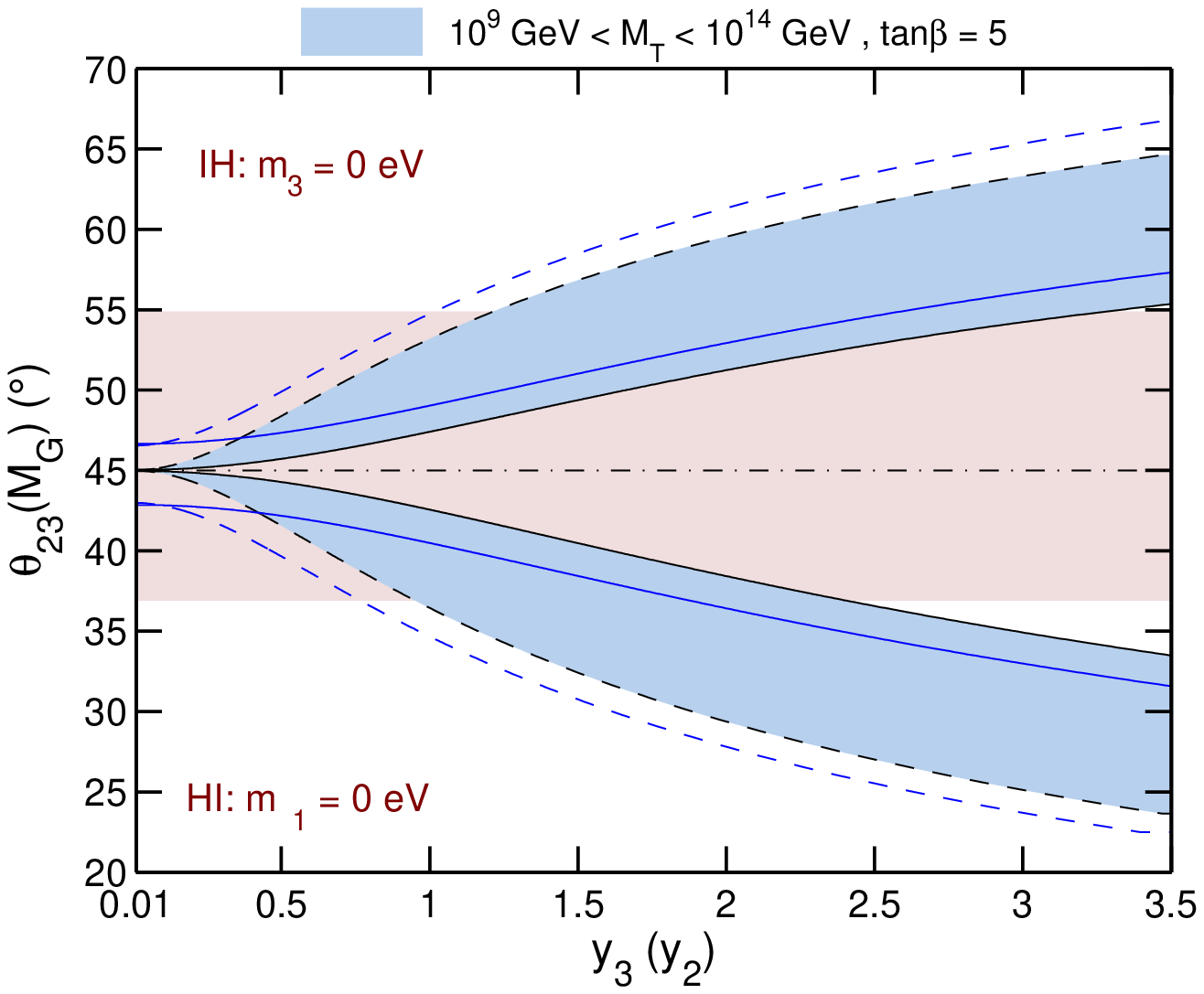}
\end{tabular}
}
In the lower (upper) part of the left plot in Fig.~\ref{fig2}, we
show the values of $r(M_G)$ ($|r(M_G)|$) as a function of the
coupling $y_3$ ($y_2$) for the HI (IH) case~\footnote{The values of
$y_3$ and $y_2$ (given at $\Lambda=M_G$) are extracted using
Eqs.~(\ref{replac}) and taking appropriate ranges for $\lambda_2$
at each value of $M_T$.}. The triplet-mass $M_T$ (always given at
$\Lambda=M_T$) is varied between $10^9$~GeV and $10^{14}$~GeV. These
limits correspond to the solid and dashed curves (in black for
$\tan\beta=5$ and in blue for $\tan\beta=50$), respectively. From
Eqs.~(\ref{rdecomp})-(\ref{rhodotT}) we obtain for the cases
under discussion:
\begin{eqnarray}
\label{rhodoHI}%
{\rm HI}:\,\dot{r}&\simeq&-\frac{y_\tau^2}{16\pi^2}r\left[1+3 \cos
(2\theta_{23})- 2\cos(2\theta_{12})s_{23}^2\right]-\frac{3\,y_3^2}
{4\pi^2}\,r(1-r)
\,,\\
\label{rhodotIH2}%
{\rm IH}:\,\dot{r}&\simeq&-\frac{y_\tau^2}{4\pi^2} (1-r)s_{23}^2
\cos(2\theta_{12}) +\frac{3\,y_1^2}{4\pi^2}\,r(1-r)\,.
\end{eqnarray}
For small $\tan\beta$ and $y_i \ll 1$ we do not expect large
RG running effects on $r$ since $\dot{r}_e$ and $\dot{r}_{\!_T}$ are
suppressed by the small couplings $y_\tau$ and $y_i$. This
is true for both the HI and IH cases, as confirmed by the
left plot of Fig.~\ref{fig2}. For small $y_i$ and
$\tan\beta=5$, the value of $r(M_G)$ is very close to
$r(m_Z)=0.032$ (see Table~\ref{Table2}), indicated by the
horizontal dash-dotted line. As the couplings $y_i$ increase,
$\dot{r}_{\!_T}$ increases and $\dot{r}$ is mainly given by the
second term on the right-hand side of Eqs.~(\ref{rhodoHI}) and
(\ref{rhodotIH2}). As expected, $r$ decreases from low to
high energies for the HI case since $\dot{r}_{\!_T}$ is negative and
$r(m_Z)>0$. In contrast, although $\dot{r}_{\!_T} <0$ also in the IH
limit, $r(m_Z)<0$ and therefore $|r|$ increases when going from
$m_Z$ to $M_G$. Notice that for $y_{3,2}\sim 1$ the value of
$r(M_G)$ is outside the $3\sigma$ low-energy allowed region
even for the largest allowed value of $M_T$. In the small
$\tan\beta$ limit, we obtain the following approximate results
\begin{eqnarray}
\label{rhoapHI}%
{\rm HI}&:&\,r(M_G)\simeq r_0\left[r_0+(1-r_0)\left(\frac{M_G}{M_T}\right)^
{\frac{3y_3^2}{4\pi^2}}\right]^{-1}
\!\!\!\!\simeq (9.1\times 10^{-3}\,,\,2.2\times 10^{-2})\,,\\
\label{rhodotIH}%
{\rm IH}&:&\,|r(M_G)|\simeq
|r_0|\left[-|r_0|+(1+|r_0|)\left(\frac{M_T}{M_G}
\right)^{\frac{3y_1^2}{4\pi^2}}\right]^{-1} \!\!\!\!\simeq (0.1
\,,\,4.9\times 10^{-2}) \,,
\end{eqnarray}
where $r_0\equiv r(m_Z)$. The numbers in parentheses correspond to
the estimates for $M_T=10^9$~GeV and $10^{14}$~GeV, respectively,
taking $y_{3,1}=1$.

For the HI case, the large $\tan\beta$ results are similar to the
small $\tan\beta$ ones, with the RG correction reaching approximately $10\%$ for
small values of $y_i$ and $\tan\beta=50$ (blue curves on the lower
part of the left plot in Fig.~\ref{fig2}). This correction is small
due to the fact that the term proportional to $y_\tau^2$ in
Eq.~(\ref{rhodoHI}) is suppressed by $r$.
Instead, a large effect is observed in the IH limit with large
$\tan\beta$ since the first term in Eq.~(\ref{rhodotIH2}) is not
suppressed by $r$. Therefore, in this case the contribution proportional to
$y_\tau^2$ is important even for small values of $y_2$. We find
$|r(M_G)|\simeq 0.2$ for $y_2 \ll 1$ and $\tan\beta =50$.

Let us now turn to the RG effects to $\theta_{23}$.
From Eqs.~(\ref{RGE23ap}) and (\ref{dott23eap}) we obtain
\begin{eqnarray}
\label{dt23apHI}%
{\rm HI}&:&\,\dot{\theta}_{23}\simeq -\frac{y_\tau^2}{32\pi^2}
\frac{1- r\cos(2\theta_{12}) +2c_{12}^2\sqrt{r}}{1-r}
\sin(2\theta_{23})-\frac{3y_3^2}{32\pi^2}(1-rc_{12}^2)\,\\
\label{dt23apIH}%
{\rm IH}&:&\,\dot{\theta}_{23}\simeq \frac{y_\tau^2}{32\pi^2}
\sin(2\theta_{23})+ \frac{3y_1^2}{32\pi^2}(1-rc_{12}^2)\,,
\end{eqnarray}
which show that $\theta_{23}$ decreases (increases) from low to high
energies for the HI (IH) neutrino mass spectrum. This can also be
seen in the right plot of Fig.~\ref{fig2} where $\theta_{23}(M_G)$
is plotted as a function of $y_{3}$ ($y_2$) for the HI (IH) case.
When $y_i\gtrsim 1$ the values of $\theta_{23}(M_G)$ are outside the
$3\sigma$ low-energy allowed interval for $\theta_{23}$ (shown in
pink). Similarly as for $r$, the running effects are negligible for
small $\tan\beta$ (black solid and dashed curves) and $y_i \ll 1$.
The value of $\theta_{23}$ at $\Lambda = M_G$ can be roughly
approximated by
\begin{equation}
\label{t23ap}%
\theta_{23}(M_G)\simeq\theta_{23}(m_Z)\mp \frac{m_\tau^2\tan^2\beta}
{32\pi^2v^2}\ln\!\left( \frac{M_G}{m_Z}\right)\mp
\frac{3\,y_{3,1}^2}{32\pi^2}\ln\!\left( \frac{M_G}{M_T}\right)\,,
\end{equation}
where the minus (plus) sign corresponds to the HI (IH) case. As can
be seen from the right plot in Fig.~\ref{fig2}, the shape of the
curves is nearly the same for different values of $\tan\beta$. The
results differ due to the second term on the right-hand side of
Eq.~(\ref{t23ap}), which increases (in absolute value) with
increasing $\tan\beta$. Notice that the overall sign of that term is
negative (positive) for the HI (IH) case, which explains the
downward (upward) displacement of the $\tan\beta=50$ curves with
respect to the $\tan\beta=5$ ones.
%
%
%
%
\section{LFV $\ell_i\rightarrow \ell_j \gamma$ decays in the TMSSM}
\label{sec5}
The observation of lepton-flavour violating processes like
$\ell_i\rightarrow \ell_j \gamma$ would definitely point towards the
existence of new sources of lepton flavour violation and/or new
physics close to the electroweak scale. So far, none of the
aforementioned processes has been observed. The current upper bounds
for their branching ratios (BRs) are:
\begin{eqnarray}
\label{muelim}%
{\rm BR}(\mu\rightarrow e \gamma)&\leq& 1.2\times 10^{-11}\,
\cite{Brooks:1999pu}\,,\\
\label{tauelim}%
{\rm BR}(\tau\rightarrow e \gamma)&\leq& 1.1\times 10^{-7}\,
\cite{Aubert:2005wa}\,\,(9.4\times 10^{-8})\,,\\
\label{taumulim}%
{\rm BR}(\tau\rightarrow \mu \gamma)&\leq& 4.5\times 10^{-8}\,
\cite{Hayasaka:2007vc}\,\,(1.6\times 10^{-8})\,,
\end{eqnarray}
where the numbers in parentheses report the results for the $\tau$
decays obtained through a combined analysis of BABAR and Belle
data~\cite{Banerjee:2007rj}.

The above limits severely constrain the MSSM LFV soft SUSY-breaking
mass matrices forcing them to be small. In the slepton sector,
flavour violation can be generated in the presence of superpotential
renormalisable interactions through RG effects. The most typical
example is having the off-diagonal slepton masses arising due to the
Dirac neutrino Yukawa couplings which participate in the type I
seesaw mechanism of neutrino mass
generation~\cite{Borzumati:1986qx}. In spite of the huge amount of
work done in the direction of establishing a direct connection
between low-energy neutrino data and lepton-flavour violation in the
SUSY type I seesaw, such goal cannot be achieved in a
model-independent way (for a discussion see {\em e.g.}
Ref.~\cite{Davidson:2004wi}). This stems from the impossibility of
reconstructing the high-energy neutrino couplings from low-energy
neutrino data. In the triplet-seesaw case the situation is improved
since the effective neutrino mass matrix is linear on the couplings
$Y_T$ (see Eq.~\ref{matchII}) and therefore, in general, the flavour
structure of $m_\nu$ is the same as the one of $Y_T$.


\subsection{Approximate predictions for the $\ell_i\rightarrow \ell_j
\gamma$ rates}
\label{sec51}

Let us consider the slepton soft SUSY-breaking lagrangian,
\begin{equation}
\label{SSBL}%
\mathcal{L}_{\rm soft} = \tilde{L}^\dag\, \mlt \tilde{L} + \tilde{e}^c
m^2_{\tilde{e^c}} \tilde{e^c}^\dag + (H_1 \tilde{e^c} A_e \tilde{L} + {\rm H.c.})\,,
\end{equation}
where $\mlt$ and $m^2_{\tilde{e^c}}$ are the SUSY-breaking masses for the
slepton doublets and singlets respectively, and $A_e$ the trilinear terms.
Starting at the scale $\Lambda > M_T$ with universal
boundary conditions $\mlt=m^2_{\tilde{e^c}}=m_0^2\openone$ and $A_e=A_0Y_e$,
at the scale $M_T$ one approximately has~\cite{Rossi:2002zb}
\begin{eqnarray}
\label{m2lap}%
(\mlt)_{ij}&\simeq& -\frac{9m_0^2+3A_0^2}{8\pi^2}(Y_T^\dag Y_T^{})_{ij}\log
\frac{\Lambda}{M_T}\,, \\
\label{m2ecap}%
(m^2_{\tilde{e^c}})_{ij}&\simeq& 0\,,  \\
\label{Aeap}%
(A_e)_{ij}&\simeq& -\frac{9}{16\pi^2}A_0(Y_eY_T^\dag Y_T^{})_{ij}\log
\frac{\Lambda}{M_T} \,,
\end{eqnarray}
where $i\neq j$ and $Y_T$ and $Y_e$ are taken at the $M_T$
scale\footnote{In this approximation the small RG running effects on
$Y_T$ and $Y_e$ between the scales $\Lambda$ and $M_T$ are
neglected.}. The above LFV terms may be large enough to generate
observable rates for LFV processes like radiative charged-lepton
decays $\ell_i \rightarrow \ell_j \gamma$. Neglecting small effects
of the RG running from $M_T$ to $m_Z$ of the LFV entries of
$(\mlt)$, the quantities on the left-hand sides of
Eqs.~(\ref{m2lap})-(\ref{Aeap}) can be identified with their values
at low-energies. Assuming that only the LFV coming from
$(\mlt)_{ij}$ is relevant, and keeping the $\tan\beta$ enhanced
contributions to the one-loop amplitudes, one can roughly
approximate ${\rm BR}(\ell_i \rightarrow \ell_j \gamma)$ by
\begin{equation}
\label{BRap}%
{\rm BR}(\ell_i \rightarrow \ell_j\gamma) \simeq \frac{48\pi^3 \alpha}{G_F^2}\,
|C_{ij}|^2 \tan^2\!\beta\, {\rm BR}(\ell_i \rightarrow \ell_j\nu_i\bar{\nu}_j)\,,
\end{equation}
where $\alpha$ and $G_F$ are the fine-structure and Fermi constants,
respectively, and ${\rm BR}(\mu \rightarrow
e\nu_\mu\bar{\nu}_e)=0.1737$ and ${\rm BR}(\tau \rightarrow \mu
\nu_\tau\bar{\nu}_\mu)=0.1784$. The coefficients $C_{ij}$ summarise
the dependence of the $\ell_i \rightarrow \ell_j \gamma$ decay rate
on the LFV entries of the slepton mass matrices and masses of the
SUSY particles running in the relevant loop diagrams. In the
simplest approximation, one has
\begin{equation}
\label{Cijap}%
C_{ij} \sim \frac{g_2^2}{16\pi^2}\frac{(\mlt)_{ij}}{m_S^4}\;,\; (i\neq j=e,\mu,\tau)\,,
\end{equation}
where $g_2$ is the SU(2) gauge-coupling constant and $m_S$ denotes a
common mass scale for the SUSY particles participating in the
process. In general, the above approximations do not
provide an accurate result for the value of ${\rm BR} (\ell_i
\rightarrow \ell_j\gamma)$. For this reason, it is convenient to
work with ratios of BRs instead of the BRs themselves. We will therefore
consider the quantities
\begin{equation}
\label{RBRS}%
R_{\tau j}\equiv \frac{{\rm BR}(\tau \rightarrow \ell_j\gamma)}
{{\rm BR}(\mu \rightarrow e\gamma)}\;,\; j=e,\mu,
\end{equation}
which in the approximation (\ref{Cijap}) are given by
\begin{equation}
\label{RBRS1}%
R_{\tau j}\simeq \left|\frac{(\mlt)_{\tau j}}{(\mlt)_{\mu e}}
\right|^2 {\rm BR}(\tau \rightarrow \ell_j \nu_\tau \bar{\nu}_j)\,,
\end{equation}
{\em i.e.}, depend only on the relative strength of LFV in the
different channels\footnote{As it will be discussed in the next
section, this statement is only valid in the limit of
quasi-degenerate masses for the three slepton generations.}. Using
the approximate expression for $(\mlt)_{ij}$ given in
Eq.~(\ref{m2lap}), the ratios of LFV entries of the slepton mass
matrix appearing in Eq.~(\ref{RBRS1}) can be expressed in terms of
the couplings $Y_T$ (taken {\em e.g.} at the scale $M_T$) as
\begin{equation}
\label{m2lYTa}%
\left|\frac{(\mlt)_{\tau j}}{(\mlt)_{\mu e}} \right|^2 \simeq
\left|\frac{(Y_T^\dag Y_T^{})_{\tau j}}{(Y_T^\dag Y_T^{})_{\mu e}}
\right|\,.
\end{equation}
Unlike the seesaw type I models, $Y_T$ at
the $M_T$ or $\Lambda$ scale can in the TMSSM be uniquely computed if values
of the neutrino parameters are specified at $m_Z$. In this section, following
the common procedure, we will set
\begin{equation}
\label{m2lYT}%
\left|\frac{(Y_T^\dag Y_T^{})_{\tau j}}{(Y_T^\dag Y_T^{})_{\mu e}}\right|^2_{\Lambda=M_T}
\!\!\!\!\simeq \left|\frac{(m_\nu^\dag m_\nu^{})_{\tau j}}{(m_\nu^\dag m_\nu^{})_{\mu e}}\right|^2
_{\Lambda=m_Z}\,,
\end{equation}
neglecting the RG running of the neutrino mass matrix between $m_Z$
and $M_T$. We will assess the quality of this approximation in the
next sections.

We now use Eq.~(\ref{diags}) and the parameterisation of the mixing
matrix $V$ adopted in (\ref{Uparam}) and (\ref{Udelta}) to express
$(\mlt)_{ij}$ in terms of the low-energy neutrino parameters:
\begin{eqnarray}
\label{m2lmue}%
|(\mlt)_{\mu e}|^2 &\propto& \frac{|\dmatm|}{4\,v_T^2}\,c_{13}^2\left[r^2c_{23}^2
\sin^2(2\theta_{12})+a^2s_{13}^2s_{23}^2+a|r|s_{13}c_\delta\sin(2\theta_{12})\sin(2\theta_{23})
\right]\,,\\
\label{m2ltaue}%
|(\mlt)_{\tau e}|^2 &\propto& \frac{|\dmatm|}{4\,v_T^2}\,c_{13}^2\left[r^2s_{23}^2\sin^2(2
\theta_{12})+a^2s_{13}^2c_{23}^2-a|r|s_{13}c_\delta\sin(2\theta_{12})\sin(2\theta_{23})\right]
\,,\\
\label{m2ltaumu}%
|(\mlt)_{\tau \mu}|^2 & \propto & \frac{|\dmatm|}{64\,v_T^2} \left\{[\,4|r|s_{13}\,c_\delta
\sin(2\theta_{12})\cos(2\theta_{23})+[2\,b\,c_{13}^2-|r|(\cos(2\theta_{23})-3)\cos(2
\theta_{12})]\right.\nonumber\\%
&&\left.\times\sin(2\theta_{23})\,]^2+16\, r^2c_\delta s_{13}^2\sin(2\theta_{12})
\sin(2\theta_{23})\right\}\,,
\end{eqnarray}
where $v_T$ and $r$ have been defined in Eqs.~(\ref{replac}) and
(\ref{rdef}). The above expressions are valid for both
the NO and IO neutrino mass spectrum with $a$ and $b$ defined as:
\begin{eqnarray}
\label{ab}%
{\rm NO:}&& \,\,a=2\,(1-|r|s_{12}^2)\simeq 2\;\;\;,\;\;\; b=-2+|r|\simeq -2\,,\\
{\rm IO:} &&\,\,a=-2\,(1+|r|s_{12}^2)\simeq -2\;\;\;,\;\;\; b=2+|r|\simeq 2\,.
\end{eqnarray}
From the results shown in (\ref{m2lmue})-(\ref{m2ltaumu}), one
immediately concludes that the LFV elements of the soft breaking
masses do not depend on the Majorana phases
$\alpha_{1,2}$~\cite{Rossi:2002zb}. Moreover, the mass of the
lightest neutrino ($m_1$ or $m_3$ depending on whether the neutrino
mass spectrum is NO or IO) does not have any impact on the LFV terms
$(\mlt)_{ij}$, at the one-loop level~\cite{Joaquim:2006uz}. The
independence of $(m_\nu^\dag m_\nu^{})_{ij}$ from the Majorana
phases and absolute neutrino mass scale does not hold once we
consider the two-loop RGEs for the soft masses. In this case, terms
of the type
\begin{eqnarray}
\label{2majme2l}%
(\mlt)_{ij}&\propto& \frac{m_0^2,A_0^2}{(16\pi^2)^2}\,[Y_T^\dag (Y_e^\dag Y_e^{})^T Y_T^{}]_{ij}=
\frac{m_0^2,A_0^2}{(16\pi^2)^2v_T^2}\sum_{k=1}^3y_{e_k}^2[m_\nu^{\dag}]_{ik}[m_\nu^{}]_{kj}\,,\\
\label{2mim2l}%
(\mlt)_{ij}&\propto& \frac{m_0^2,A_0^2}{(16\pi^2)^2}\,{\rm Tr}(Y_T^\dag Y_T^{})[Y_T^\dag Y_T^{}]_{ij}=
\frac{m_0^2,A_0^2}{(16\pi^2)^2v_T^4}[m_\nu^{\dag}m_\nu^{}]_{ij}\sum_{k=1}^3 m_k^2\,,\\
\label{2mim2l4}%
(\mlt)_{ij}&\propto& \frac{m_0^2,A_0^2}{(16\pi^2)^2}\,[Y_T^\dag Y_T^{}Y_T^\dag Y_T^{}]_{ij}=
\frac{m_0^2,A_0^2}{(16\pi^2)^2v_T^4}[m_\nu^{\dag}m_\nu^{}m_\nu^{\dag}m_\nu^{}]_{ij}
\end{eqnarray}
will be generated. The contributions of the form (\ref{2majme2l})
do depend on $\alpha_{1,2}$ and those of the form (\ref{2mim2l}) and (\ref{2mim2l4})
depend on the mass of the lightest neutrino mass. However, being a
two-loop effect, all those terms are negligible when compared with
the one-loop ones. Hence, the quantities $R_{\tau j}$ will be
mainly sensitive to $\theta_{13}$ and $\delta$.

The results for the ratios $R_{\tau\mu}$ and $R_{\tau e}$ obtained within the
approximation described above and using the latest neutrino oscillation data
summarised in Table~\ref{Table3} are shown in Fig.~\ref{fig3} as functions
of $s_{13}$ for both NO (left plots) and IO (right plots) cases and for the
entire possible range of $\delta$ (see also Ref.~\cite{Rodejohann:2008xp}).
\FIGURE[!ht]{ \label{fig3} \caption{Allowed regions for
$R_{\tau\mu}$ (upper plots) and $R_{\tau e}$ (lower plots) given in
Eqs.~(\ref{RBRS}) and (\ref{m2lYT}) as a function of $s_{13}$ and
$\delta$, for both the NO (left plots) and the IO (right plots)
neutrino mass spectra. In dark green (light green) we show the
$3\sigma$ (best-fit) allowed regions obtained by varying the
CP-violating phase $\delta$ in the interval $[0,2\pi]$ and using the
neutrino data displayed in Table~\ref{Table3}. The black solid (blue
dashed) [red dash-dotted] line delimits the $3\sigma$ region for
$\delta=0$ ($\delta=\pi$) [$\delta=\pi/4$].}
\begin{tabular}{cc}
\includegraphics[width=7.2cm]{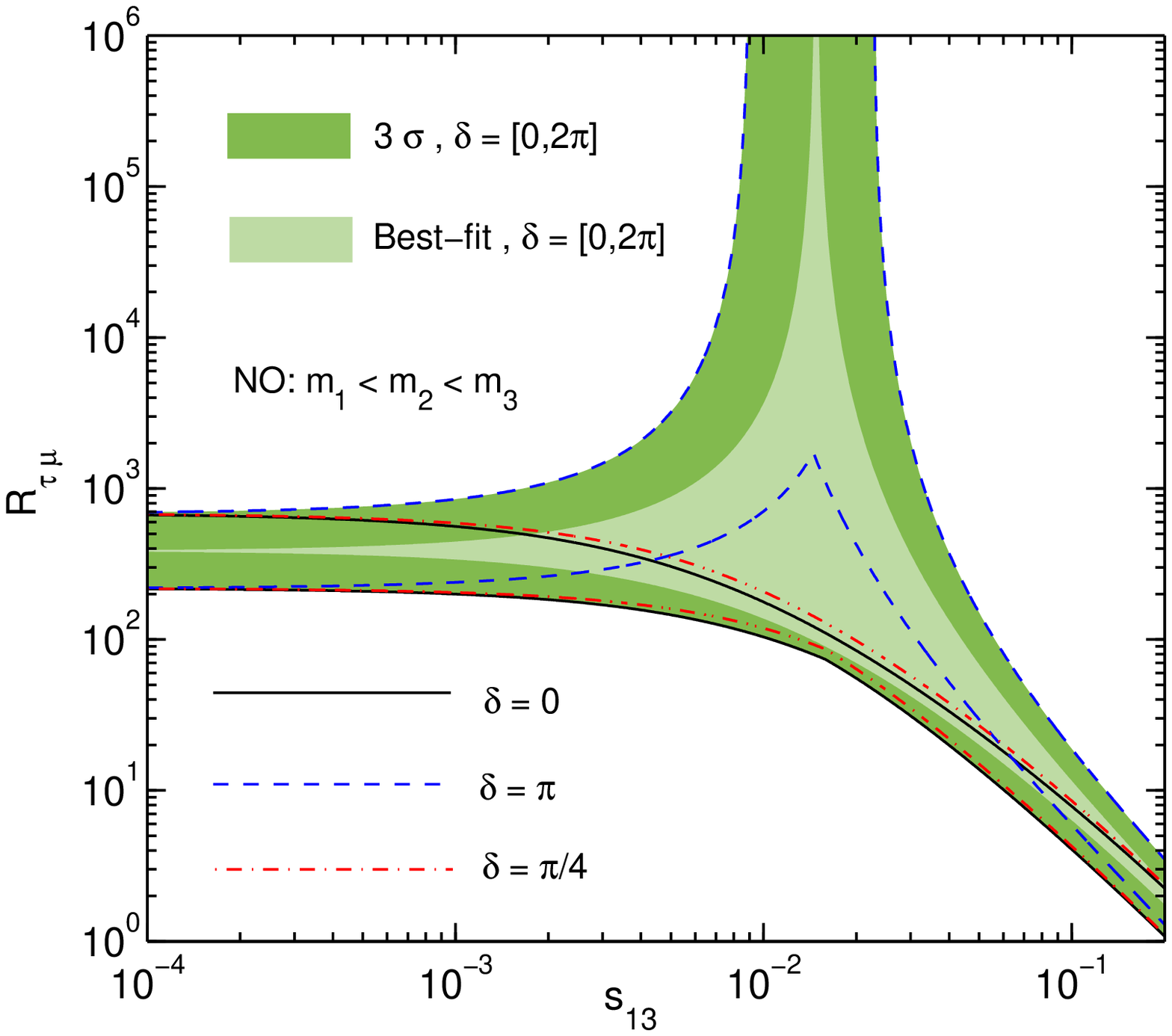} &
\includegraphics[width=7.2cm]{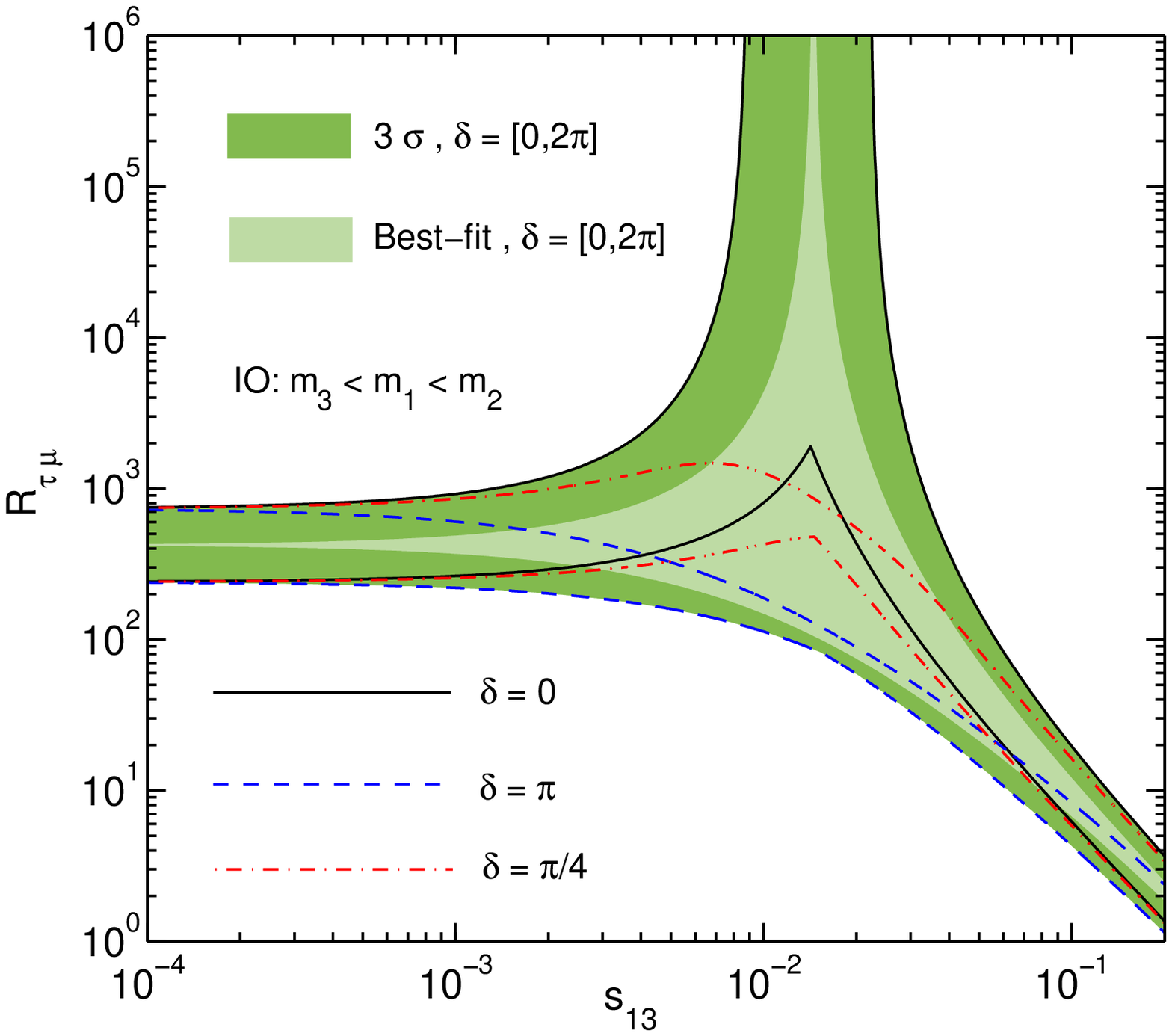}\\
\includegraphics[width=7.2cm]{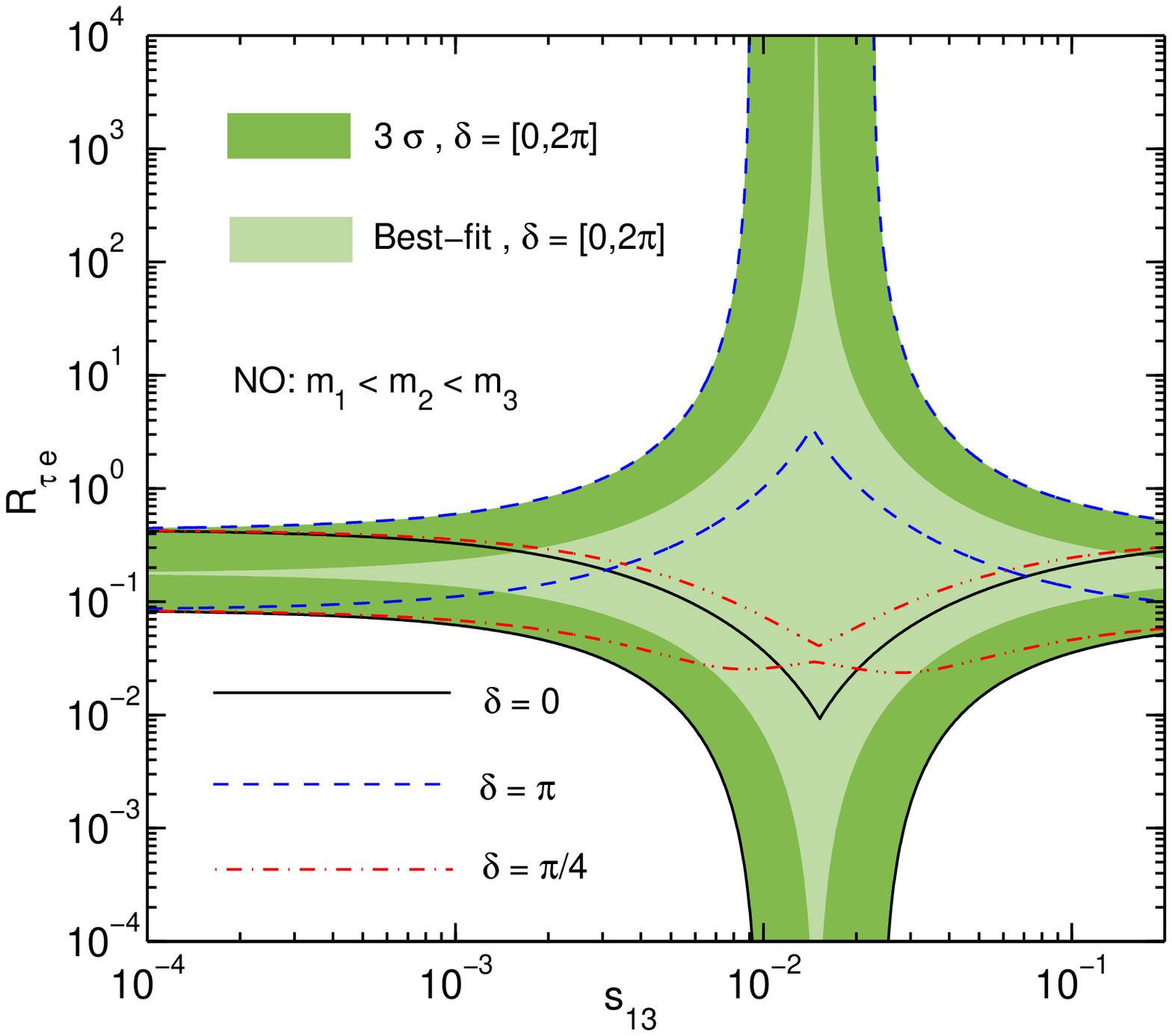} &
\includegraphics[width=7.2cm]{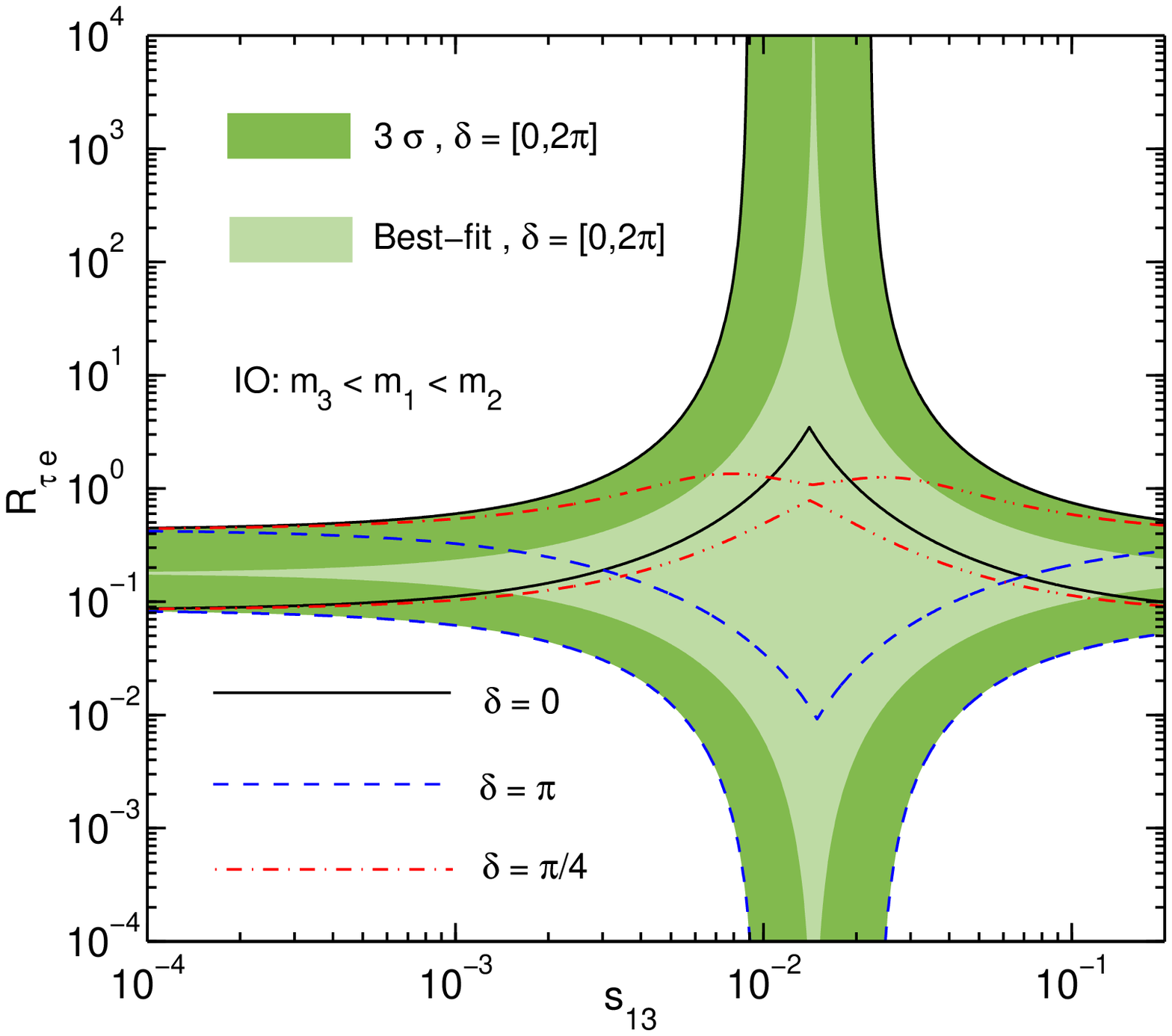}
\end{tabular}
}

For $s_{13}\rightarrow 0$, the ratios $R_{\tau\mu}$ and $R_{\tau e}$ are given by
(the quoted numbers are for the best-fit values of $\theta_{12}$, $\theta_{23}$ and
$r$, given in Table~\ref{Table3})
\begin{eqnarray}
\label{R32s130}%
R_{\tau\mu} &=& \frac{4(1\mp|r| c_{12}^2)^2s_{23}^2}{r^2
\sin^2(2\theta_{12})}\,{\rm BR}(\tau \rightarrow \mu\nu_\tau
\bar{\nu}_{\mu})=404.0^{+18.1\,
{\rm (IO)}}_{-17.7\,{\rm (NO)}}\,,\\
\label{R31s130}%
R_{\tau e} &=& \tan^2\theta_{23}\,{\rm BR}(\tau \rightarrow
e\nu_\tau  \bar{\nu}_{e})=0.18\,,
\end{eqnarray}
where the minus (plus) sign in the first equality of (\ref{R32s130})
corresponds to the case of a NO (IO) neutrino mass spectrum. Taking
into account the uncertainties on the neutrino parameters reported
in Table~\ref{Table3}, the following $3\sigma$ ranges are obtained
\begin{eqnarray}
R_{\tau\mu}=[260\,(238),696\,(751)]\,, R_{\tau e}=[0.8,4.5]\times 10^{-1}\,,
\end{eqnarray}
where the numbers in parentheses correspond to the IO case.

The fact that $R_{\tau\mu}$ is for $s_{13}=0$ larger than $R_{\tau
e}$  is due to the $r^2$ suppression present in both $|(\mlt)_{\mu
e}|^2$ and $|(\mlt)_{\tau e}|^2$, but absent from $|(\mlt)_{\tau
\mu}|$ (see Eqs.~(\ref{m2lmue})-(\ref{m2ltaumu})). The above results
also show that in the limit $s_{13}\rightarrow 0$ the ratio
$R_{\tau\mu}$ is different for the NO and IO cases. This is due to
the fact that, although $|(\mlt)_{e\mu}|$ is the same in both
scenarios, in the IO case $|(\mlt)_{\tau\mu}|$ is slightly larger ,
as can be seen from Eqs.~(\ref{m2ltaumu}) and (\ref{ab}).

The parameters $\theta_{13}$ and $\delta$ turn out to be crucial in
determining the rates of the $\mu e$ and $\tau e$ LFV transitions.
As $\theta_{13}$ increases, $|(\mlt)_{\mu e}|$ and $|(\mlt)_{\tau
e}|$ are either suppressed or enhanced, depending on the sign of
$\cos\delta$. This can be seen from Eqs.~(\ref{m2lmue}) and
(\ref{m2ltaue}) where, for instance, cancelations among different
terms are possible for specific values of $\theta_{13}$. The
condition $(\mlt)_{ij}\rightarrow 0$ automatically implies no
Dirac-type CP violation in the neutrino sector: the Jarlskog CP
invariant $\mathcal{J}$ is proportional to ${\rm Im}[(m_\nu^\dag
m_\nu^{})_{12}(m_\nu^\dag m_\nu^{})_{13}(m_\nu^\dag m_\nu^{})_{23}]$
and since $(m_\nu^\dag m_\nu^{})_{ij}\propto (\mlt)_{ij}$,
$(\mlt)_{ij} \rightarrow 0$ implies $\mathcal{J} \rightarrow
0$~\cite{Branco:2002xf}. However, by itself $\mathcal{J}\rightarrow
0$ does not imply the absence of $\mu-e$ and $\tau-e$ transitions
because the vanishing of $(\mlt)_{\mu e}$ and $(\mlt)_{\tau e}$ for
$\delta=0$ require respectively\footnote{The necessary conditions
for the cancelation of $(m_{\nu}^\dag m_{\nu})_{ij}$ have also been
discussed in Refs.~\cite{Hagedorn:2005kz}, albeit in a different
context.}:
\begin{eqnarray}
\label{s130}%
s_{13}=\mp\frac{1}{2}\frac{|r|\cot\theta_{23}
\sin{2\theta_{12}}}{1\mp|r|s_{12}^2}\;,\;
s_{13}=\pm\frac{1}{2}\frac{|r|\tan\theta_{23}
\sin{2\theta_{12}}}{1\mp|r|s_{12}^2}\,,
\end{eqnarray}
where the upper (lower) sign corresponds to the NO (IO) neutrino
mass spectrum case. This shows clearly that simultaneous suppresion
of both $\mu-e$ and $\tau-e$ LFV transitions cannot occur: the sign
of $s_{13}$ in (\ref{s130}) required to suppress the former is
always the opposite than that required to suppress the latter.
Furthermore, inserting the best-fit values of Table~\ref{Table3} in
(\ref{s130}) one finds that the value of $|s_{13}|\simeq 0.015$ for
which one of these two transitions is suppressed is far beyond the
sensitivity of future reactor neutrino experiments like Daya
Bay~\cite{Guo:2007ug}, Double Chooz~\cite{Ardellier:2006mn} or
Reno~\cite{Kim:2008zzb}. Regarding the $\tau\mu$ sector, we conclude
that $(\mlt)_{\tau\mu}$ shows a very weak dependence on
$\theta_{13}$ and $\delta$ since the dominant term in
Eq.~(\ref{m2ltaumu}) is proportional to
$b^2c_{13}^4\sin^2(2\theta_{23})\simeq 4$, implying
\begin{equation}
\label{m2ltaumuap}%
|(\mlt)_{\tau\mu}| \simeq 
\frac{9m_0^2+3A_0^2}{16\pi^2}
\frac{|\dmatm|}{v_T}\log\frac{\Lambda}{M_T}\,.
\end{equation}
Moreover, the $\tau-\mu$ transition cannot be suppressed because the limit
$|(\mlt)_{\tau\mu}|\rightarrow 0$ would require $s_{13}\simeq 1$,
which is excluded by reactor neutrino experiments.


\subsection{Large $\tan\beta$ effects on the ratios of branching ratios}
\label{sec52}
In computing the ratios $R_{\tau j}$ in Section~\ref{sec51} several
simplifications were made. Firstly, the running of $m_\nu$ from
$m_Z$ to $M_T$ was neglected. Secondly, the running of $(m^2_{\tilde
L})_{ij}$ between $M_G$ and $M_T$ was treated in a simplified way
and the running of $(m^2_{\tilde L})_{ij}$ (including its diagonal
entries) between $M_T$ and $m_Z$ (or $m_S$ - the SUSY scale) was
neglected. Finally, the ratios $R_{\tau j}$ were computed with the
help of the simplified formula (\ref{RBRS1}). In this section we
will analyse scenarios in which some of these simplifications lead
to incorrect results.

We first improve the approximate calculation of Section~\ref{sec51}
by taking into account the running of $(\mlt)_{ij}$ between $M_G$
and $M_T$ exactly. To this end we correct Eq.~(\ref{m2lYTa})
replacing it by
\begin{equation}
\label{m2lYTcor}%
\left|\frac{(Y_T^\dag Y_T^{})_{\tau j}}{(Y_T^\dag Y_T^{})_{\mu
e}}\right|_ {\Lambda=M_T}^2\simeq \left|\frac{(m_\nu^\dag
m_\nu^{})_{\tau j}}{(m_\nu^\dag m_\nu^{})_{\mu e}}\right|^2
_{\Lambda=M_T}\,,
\end{equation}
with the right-hand side obtained by evolving $m_\nu$ from $m_Z$ to
$M_T$ with the help of the RGE (\ref{rgemnu}). Alternatively, one
can use Eqs.~(\ref{m2lmue})-(\ref{m2ltaumu}) provided we take the
values of all the neutrino parameters in these formulae at the scale
$\Lambda=M_T$. We will illustrate the effects of this improvement by
comparing the ratios $|(\mlt)_{\tau \mu}/(\mlt)_{\mu e}|$ and
$|(\mlt)_{\tau e}/(\mlt)_{\mu e}|$ at the scale $m_Z$ obtained using
Eqs.~(\ref{m2lYTa}) and (\ref{m2lYTcor}) with the ones resulting
from the exact numerical computation of $(\mlt)_{ij}$ at the scale
$m_Z$. The latter results are obtained as follows. We perform
numerically the RG running of $m_\nu$ from $\Lambda=m_Z$ up to the
$M_T$ scale, extract the couplings $Y_T$ and run them up to
$\Lambda=M_G$ using the TMSSM RGEs. At this scale, we impose
universal boundary conditions on the SUSY-breaking terms and run all
the couplings and masses down to low energies. We expect deviations
between the approximation of Section~\ref{sec51} and the improved
and exact approaches to increase with $\tan\beta$ because the
running of the neutrino mass matrix is stronger for larger
$\tan\beta$ values.

In Fig.~\ref{fig5} we show $|(\mlt)_{\tau j}/(\mlt)_{\mu e}|^2$ with
$j=e,\mu$ as a function of $\tan\beta$ for the values of the input
parameters $m_0$, $m_{1/2}$ and $A_0$ specified in the plots. We
consider two benchmark values of $s_{13}(m_Z)$: $s_{13}=0$ (upper
plots) and $s_{13}=0.2$ (lower plots). The results obtained by using
the full numerical procedure are shown by the black-solid lines,
while the red-dashed curves correspond to the approximations
(\ref{m2lYTcor}). The results extracted by means of the
approximation of Section~\ref{sec51} are shown by the black
dash-dotted line which, of course, does not change with $\tan\beta$.

Let us first analyse the case $s_{13}(m_Z)=0$. The deviations
between the exact and the approximate results increase with
increasing $\tan\beta$ due to stronger RG effects. However, since we
are considering the case of HI neutrino masses we would naively not
expect such large effects even for large values of $\tan\beta$
because the neutrino parameters run very little in this case.
Although this is true for $\theta_{12}$, $\theta_{23}$ and $r$, the
same does not hold for $\theta_{13}$. Starting with a low-energy
value $\theta_{13}(m_Z)$, we have at $\Lambda=M_T$:
\begin{equation}
\label{t12mT}%
\theta_{13}(M_T)\simeq \theta_{13}(m_Z)-\frac{y_\tau^2}{32\pi^2}
\frac{r+\sqrt{r}}{1-r}\sin(2\theta_{23})
\sin(2\theta_{12})\ln\left(\frac{M_T}{m_Z}\right)\,.
\end{equation}
Taking $\theta_{13}(m_Z)=0$ and neglecting the RG effects on the parameters
$r$, $\theta_{12}$ and $\theta_{23}$, we obtain the following estimate for the value
of $\theta_{13}$ at the scale $M_T$
\begin{equation}
\label{t12mT0}%
\theta_{13}(M_T)\simeq -6.2\times 10^{-8}\tan^2\beta
\ln\left(\frac{M_T}{m_Z}\right)\,,
\end{equation}
\FIGURE[!ht]{ \label{fig5} \caption{Ratios
$|(\mlt)_{\tau\mu}/(\mlt)_{\mu e}|^2$ (left plots) and
$|(\mlt)_{\tau e}/(\mlt)_{\mu e}|^2$ (right plots) as a function of
$\tan\beta$ for $s_{13}(m_Z)=0$ (upper plots) and $s_{13}(m_Z)=0.2$
(lower-plots). The black-solid lines correspond to the exact
numerical result while the red-dashed ones were obtained using
Eq.~(\ref{m2lYTcor}) with $Y_T$ taken at $\Lambda=M_T$. The
horizontal dashed-dotted line indicates the value obtained using
Eq.~(\ref{m2lYT}) and the low-energy best-fit values for the
neutrino parameters given in Table~\ref{Table3} (all the
CP-violating phases are set to zero). The red-dashed lines denote
the values obtained using Eq.~(\ref{m2lYTcor}) with $Y_T$ taken at
$\Lambda=M_T$. Filled in yellow are the regions (delimited by
black-solid curves) (delimited by the black-solid curves) allowed
for the $M_T$ values indicated in each plot. The two red-dashed
lines have been obtained for the two limiting values of such
interval. All the plots have been obtained for
$m_0=m_{1/2}=300$~GeV, $A_0=0$ and assuming a HI neutrino mass
spectrum ($m_1=0$~eV). For each value of $M_T$ the value of
$\lambda_2$ has been chosen in such a way that for $\tan\beta=50$,
${\rm BR}(\mu\rightarrow e \gamma)\simeq 1.2\times 10^{-11}$. }
\begin{tabular}{cc}
\includegraphics[width=7.2cm]{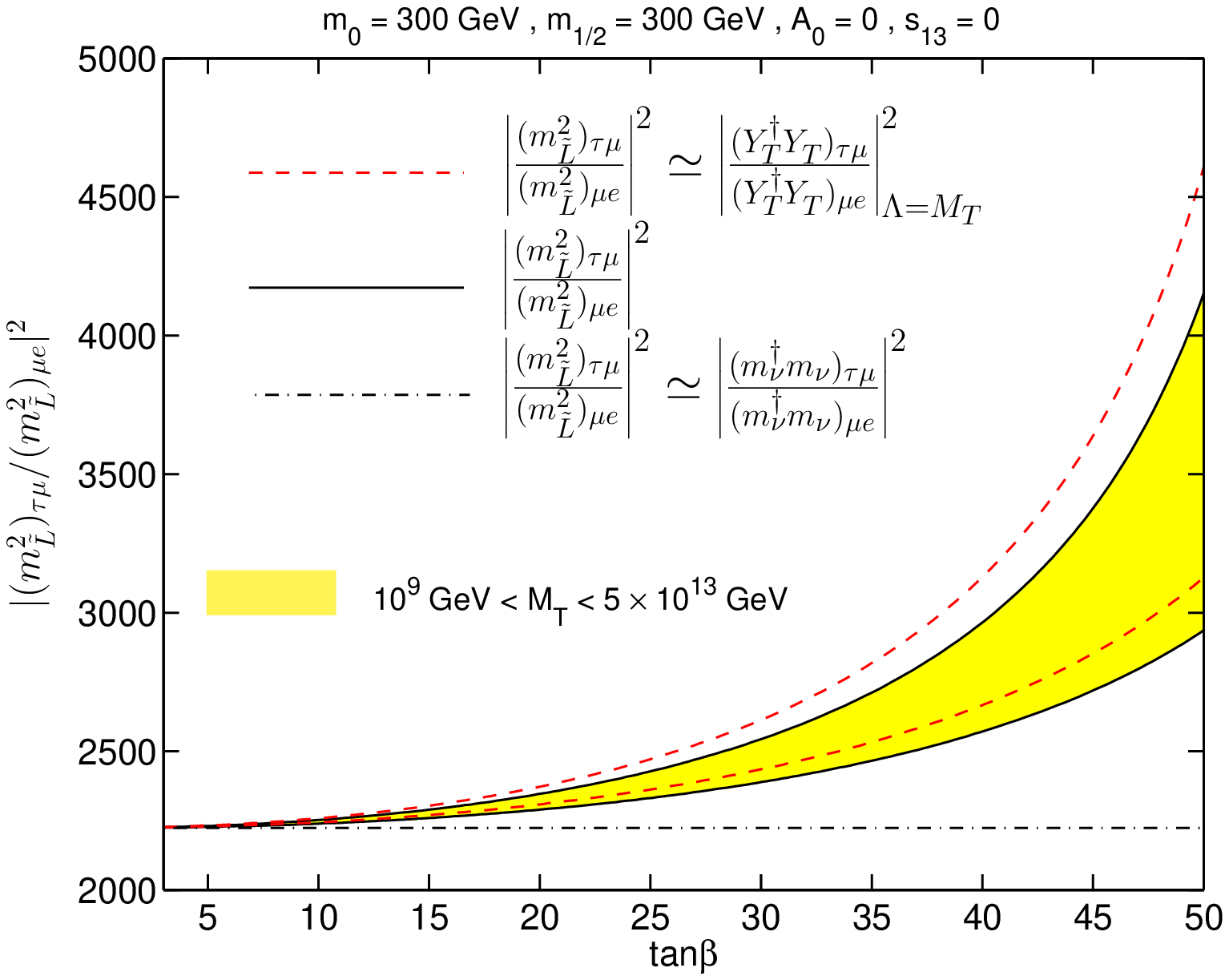} &
\includegraphics[width=7.2cm]{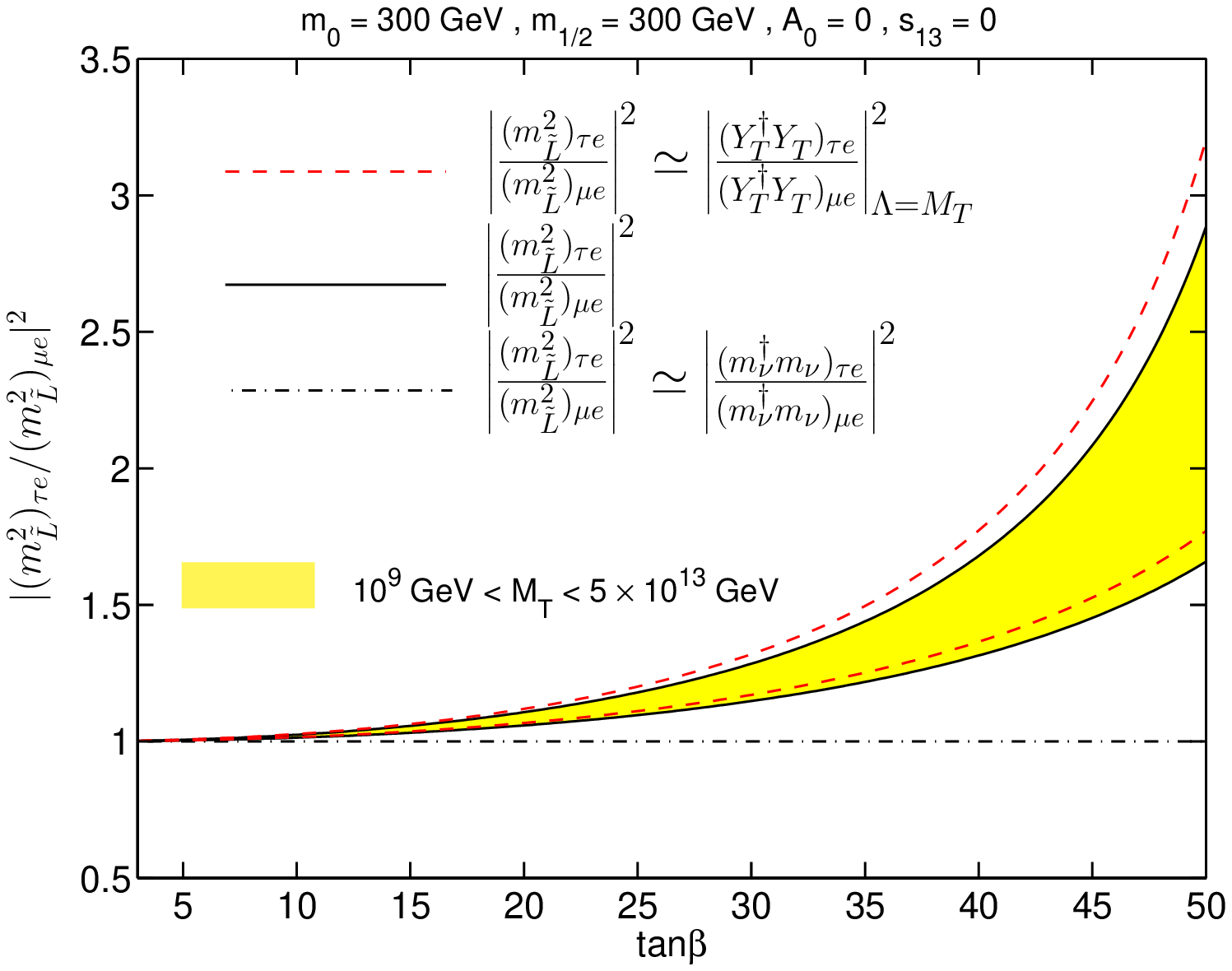}\\
\includegraphics[width=7.2cm]{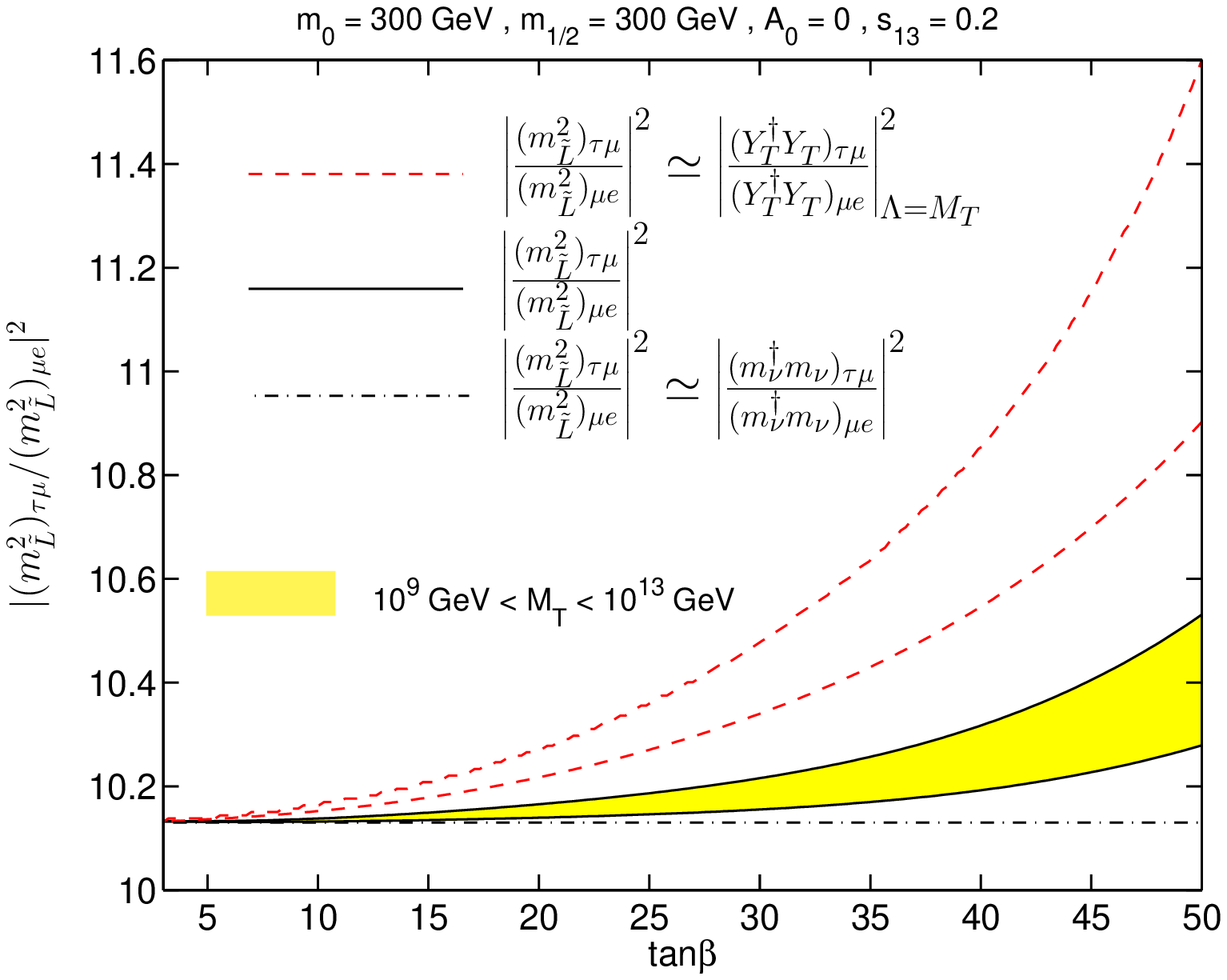} &
\includegraphics[width=7.2cm]{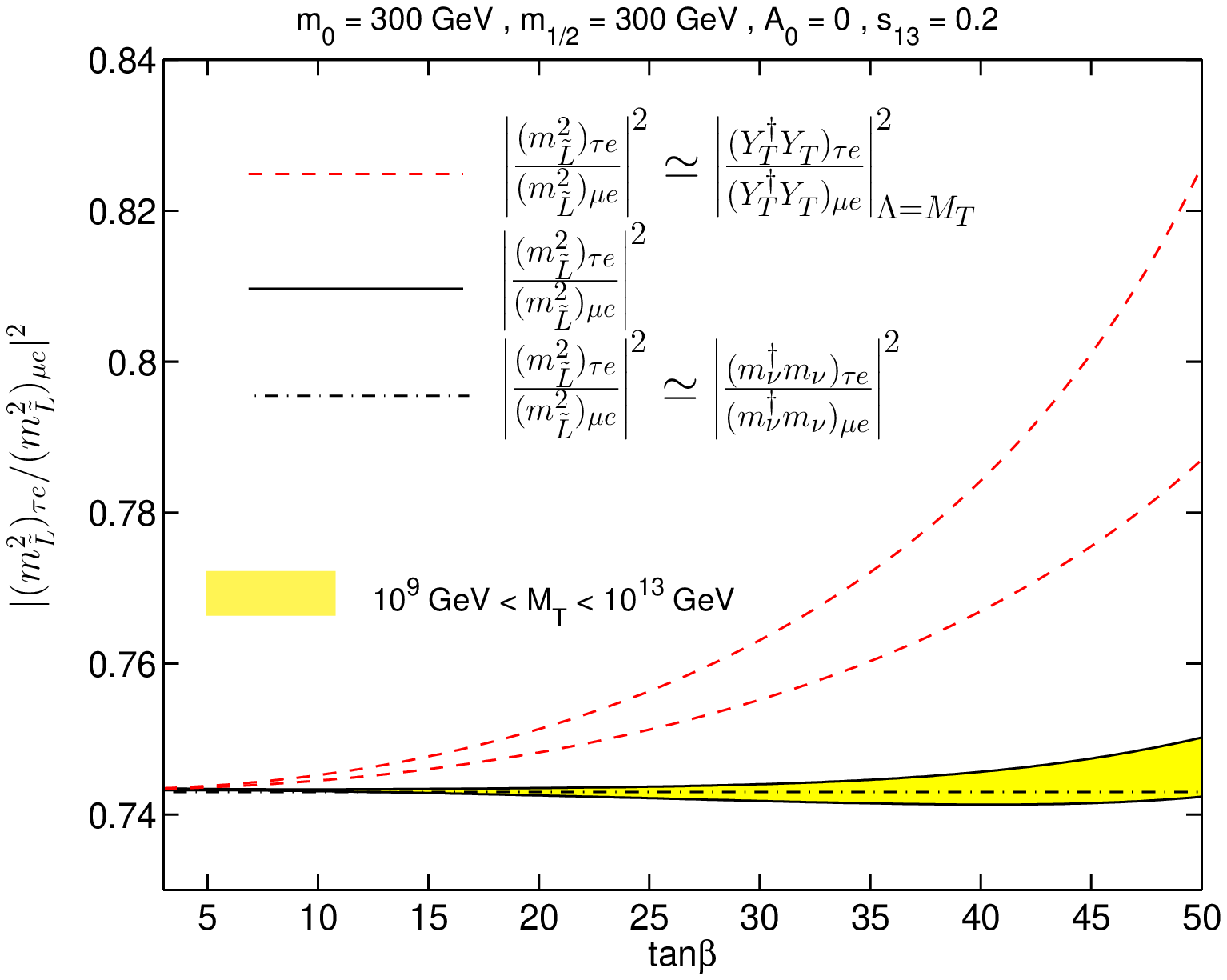}
\end{tabular}
}
which reasonably agrees with the exact numerical result, even for
large $\tan\beta$. Although small, these values of $\theta_{13}$ at
$M_T$ may have some impact on the values of $(Y_T^\dag Y_T^{})_{ij}$
at that scale. Clearly, the effect will be stronger for larger
values of $\tan\beta$ and/or $M_T$. The fact that $\theta_{13}(M_T)$
is negative leads to a suppression of $|(Y_T^\dag Y_T^{})_{\mu e}|$
with respect to the value extracted using $\theta_{13}=0$. This can
be understood taking into account that the last term of
Eq.~(\ref{m2lmue}) becomes negative for $\theta_{13}<0$. The
opposite occurs for $|(Y_T^\dag Y_T^{})_{\tau e}|$ since the last
contribution in Eq.~(\ref{m2ltaue}) is now positive, implying an
enhancement of LFV in the $\tau e$ sector. Since $|(Y_T^\dag
Y_T^{})_{\tau \mu}|$ is practically insensitive to $\theta_{13}$,
the RG effects on this quantity can be safely neglected. Therefore,
we expect that the values of $|(Y_T^\dag Y_T^{})_{\tau
\mu}/(Y_T^\dag Y_T^{})_{\mu e}|^2$ and $|(Y_T^\dag Y_T^{})_{\tau
e}/(Y_T^\dag Y_T^{})_{\mu e}|^2$ at $\Lambda=M_T$ are larger than
the ones predicted for $\theta_{13}=0$. This enhancement should be
more significant in the latter case since $|(Y_T^\dag Y_T^{})_{\tau
e}|$ grows and $|(Y_T^\dag Y_T^{})_{\mu e}|$ is suppressed. This is
shown in Fig.~\ref{fig5} (upper plots) where one can see that the
true values of $|(m^2_{\tilde{L}})_{\tau\mu}/(m^2_{\tilde{L}})_{\mu
e}|^2$ (left plot) and $|(m^2_{\tilde{L}})_{\tau
e}/(m^2_{\tilde{L}})_{\mu e}|^2$ (right plot) increase with
$\tan\beta$. For $\tan\beta=50$, the deviation of the exact result
with respect to the one obtained in the crude approximation of
Section~\ref{sec51} (horizontal dashed-dotted lines) amounts to
about $30\%$ for $M_T=10^9$~GeV (lower black-solid line) and $90\%$
for $M_T=10^{14}$~GeV (upper black-solid line) in the case of
$|(m^2_{\tilde{L}})_{\tau\mu}/(m^2_{\tilde{L}})_{\mu e}|^2$. For
$|(m^2_{\tilde{L}})_{\tau e}/(m^2_{\tilde{L}})_{\mu e}|^2$, one
observes deviations of the order of $70\%$ and $190\%$, for the same
values of $M_T$ and $\tan\beta$. Notice that the exact results
(black-solid curves) are slightly lower than the ones obtained using
Eq.~(\ref{m2lYTcor}) (red-dashed curves). This deviation is due to
the running of $(m^2_{\tilde{L}})_{ij}$ between $M_T$ and $m_Z$.
Considering the RGE of $m^2_{\tilde{L}}$, it can be shown that
$(m^2_{\tilde{L}})_{ij}(m_Z)$ is approximately given by:
\begin{equation}
\label{m2lijye}%
(m^2_{\tilde{L}})_{ij}(m_Z)\simeq\left[1-\frac{y_{e_i}^2+y_{e_j}^2}
{16\pi^2}\ln\left(\frac{M_T}{m_Z}\right)\right](m^2_{\tilde{L}})_{ij}(M_T)
\,,(i\neq j=e,\mu,\tau)\,,
\end{equation}
where $y_{e_i}$ are the charged-lepton Yukawa couplings. It is clear
that this effect in $(m^2_{\tilde{L}})_{\mu e}$ can be neglected
while for $(m^2_{\tilde{L}})_{\tau\mu,\tau e}$ one has:
\begin{equation}
\label{m2lijytau}%
(m^2_{\tilde{L}})_{\tau\mu,\tau e}(m_Z)\simeq\left[1-\frac{y_\tau^2}
{16\pi^2}\ln\left(\frac{M_T}{m_Z}\right)\right](m^2_{\tilde{L}})_{\tau\mu,\tau
e}(M_T)\,.
\end{equation}
Consequently, the ratios $|(m^2_{\tilde{L}})_{\tau
j}/(m^2_{\tilde{L}})_{\mu e}|^2$ obtained by solving the exact RGEs
are enhanced with respect to those obtained in the improved
approximation. Obviously, this effect is more relevant for large
$\tan\beta$. Combining Eqs.~(\ref{m2lijye}) and (\ref{m2lijytau}) we
get the relation
\begin{equation}
\label{m2lyecor}%
\left|\frac{(\mlt)_{\tau j}}{(\mlt)_{\mu e}} \right|_{\Lambda=m_Z}^2
\simeq \left[1-\frac{y_\tau^2}
{8\pi^2}\ln\left(\frac{M_T}{m_Z}\right)
\right]\left|\frac{(\mlt)_{\tau j}}{(\mlt)_{\mu e}}
\right|_{\Lambda=M_T}^2\;,\;(j=\mu,e)\,,
\end{equation}
which explains the deviation between the solid and dashed curves in
Fig.~\ref{fig5}.

When $s_{13}(m_z)=0.2$ (lower plots in Fig.~\ref{fig5}), the running
of the neutrino parameters does not affect much the quantities
$|(m^2_{\tilde{L}})_{\tau j}/ (m^2_{\tilde{L}})_{\mu e}|^2$. This
stems from the fact that now the RG correction induced on
$\theta_{13}$ is negligible when compared with the low-energy value
of this angle, as can be seen from Eqs.~(\ref{t12mT}) and
(\ref{t12mT0}). The deviations with respect to the approximate
results are of the order of $10\%$ (for the largest value of $M_T$),
as confirmed by comparing the red-dashed and dashed-dotted lines.
Once more, the difference between the red-dashed and solid-black
lines is due to the suppression factor shown in (\ref{m2lyecor}).

In Fig.~\ref{fig6} we confront the ratios $R_{\tau \mu}$ and
$R_{\tau e}$ obtained (for the same set of parameters as in
Fig.~\ref{fig5}) by inserting into the simplified formula
(\ref{RBRS1}) exact values of $|(m^2_{\tilde{L}})_{\tau
j}/(m^2_{\tilde{L}})_{\mu e}|^2$ obtained from the RG procedure with
the ratios $R_{\tau \mu}$ and $R_{\tau e}$ obtained by the exact
one-loop calculation of the individual branching fractions ${\rm
BR}(\ell_i \rightarrow \ell_j\gamma)$ as {\em e.g.} in
Ref.~\cite{Hisano:1995cp}.
\FIGURE[!ht]{ \label{fig6} \caption{Ratios
$R_{\tau\mu}$ (left plots) and $R_{\tau e}$ (right plots) as defined
in Eq.~(\ref{RBRS1}) for $s_{13}(m_Z)=0$ (upper plots) and
$s_{13}(m_Z)=0.2$ (lower plots). The black-solid lines refer to the
exact numerical result obtained performing the numerical RG running,
full calculation of SUSY spectrum and exact computation of the
one-loop ${\rm BR}(\ell_i \rightarrow \ell_j\gamma)$. The red-dashed
curves indicate the result obtained by means of Eq.~(\ref{RBRS1})
using the exact values of $|(m^2_{\tilde{L}})_{ij}/
(m^2_{\tilde{L}})_{\mu e}|^2$, while the dash-dotted horizontal line
corresponds to the approximation of Section~\ref{sec51}.}
\begin{tabular}{cc}
\includegraphics[width=7.0cm]{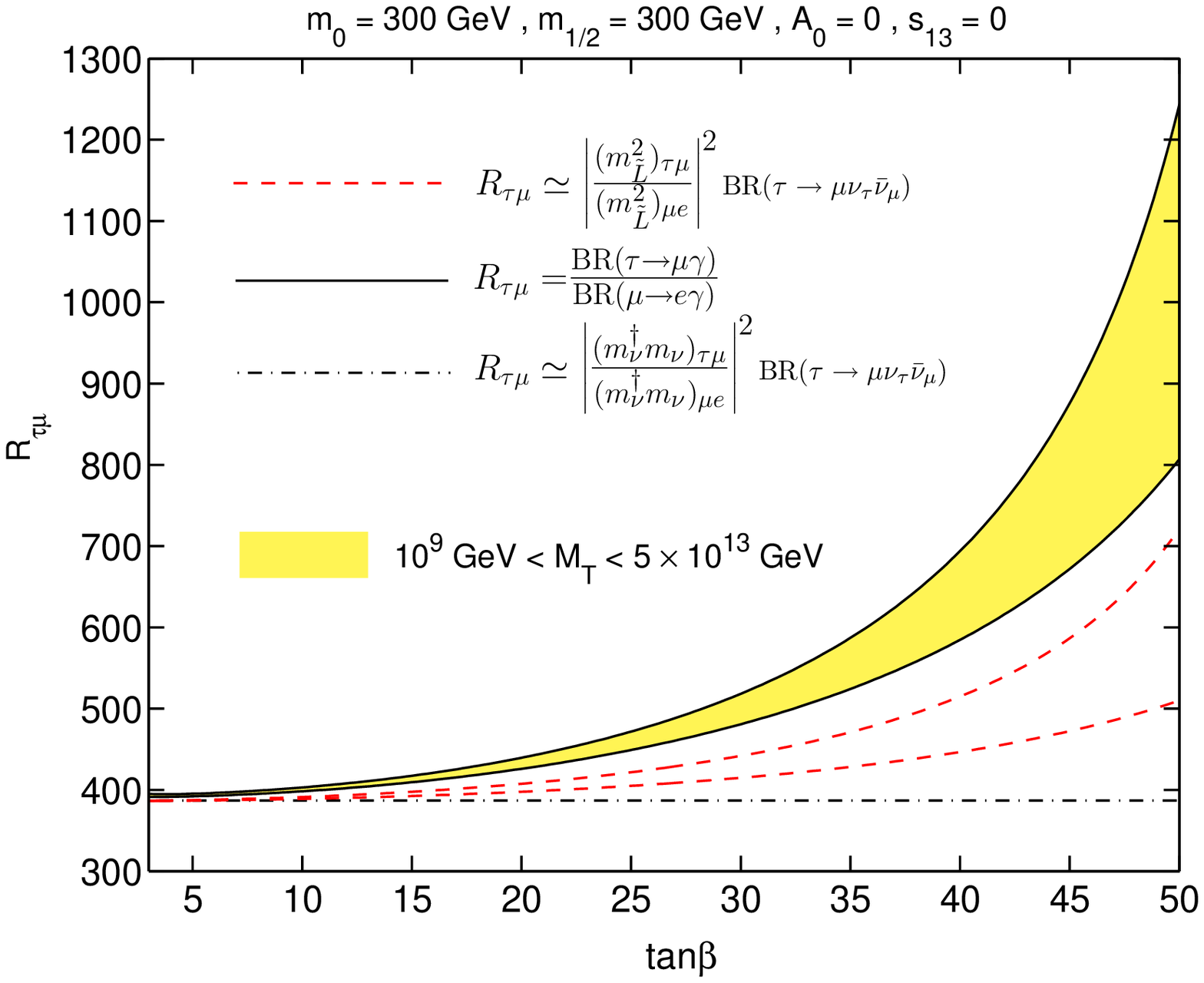} &
\includegraphics[width=7.0cm]{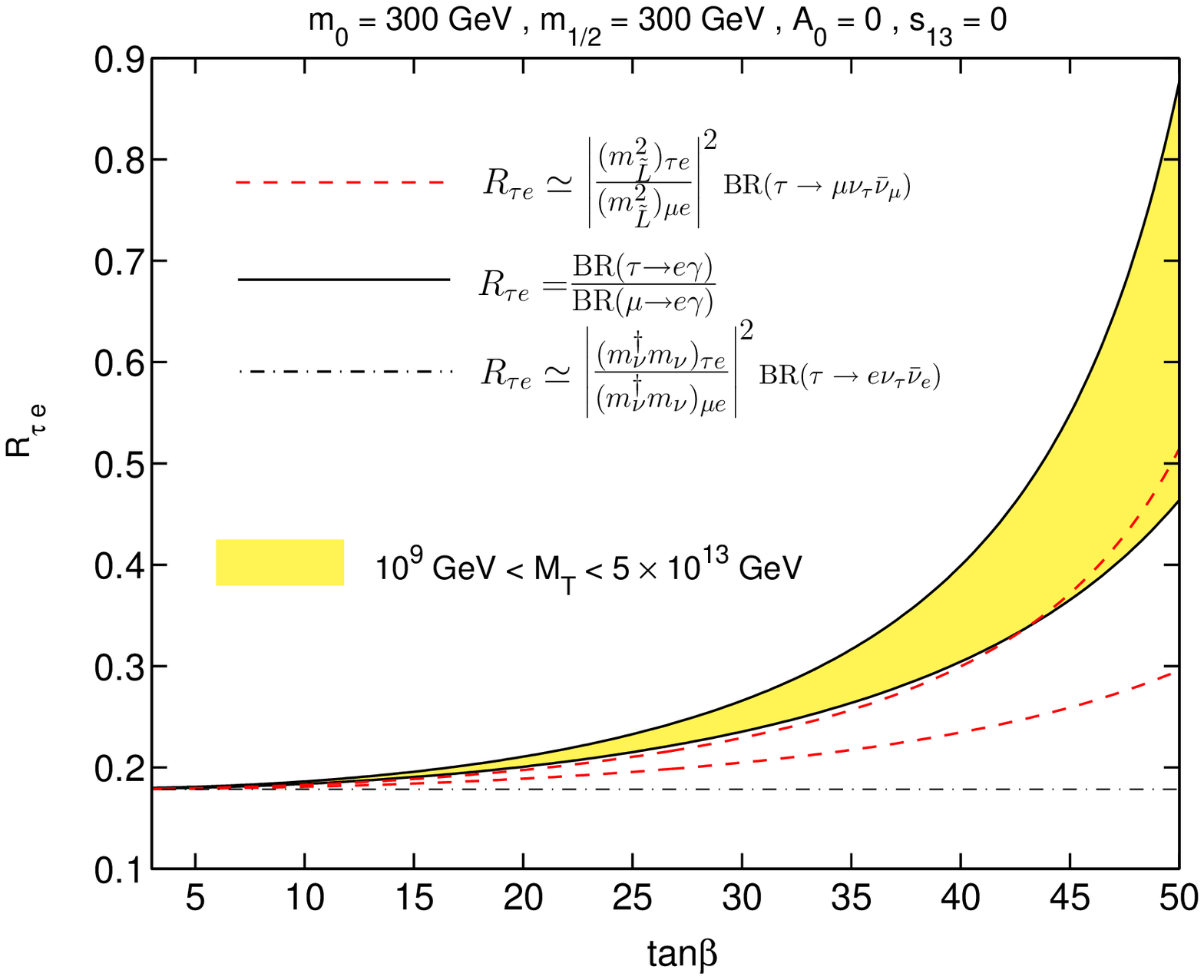}\\
\includegraphics[width=7.0cm,height=6.0cm]{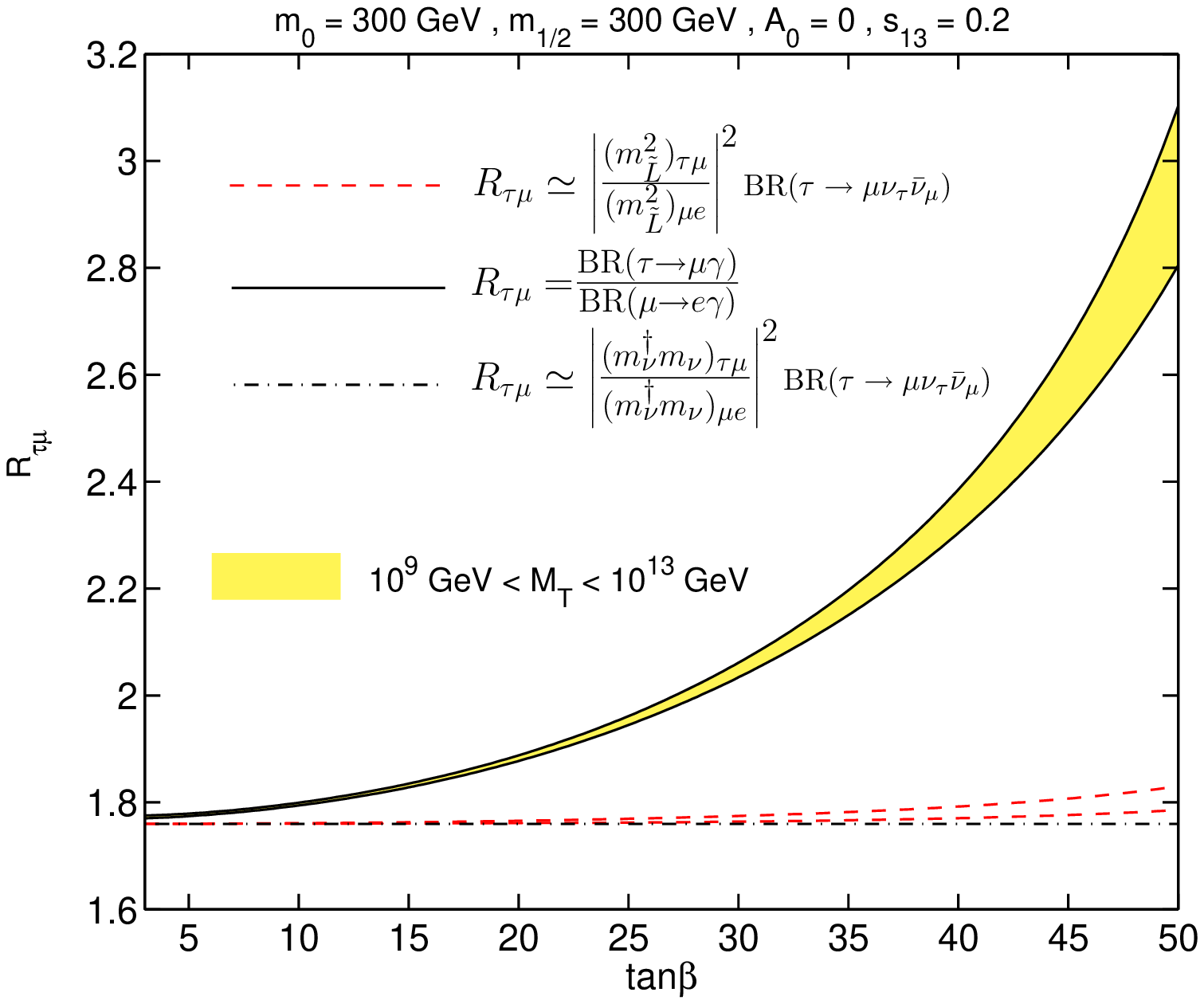} &
\includegraphics[width=7.2cm,height=6.0cm]{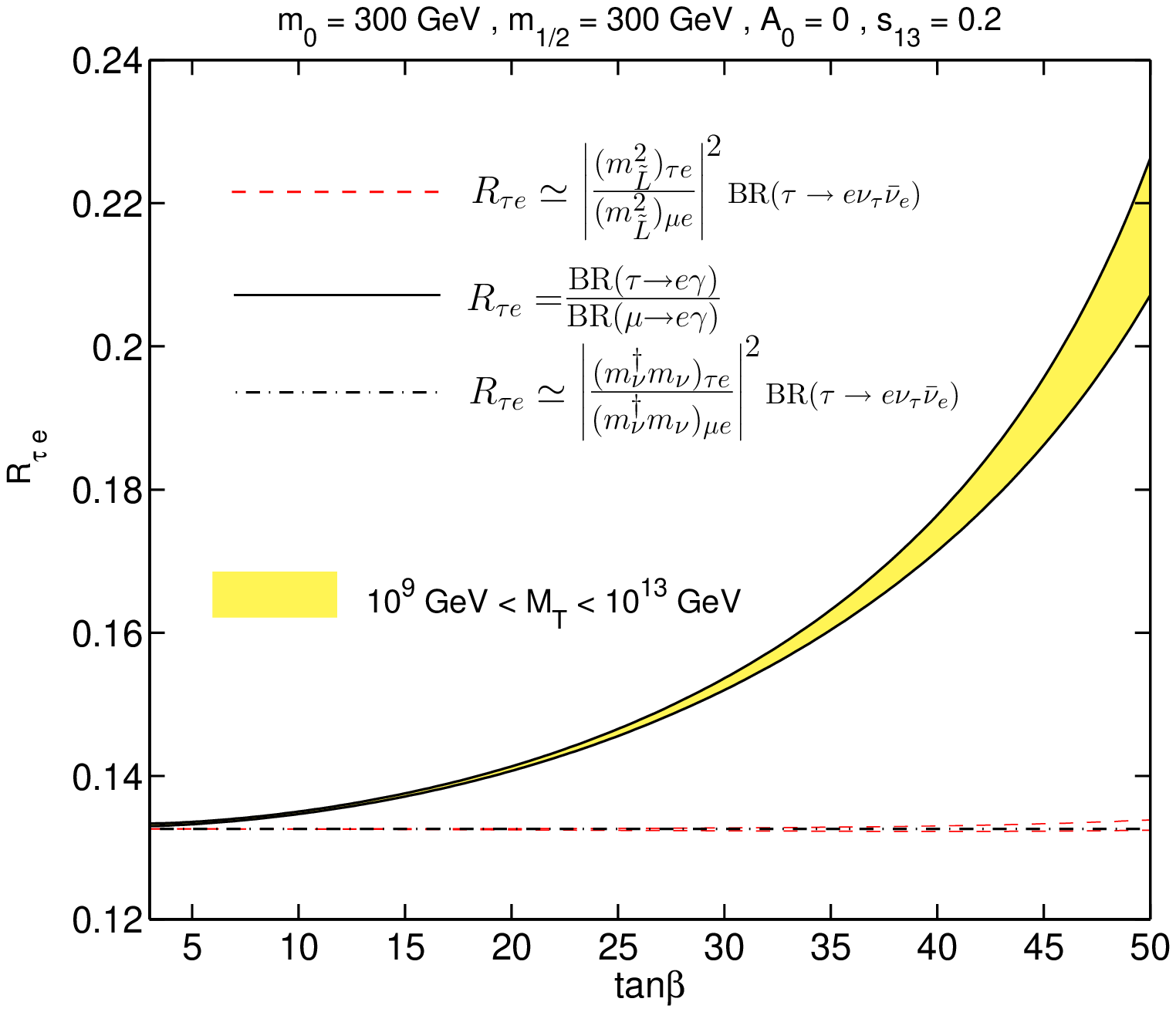}
\end{tabular}
}
\FIGURE[!ht]{ \label{fig7} \caption{Example of
one-loop (chargino-exchange) diagrams for the decays $\tau\rightarrow
\mu \gamma$ (left) and $\mu\rightarrow e\gamma$ (right) in the mass
insertion approximation. The crossed circles denote the
corresponding LFV soft masses.}
\begin{tabular}{cc}
\includegraphics[width=6.0cm]{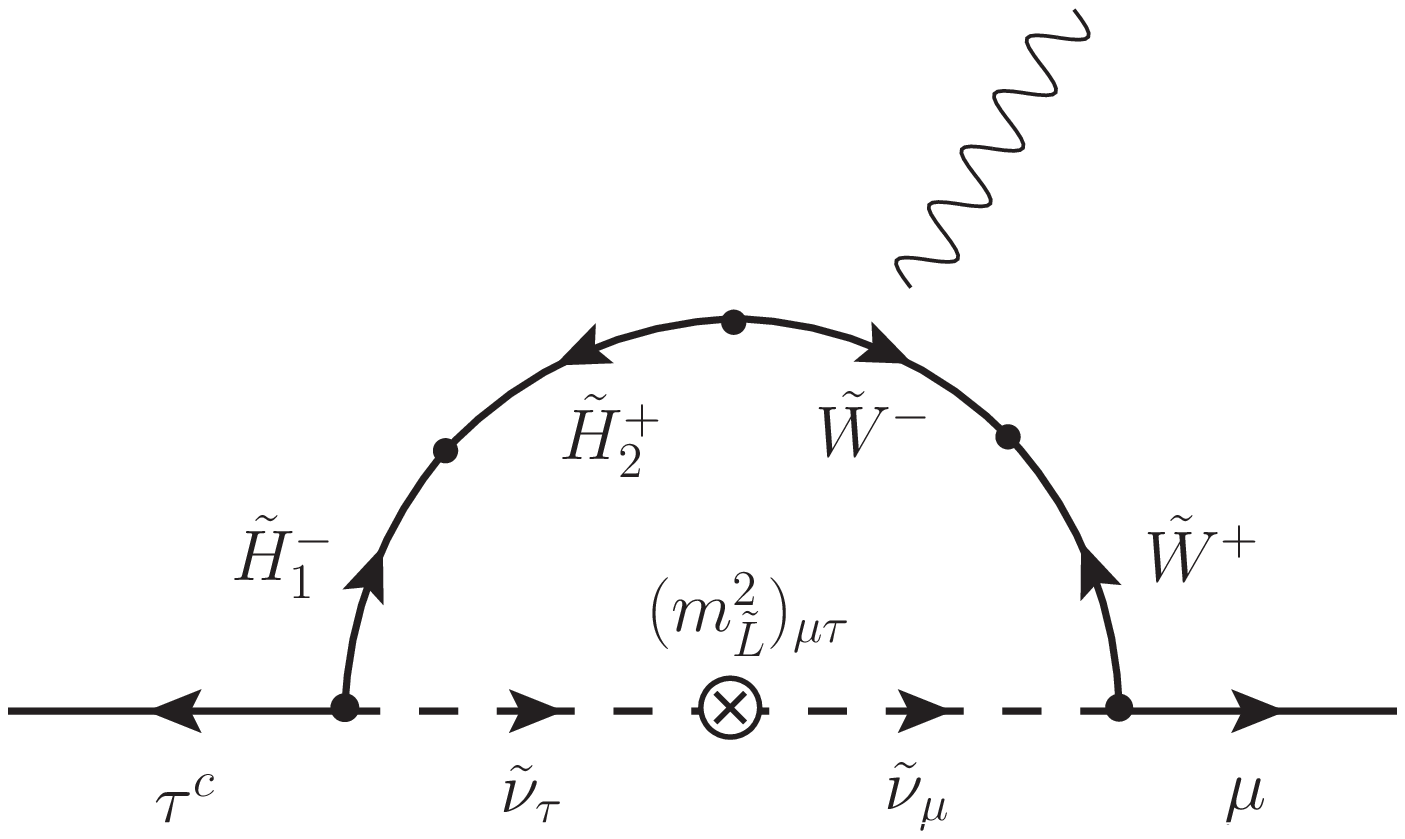}&
\includegraphics[width=6.0cm]{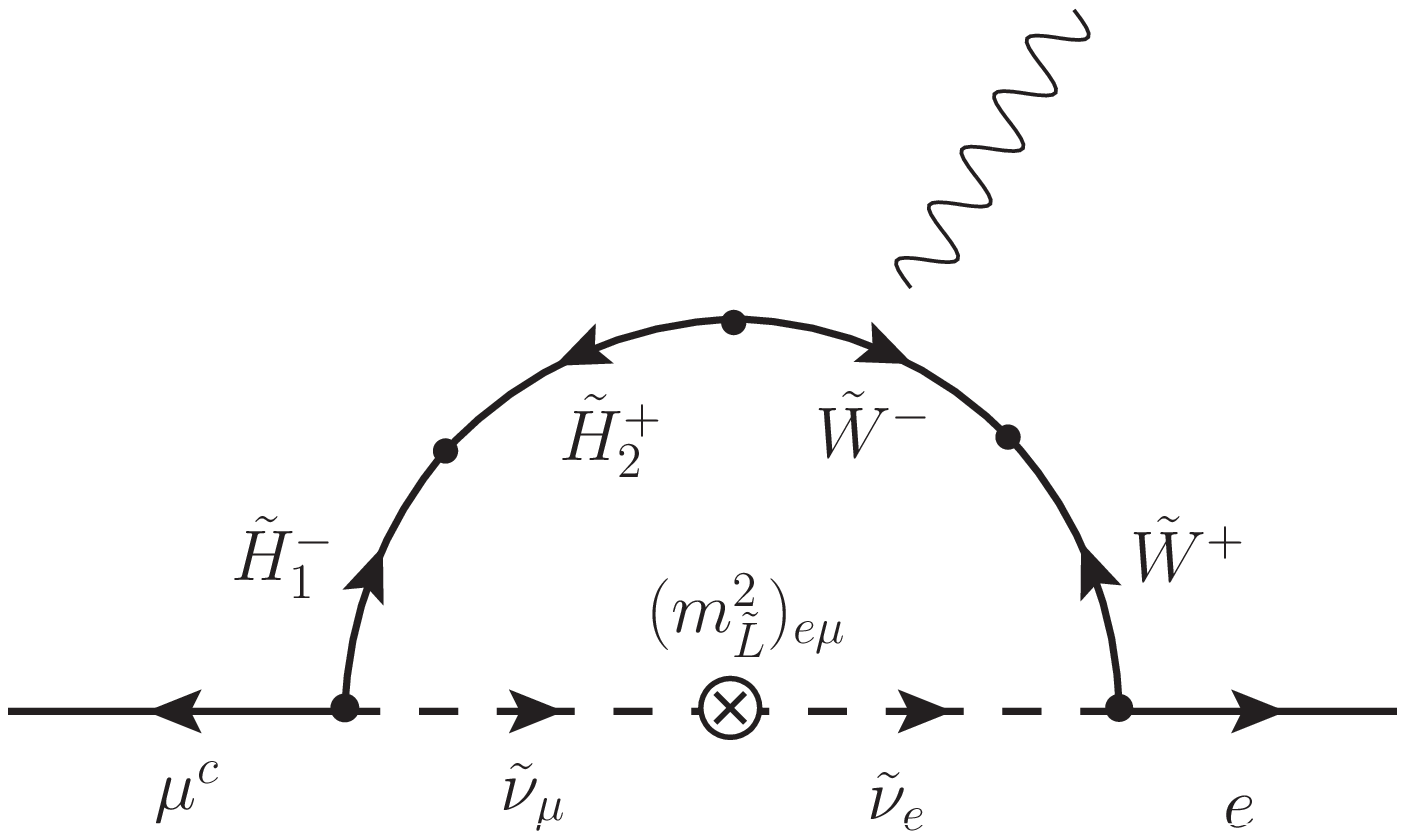}
\end{tabular}
}
Naively one would expect a good agreement between the red-dashed and
solid-black curves since the former were obtained inserting the
exact numerical results for the quantities $|(m^2_{\tilde{L}})_{ij}/
(m^2_{\tilde{L}})_{\mu e}|^2$ into Eq.~(\ref{RBRS}). However, this
is not the case as can be observed in all plots in Fig.~\ref{fig6}.
The observed discrepancy is due to the fact that the formulae
Eqs.~(\ref{BRap}) and (\ref{Cijap}) with the same mass $m_S$ can be
a good approximation only if the slepton mass spectrum is not too
much split: only then can the masses of the sleptons circulating in
loop diagrams for the processes $\ell_i \rightarrow \ell_j\gamma$
shown in Fig.~\ref{fig7} be replaced by an average mass $m_S$. In
writing Eq.~(\ref{RBRS}) we have considered that $m_S$ is the same
for $\tau \rightarrow \ell_j\gamma$ ($\ell_j=e,\mu$) and $\mu
\rightarrow e\gamma$. This is a valid approximation if the slepton
masses are nearly degenerate for all three generations.

In the TMSSM with universal boundary conditions for the
SUSY-breaking terms at the GUT scale, the soft masses for the first
two generations are approximately given by:
\begin{equation}
\label{msferm1}%
(m_{\tilde{f}}^2)_{ii} (m_Z) \simeq m_0^2+\Delta M_{\tilde{f}}^2\,,
\end{equation}
where $\Delta M_{\tilde{f}}^2$ is the contribution due to the
running of $(m_{\tilde{f}}^2)_{ii}$ from the GUT scale down to low
energies, which is mainly controlled by the terms $g_a^2|M_a|^2$
(where $M_a$ is the $G_a$ gaugino mass) present in the RGEs of the
soft scalar masses. Performing the (one-loop) integration in the
limit of vanishing Yukawa couplings, we obtain:
\begin{equation}
\label{msferm2}%
\Delta M_{\tilde{f}}^2=2\,\sum_{a=1}^3 C_a^{f}m_{1/2}^2 \left\{
\frac{1}{b_a^\prime}\left[1-\frac{g_a^4(M_T)}{g_a^4(M_G)}\right]+
\frac{1}{b_a}\frac{g_a^4(M_T)}{g_a^4(M_G)}\left[1-\frac{g_a^4(m_Z)}
{g_a^4(M_T)}\right]\right\}\,,
\end{equation}
where $C_a^{f}$ is the quadratic Casimir invariant of the
representation of $f$ under the gauge group $G_a$. The
ratios of the gauge couplings appearing in the above equation are
given by:
\begin{equation}
\label{garat1}%
\frac{g_a^4(M_T)}{g_a^4(M_G)}=\left[1-\frac{g_a^2(M_T)}{8\pi^2}
b_a^\prime\ln\left(\frac{M_G} {M_T}\right)\right]^2\;,\;
\frac{g_a^4(m_Z)}{g_a^4(M_T)}=\left[1-\frac{g_a^2(m_Z)}{8\pi^2}
b_a\ln\left(\frac{M_T} {m_Z}\right)\right]^2\,,
\end{equation}
where $b_a=(33/5,1,-3)$ and $b_a^\prime= b_a+7$ are the
$\beta$-function coefficients in the RGEs of $g_a$ for the MSSM and
TMSSM cases, respectively~\cite{Rossi:2002zb}. The sneutrino masses
of the first two generations (neglecting the D-term contributions
and generation mixing) are given by $m_{\tilde{\nu}_{e,\mu}}^2\simeq
m_0^2+\Delta M_{\tilde{L}}^2$ with $C_a^{\tilde{L}}=(3/20,3/4,0)$.
For large $\tan\beta$, the effects of the running due to the large
$\tau$ Yukawa coupling induce a nonnegligible splitting between the
masses of $\tilde{\nu}_\tau$ and $\tilde{\nu}_{e,\mu}$. In the limit
of small $Y_T$ couplings, one can show that
\begin{equation}
\label{msferm2}%
m_{\tilde{\nu}_{\tau}}^2=m_{\tilde{\nu}_{e,\mu}}^2\!\!\!-y_\tau^2
\frac{3\,m_0^2+A_0^2}{8\pi^2} \ln\left(\frac{M_G}{m_Z}\right)\,,
\end{equation}
where, for simplicity, we have approximated $y_\tau$
between $M_G$ and $m_Z$ by a constant. The term proportional to
$y_\tau^2$ is always negative making the $\tau$-sneutrino lighter
than the remaining two. In order to account for the effect of the
sneutrino mass splitting on the quantities $R_{\tau j}$, we correct
Eq.~(\ref{RBRS1}) with a factor $\eta$
\begin{equation}
\label{RBRS2}%
R_{\tau j} \simeq \eta\left|\frac{(\mlt)
_{\tau j}}{(\mlt)_{\mu e}} \right|^2 {\rm BR}(\tau \rightarrow
\ell_j\nu_\tau\bar{\nu}_j)\;, \;
\eta=\left[\frac{F(m_{\tilde{\nu}_{\tau}}^2,m_{\tilde{\nu}
_j}^2,m_S^i)}{F(m_{\tilde{\nu} _{\mu}}^2,m_{\tilde{\nu}
_e}^2,m_S^i)}\right]^2\,,
\end{equation}
where $F$ denotes the loop function of the corresponding decay
amplitude and $m_S^i$ the masses of the non-sleptonic particles
running inside the loops.

Deviations\footnote{These deviations are negligible for the ratio
${\rm BR}(\tau\rightarrow \mu \gamma) /{\rm BR}(\tau\rightarrow e
\gamma)$ since $\eta \simeq 1$, irrespective of the value
$\tan\beta$.} from the limit $\eta=1$ are expected to grow with
increasing $\tan\beta$ because the mass splitting between
$\tilde{\nu}_{e,\mu}$ and $\tilde{\nu}_{\tau}$
increases\footnote{One should recall that the dependence of $\eta$
on the initial conditions at $M_G$ is non-trivial. Although we will
not address the analytical treatment of this subject here, we will
show some numerical examples in the following. The loop functions
for the various LFV operators in the two-generation limit can be
found in Ref.~\cite{Brignole:2004ah}} with $\tan\beta$. This can be
observed in Fig.~\ref{fig6}. The differences between the exact
numerical results for $R_{\tau j}$ (black-solid curves) and the ones
obtained by using Eq.~(\ref{RBRS}) together with the real values of
the LFV soft masses (red-dashed lines) increase with $\tan\beta$.
Moreover, $\eta$ depends on the initial values of $m_0$, $m_{1/2}$
and $A_0$ which for large $\tan\beta$ leads to a variation of
$R_{\tau\mu}$ and $R_{\tau e}$ in the SUSY parameter space, even for
fixed values of $M_T$, $\lambda_2$ and the neutrino parameters.
Finally, it is worth stressing that this effect does not depend on
the way through which LFV in the soft masses is generated. Hence, it
should also be present in the SUSY version of the type-I seesaw.
\FIGURE[!ht]{ \label{fig8} \caption{Variation of $R_{\tau \mu}$
(left plots) and $R_{\tau e}$ (right plots) in the $(m_{1/2},m_0)$
parameter space for $s_{13}(m_Z)=0$ (upper plots) and 0.2 (lower
plots) and $\tan\beta=10$ (the values of $M_T$, $\lambda_2$ and
$A_0$ are indicated on the top of each plot and the remaining
low-energy parameters are taken at their best-fit points). The mass
of the lightest slepton $m_{\tilde{\ell}_1}$ is below 100 GeV inside
the regions filled in cyan and the black dashed-dotted line
corresponds to $m_{\tilde{\ell}_1}=200\,{\rm GeV}$. To the left of
the black-dashed line the lightest-chargino mass is below the LEP
bound. The black-hatched region is excluded by the MEGA bound ${\rm
BR}(\mu \rightarrow e \gamma)< 1.2\times
10^{-11}$~\cite{Brooks:1999pu} and the blue contours correspond to
${\rm BR}(\mu \rightarrow e \gamma)=10^{-12},10^{-13}$. The
variation of $R_{\tau j}$ follows the colour bars shown on the right
of each panel.}
\begin{tabular}{cc}
\includegraphics[width=7.2cm,height=5.8cm]{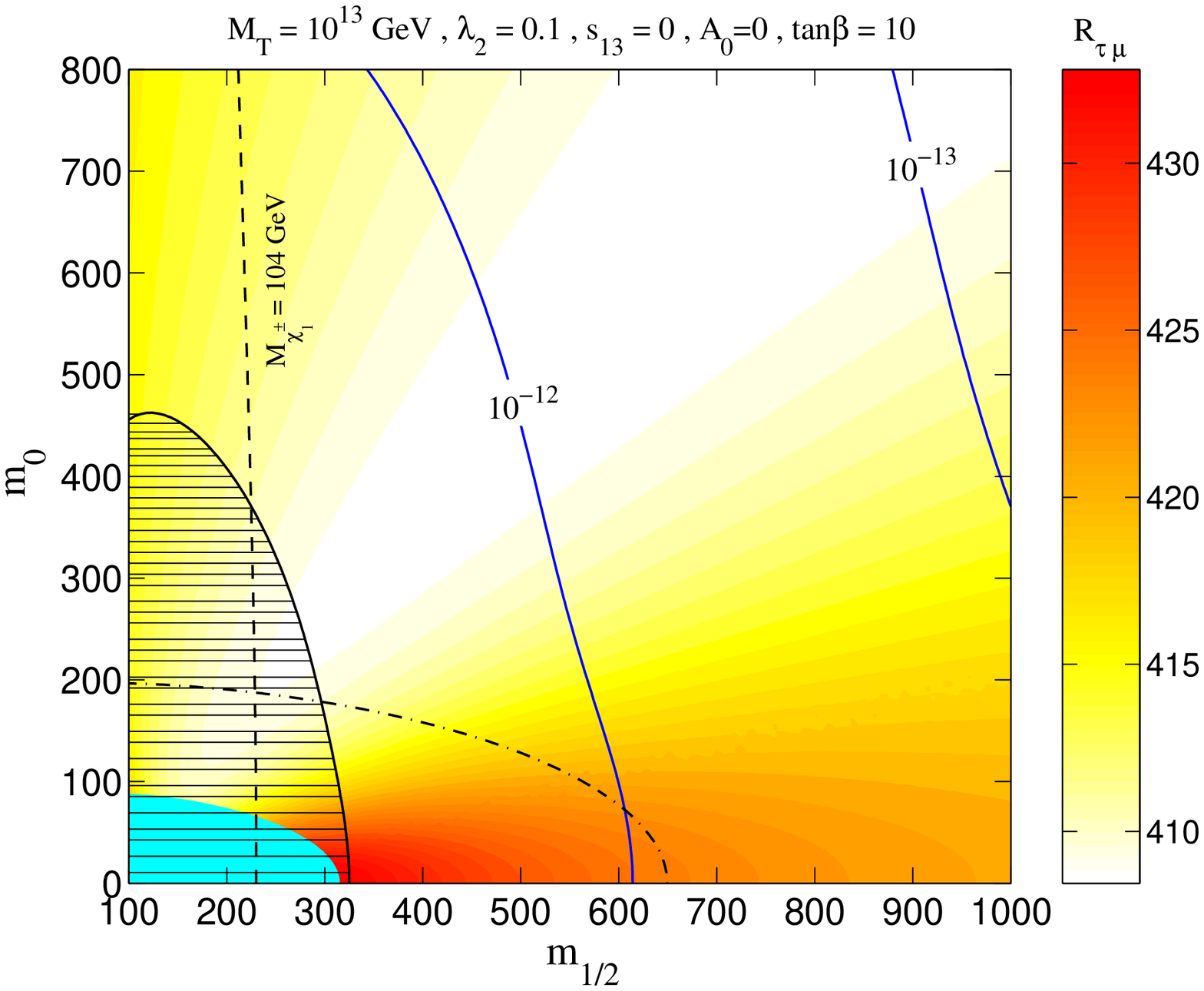} &
\hspace*{-0.1cm}\includegraphics[width=7.3cm]{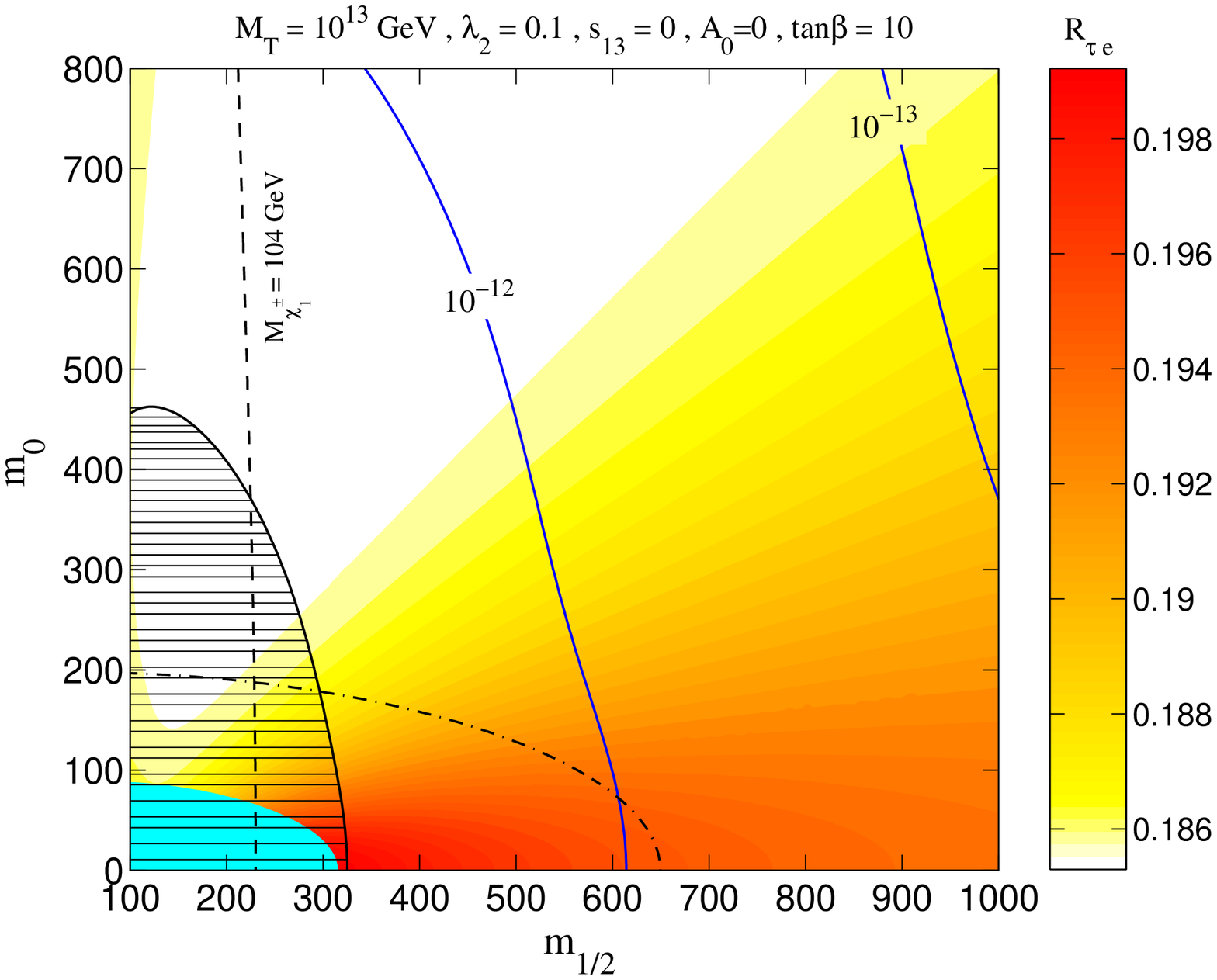} \\
\includegraphics[width=7.2cm]{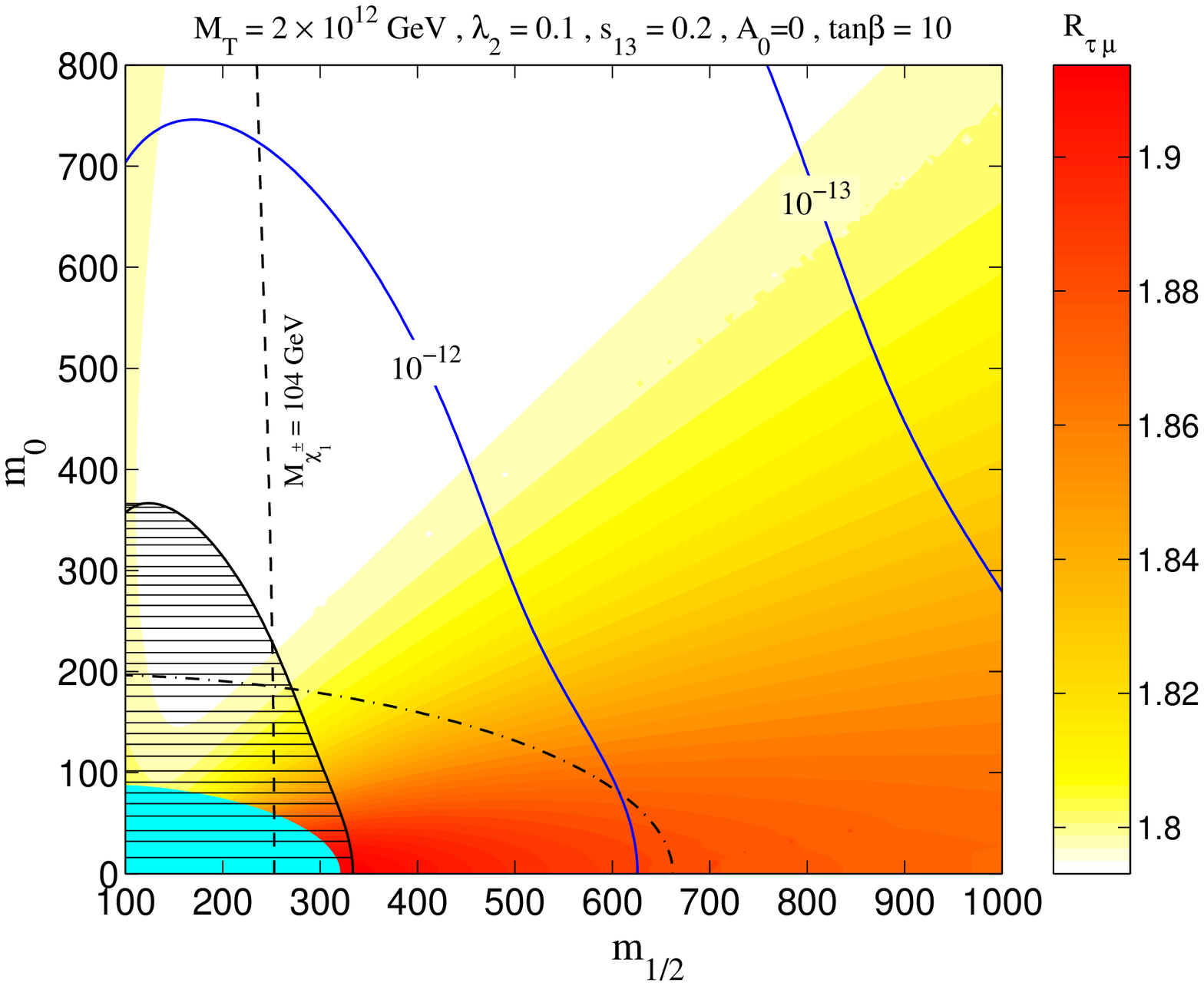} &
\hspace*{-0.1cm}\includegraphics[width=7.3cm]{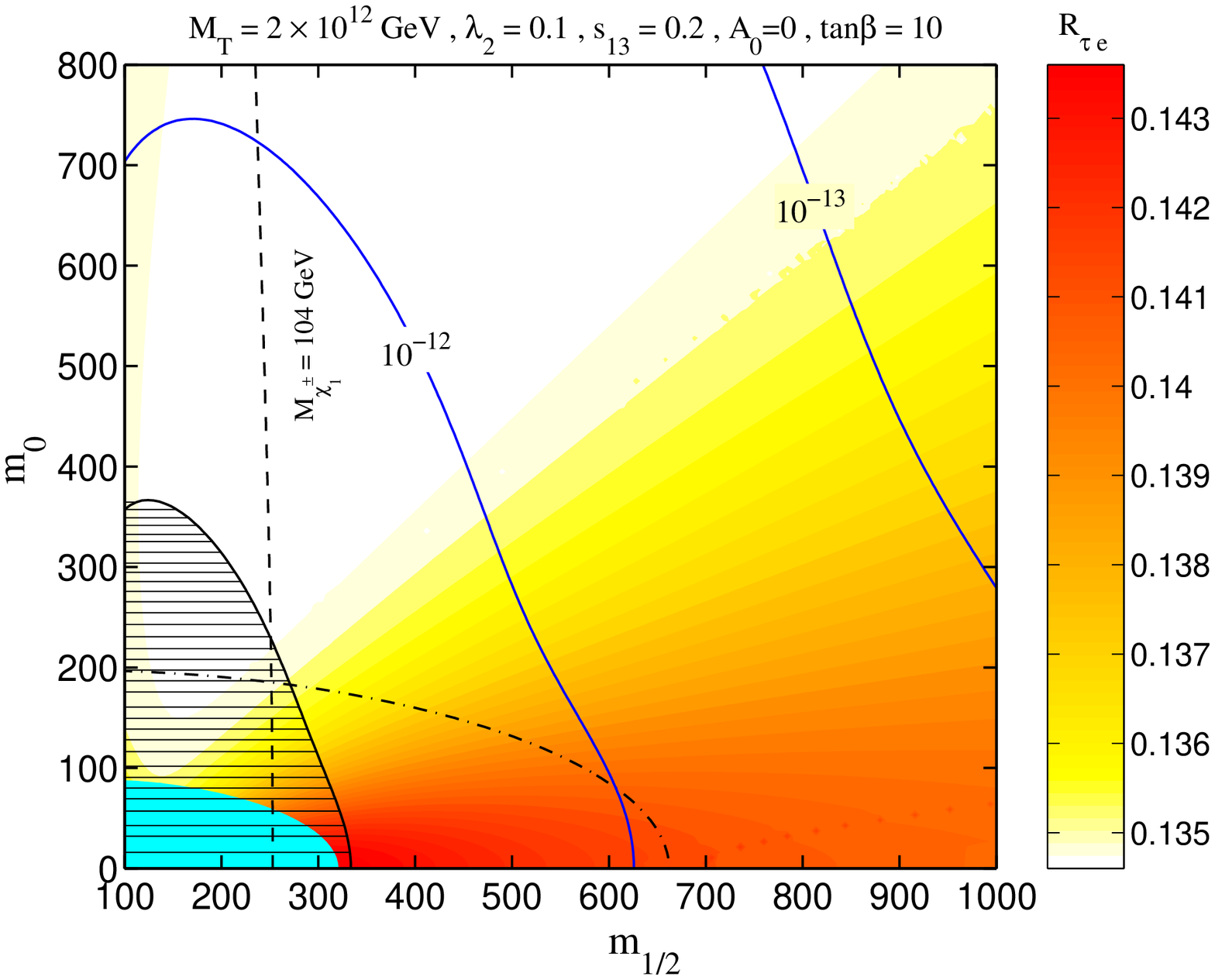}
\end{tabular}
}

In order to study the behaviour of $R_{\tau j}$ in the SUSY
$(m_{1/2},m_0)$ parameter space, we present two examples in
Figs.~\ref{fig8} and \ref{fig9} for $\tan\beta=10$ and 50,
respectively. Both figures show the variation of $R_{\tau\mu}$ (left
plots) and $R_{\tau e}$ (right plots) for $s_{13}(m_Z)=0$ (upper
plots) and $s_{13}(m_Z)=0.2$ (lower plots). The blue-solid lines
indicate the contours corresponding to ${\rm BR}(\mu \rightarrow e
\gamma)=10^{-12},10^{-13}$ and the hatched regions are excluded by the
MEGA bound ${\rm BR}(\mu \rightarrow e \gamma)< 1.2\times
10^{-11}$~\cite{Brooks:1999pu}. In cyan and light grey we show the
regions where the mass of the lightest slepton $m_{\tilde{\ell}_1}$
is below 100 GeV (the dashed-dotted line corresponds to
$m_{\tilde{\ell}_1}=200\,{\rm GeV}$) and the lightest supersymmetric
particle (LSP) is charged, respectively. To the left of the
black-dashed line the lightest chargino mass violates the LEP bound
$m_{\chi^{\pm}_1}>104$ GeV~\cite{Amsler:2008zzb}.

Fig.~\ref{fig8} shows that for $\tan\beta=10$ (moderate RG running
of the neutrino sector parameters and slepton mass splitting) the
ratios $R_{\tau\mu}$ and $R_{\tau e}$ vary in the range of $10\,\%$
at most, in the whole SUSY parameter space. Moreover, the
approximate estimates of these ratios given by Eqs.~(\ref{R32s130})
and (\ref{R31s130}) are in good agreement with the exact results.
\FIGURE[!ht]{ \label{fig9}\caption{The same as in Fig.~\ref{fig8}
for $\tan\beta=50$. Inside the light-grey regions the LSP is
charged.}
\begin{tabular}{cc}
\includegraphics[width=7.5cm]{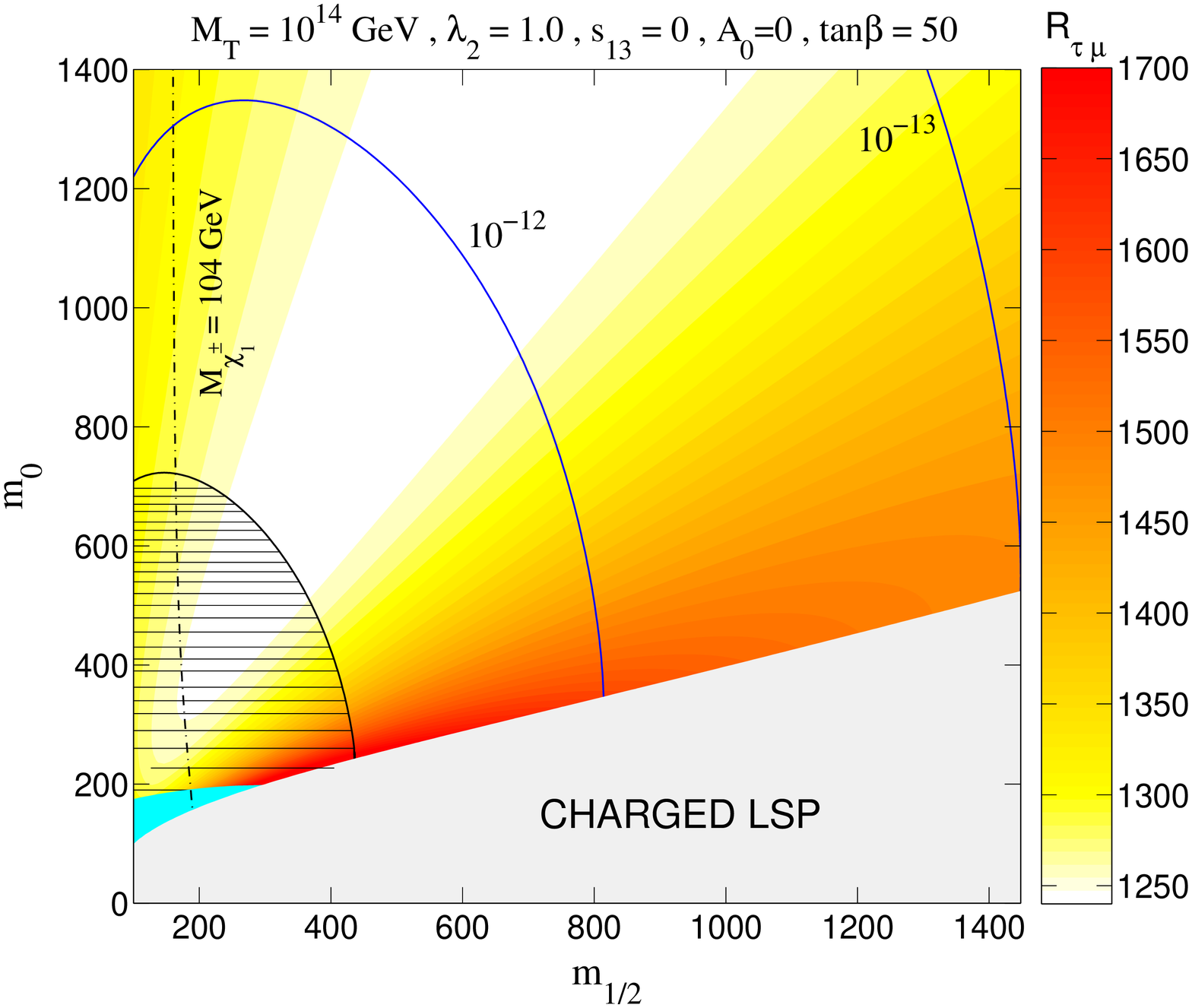}&
\hspace*{-0.5cm}\includegraphics[height=6.3cm,width=7.5cm]{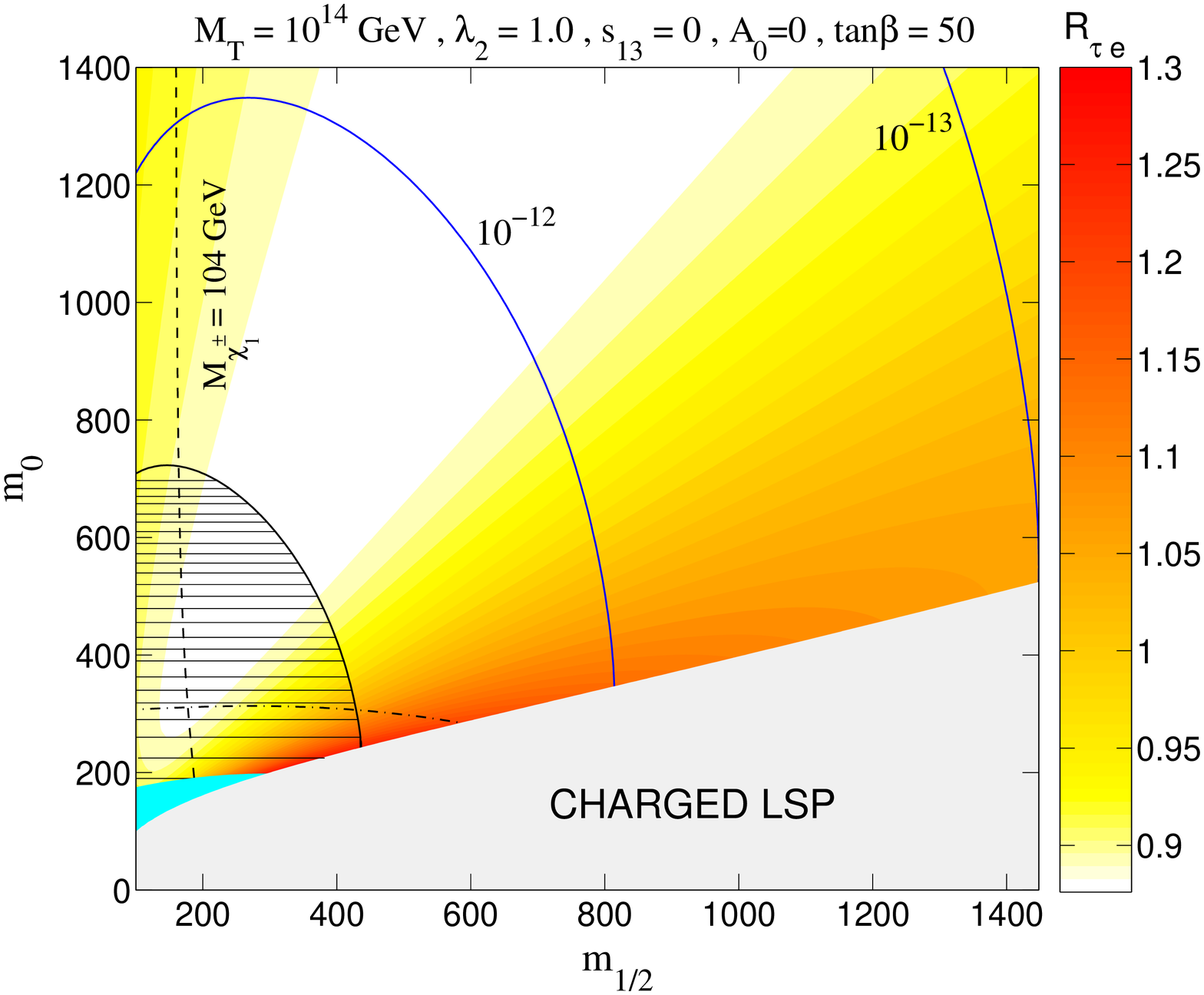}\\
\includegraphics[width=7.5cm]{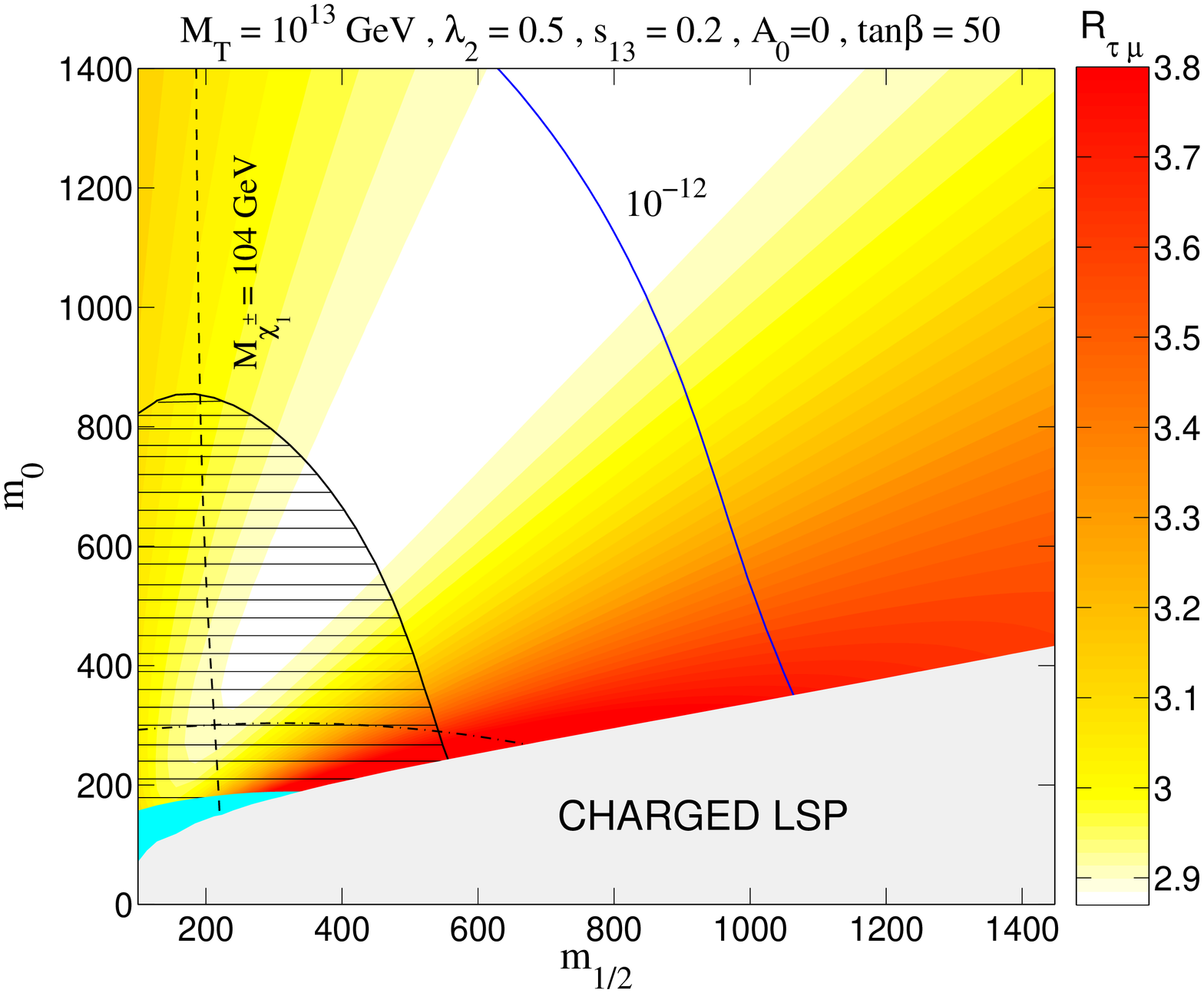}&
\hspace*{-0.5cm}\includegraphics[width=7.5cm]{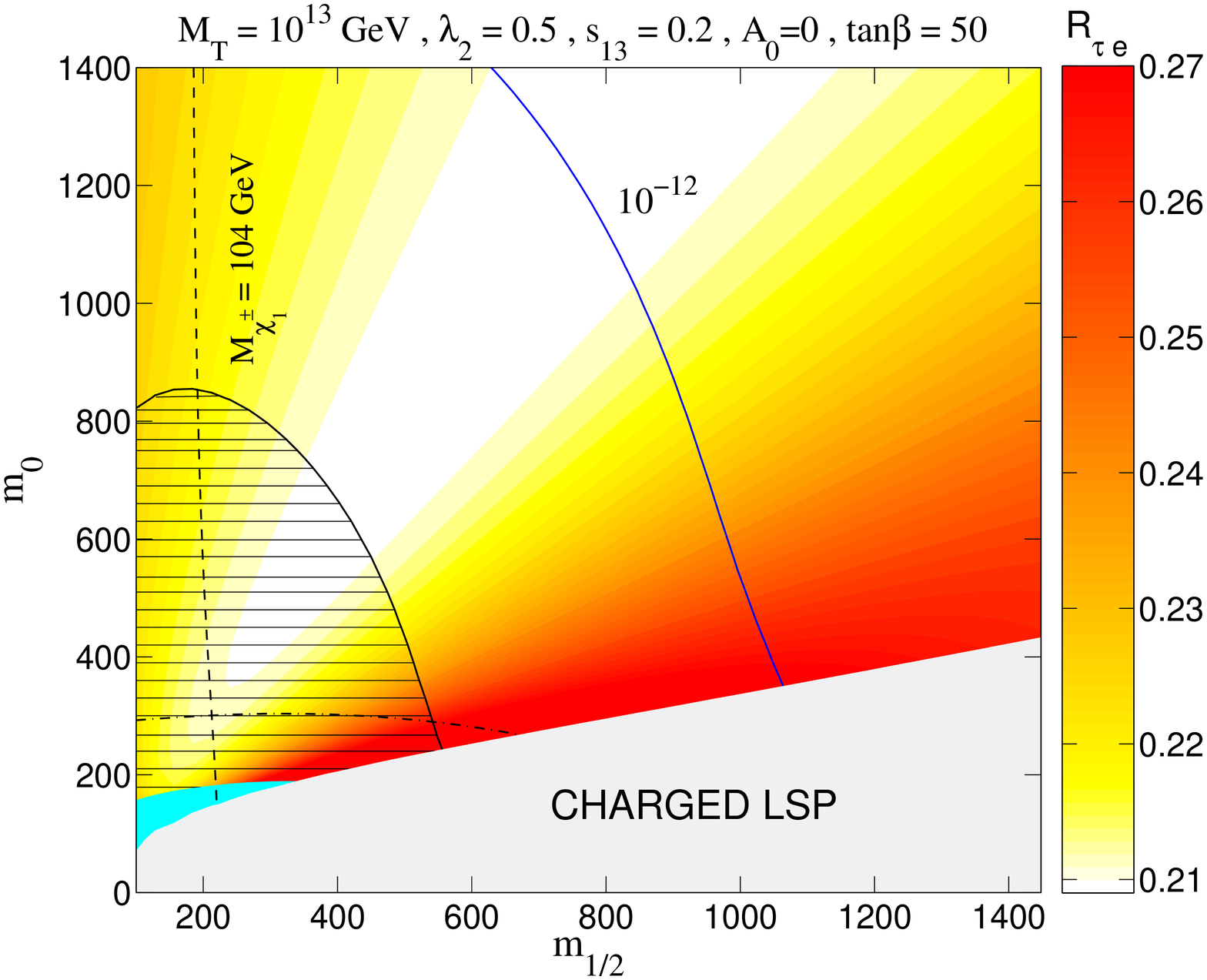}
\end{tabular}
} In contrast, for $\tan\beta=50$ (Fig.~\ref{fig9}) we see that the
ratios of BRs $R_{\tau j}$ are considerably enhanced for
$s_{13}(m_Z)=0$ due to the RG effects on $\theta_{13}$, as
previously discussed. Also, the variation of $R_{\tau\mu}$ and
$R_{\tau e}$ in the SUSY parameter space is now more pronounced due
to the larger mass-splitting induced in the slepton masses (see
discussion above). The observed enhancement with respect to the case
in which all sleptons are degenerate is due to the factor $\eta$
defined in Eq.~(\ref{RBRS2}) which depends on $m_{1/2}$ and $m_0$,
as seen in Fig.~\ref{fig9}.
%
%
%

\section{Summary and concluding remarks}
\label{sec6}

It is well known that RG effects may induce important corrections to
neutrino masses and mixing. This subject has been extensively
studied in the literature in the effective-theory framework and also
in the context of the type I seesaw mechanism. In this work we have
addressed this problem in a supersymmetric scenario where heavy
triplet states are added to the MSSM field content. We started by
obtaining the RGE for the effective neutrino mass operator in an
SU(5) model in which the triplet superfields can be accommodated in a
15-dimensional representation. The general expressions for the RGEs
of the neutrino mixing angles, masses and CP-violating phases were
also derived. Taking the pure type II seesaw case, we have analysed
both analytically and numerically the effect of the couplings $Y_T$
on the RG flow of the neutrino masses and mixing. Our results can be
summarised as follows:

\begin{itemize}

\item{We have pointed out some differences between the present results
and those previously obtained in Ref.~\cite{Schmidt:2007nq}. Apart
from discrepancies in the RGEs, we have shown that for energies
above $M_T$ the RG flow of the neutrino mixing angles and CP phases
does depend on the Majorana phases.}

\item{The RG-induced effects on the neutrino masses and mixing angles
due to the presence of the heavy triplet states become more relevant
as the size of the couplings $Y_T$ increases. These new effects are
not sensitive to the value of $\tan\beta$ and may be equally
relevant in a non-SUSY case. The running contributions controlled by
$Y_T$ are more important for the parameter $r$ (the ratio between
the solar and atmospheric neutrino mass squared differences) and the
atmospheric neutrino mixing angle. Regarding the running of the
CP-violating phases $\delta$ and $\alpha_{1,2}$, we have shown that
the RG effects induced by the couplings $Y_T$ are negligible.}

\item{Within the bottom-up approach, we have worked out some numerical
examples (for both HI and IH neutrino mass spectra) with the purpose
of quantifying the running effects on $r$ and $\theta_{23}$. We have
shown that if $y_i \sim \mathcal{O}(1)$, then the values of
$\theta_{23}$ and $r$ at the GUT scale are outside their present
low-energy $3\sigma$ intervals. This means that type II seesaw-based
models for neutrino masses and mixing with large $Y_T$ couplings
should not predict neutrino mass and mixing parameters at a high
scale in agreement with the low-energy data and, therefore, a
consistent RG analysis is demanded.}
\end{itemize}

The second part of this work has been devoted to the analysis of the
LFV charged-lepton radiative decays $\ell_i \rightarrow \ell_j
\gamma$ and their connection to low-energy neutrino data in the
framework of the TMSSM with universal boundary conditions for the
soft SUSY-breaking terms. After having updated the approximate
predictions for the ratios of BRs $ R_{\tau j} =\BR (\ell_i
\rightarrow \ell_j \gamma)/\BR (\mu \rightarrow e \gamma)$, we
compared them with the exact numerical results. The predictions
depend on the value of $\tan\beta$ and, of course, on the value of
the yet unknown neutrino parameters $\theta_{13}$ and $\delta$. To
summarise the results of this part, let us review the two extreme
limits of low and high $\tan\beta$.

\begin{itemize}
\item{{\bf Small tan}$\boldsymbol{\beta}$\\
If $\tan\beta$ is small, the RG-induced effects on the neutrino mass
and mixing parameters can be safely neglected, and the ratios of
branching ratios $R_{\tau j}$ $(j=e,\mu)$ only depend on two
parameters: $\delta$ and $\theta_{13}$. Furthermore, the approximate
results are in good agreement with the exact numerical ones. For
$\theta_{13}=0$, we have shown that the $3\sigma$ allowed interval
for $R_{\tau\mu}$ is $R_{\tau\mu}=[260\,(238) ,696\,(751)]$ (where
the numbers in parentheses refer to the IO neutrino mass spectrum
case). This means that, if $\BR(\mu\rightarrow e\gamma)\sim
10^{-11}$ (close to the present upper bound), then $\tau\rightarrow
\mu\gamma$ could be observed with a future sensitivity of $10^{-9}$,
reachable at the SuperKEKB~\cite{Akeroyd:2004mj} upgrade and
SUPERB~\cite{Bona:2007qt} flavour factories. However, if the bound
on $\BR(\mu\rightarrow e\gamma)$ is lowered to $10^{-12}$, then an
observation of $\tau\rightarrow \mu\gamma$ at the level of $10^{-9}$
or above would be in conflict with the predictions of the model for
small $\tan\beta$ and $\theta_{13}=0$. For the $\tau\rightarrow
e\gamma$ decay, its BR is too small to allow for its future
observation.

Considering now values of $\theta_{13}$ close to the experimentally
allowed upper bound ($\theta_{13}\simeq 0.2$), the predictions for
$R_{\tau \mu}$ show that an observation of
$\tau\rightarrow\mu\gamma$ above $10^{-9}$ would exclude the TMSSM
with universal boundary conditions for the SUSY breaking terms at
the GUT scale. This stems from the fact that for $\theta_{13}\gtrsim
0.1$ (and considering $\BR(\mu\rightarrow e\gamma) < 10^{-11}$),
${\rm BR}(\tau\rightarrow\mu\gamma) \lesssim 10^{-10}$. For the
$\tau\rightarrow e\gamma$ decay, its observation with
$\BR(\tau\rightarrow e\gamma) \gtrsim 10^{-9}$ would also be in
conflict with the TMSSM if $\theta_{13}$ is close to its upper
bound. Nevertheless, for $\theta_{13} \sim 0.015$ flavour
suppressions in the $\mu e$ and $\tau e$ slepton masses may occur
and, under special conditions, all the three LFV decays could be
observed.}
\item{{\bf Large tan}$\boldsymbol{\beta}$\\
We have concluded that if $\tan\beta$ is large the values of
$R_{\tau j}$ may deviate considerably from the approximate ones.
Moreover, in this $\tan\beta$ regime the splitting between the third
generation slepton masses and the remaining two introduces a
non-trivial dependence of $R_{\tau j}$ on the soft SUSY-breaking
masses. This is in contrast with the low $\tan\beta$ limit in which
the slepton mass splitting is much less pronounced. We have shown
that the running of the unknown mixing angle $\theta_{13}$ from the
electroweak to the heavy-triplet decoupling scale may affect the
relative magnitude of various LFV soft slepton masses, especially in
the case of a very small $\theta_{13}$. In short, this means that
ignoring the RG running in estimating the rates of the LFV processes
on the basis of the neutrino data may lead to misleading predictions
for the rates of the LFV decays. Therefore, for large $\tan\beta$,
the only way to obtain reliable results is to perform a complete
numerical calculation. In particular, $R_{\tau \mu}$ may be enhanced
by a factor of four (or even larger) when going from $\tan\beta=10$
to $\tan\beta=50$, for the same set of initial conditions (see
Figs.~\ref{fig8} and \ref{fig9}). Consequently, if $\tan\beta$ is
large (and $\theta_{13}$ is close to zero) ${\rm BR}
(\tau\rightarrow\mu\gamma)$ can reach the value of $10^{-9}$ (and,
therefore, be experimentally detectable) even if $\BR(\mu\rightarrow
e\gamma)$ is as low as $10^{-12}$. Still, the simultaneous
observation of $\mu\rightarrow e\gamma$ and any of the two radiative
$\tau$ decays would strongly disfavour the present scenario for
large $\theta_{13}$, even in the high $\tan\beta$ regime.}

\end{itemize}

In conclusion, we have shown that the RG corrections in the
framework of the TMSSM may be important when making predictions
regarding neutrino masses and mixings. These radiative effects
become more relevant when the couplings between the lepton doublets
and the heavy triplet are large. On the other hand, provided that
$\tan\beta$ is large, important effects may be observed on the LFV
decay rates even if the couplings $Y_T$ are small. In such a
situation, the knowledge of all low-energy neutrino parameters and
SUSY mass spectrum is crucial for an accurate prediction of the LFV
decay rates. Nevertheless, the approximate results for the ratios of
BRs in distinct channels are reliable if $\tan\beta$ is not too
large.

\acknowledgments

We are grateful to M. Maltoni and C. Savoy for discussions and to A.
Rossi and A.~M.~Teixeira for the careful reading of the manuscript
and the numerous comments and suggestions. The author also thanks M.
A. Schmidt for several private communications regarding the results
obtained in Ref.~\cite{Schmidt:2007nq}.


\end{document}